\documentclass[superscriptaddress,showpacs,amsmath,amssymb,aps,prd,twocolumn,floatfix,nofootinbib,letterpaper]{revtex4-1}

\usepackage{graphicx}
\graphicspath{{EVD/} {Pi0Plots/} {1e1p_plots/} {1e1p_plots/AnalysisRange_v48/} {1e1p_plots/Misc/} {1m1p_plots/} {Michels/} {NumuBkgFit/} {StatPlots/}{SysPlots/} {1e1p_plots/AnalysisRange_v48/AfterRefereeComments/}}
\usepackage{dcolumn}
\usepackage{bm}
\usepackage{color}
\usepackage{amsmath}
\usepackage{amssymb}
\usepackage{bbm}
\usepackage{subfig}
\usepackage{hyperref}
\usepackage{hhline}
\usepackage[percent]{overpic}
\usepackage{layouts}
\usepackage{float}
\usepackage{comment}
\usepackage{graphicx}
\usepackage{adjustbox}
\usepackage{placeins}
\usepackage{amssymb}
\usepackage{siunitx}
\usepackage{rotating}
\usepackage[justification=raggedright]{caption}

\newcommand{\sssnet}{\textsc{SparseSSNet}}
\usepackage{multirow}
\usepackage{float}
\usepackage{hyperref} 
\usepackage{comment}
\usepackage[normalem]{ulem}
\usepackage{lineno}

\begin{document}

\title{Search for an anomalous excess of charged-current quasi-elastic $\nu_e$ interactions with the MicroBooNE experiment using Deep-Learning-based reconstruction}

\newcommand{\Bern}{Universit{\"a}t Bern, Bern CH-3012, Switzerland}
\newcommand{\BNL}{Brookhaven National Laboratory (BNL), Upton, NY, 11973, USA}
\newcommand{\UCSB}{University of California, Santa Barbara, CA, 93106, USA}
\newcommand{\Cambridge}{University of Cambridge, Cambridge CB3 0HE, United Kingdom}
\newcommand{\CIEMAT}{Centro de Investigaciones Energ\'{e}ticas, Medioambientales y Tecnol\'{o}gicas (CIEMAT), Madrid E-28040, Spain}
\newcommand{\Chicago}{University of Chicago, Chicago, IL, 60637, USA}
\newcommand{\Cincinnati}{University of Cincinnati, Cincinnati, OH, 45221, USA}
\newcommand{\CSU}{Colorado State University, Fort Collins, CO, 80523, USA}
\newcommand{\Columbia}{Columbia University, New York, NY, 10027, USA}
\newcommand{\Edinburgh}{University of Edinburgh, Edinburgh EH9 3FD, United Kingdom}
\newcommand{\FNAL}{Fermi National Accelerator Laboratory (FNAL), Batavia, IL 60510, USA}
\newcommand{\Granada}{Universidad de Granada, Granada E-18071, Spain}
\newcommand{\Harvard}{Harvard University, Cambridge, MA 02138, USA}
\newcommand{\IIT}{Illinois Institute of Technology (IIT), Chicago, IL 60616, USA}
\newcommand{\KSU}{Kansas State University (KSU), Manhattan, KS, 66506, USA}
\newcommand{\Lancaster}{Lancaster University, Lancaster LA1 4YW, United Kingdom}
\newcommand{\LANL}{Los Alamos National Laboratory (LANL), Los Alamos, NM, 87545, USA}
\newcommand{\Manchester}{The University of Manchester, Manchester M13 9PL, United Kingdom}
\newcommand{\MIT}{Massachusetts Institute of Technology (MIT), Cambridge, MA, 02139, USA}
\newcommand{\Michigan}{University of Michigan, Ann Arbor, MI, 48109, USA}
\newcommand{\Minnesota}{University of Minnesota, Minneapolis, MN, 55455, USA}
\newcommand{\NMSU}{New Mexico State University (NMSU), Las Cruces, NM, 88003, USA}
\newcommand{\Oxford}{University of Oxford, Oxford OX1 3RH, United Kingdom}
\newcommand{\Pitt}{University of Pittsburgh, Pittsburgh, PA, 15260, USA}
\newcommand{\Rutgers}{Rutgers University, Piscataway, NJ, 08854, USA}
\newcommand{\SLAC}{SLAC National Accelerator Laboratory, Menlo Park, CA, 94025, USA}
\newcommand{\SDSMT}{South Dakota School of Mines and Technology (SDSMT), Rapid City, SD, 57701, USA}
\newcommand{\Maine}{University of Southern Maine, Portland, ME, 04104, USA}
\newcommand{\Syracuse}{Syracuse University, Syracuse, NY, 13244, USA}
\newcommand{\TelAviv}{Tel Aviv University, Tel Aviv, Israel, 69978}
\newcommand{\Tennessee}{University of Tennessee, Knoxville, TN, 37996, USA}
\newcommand{\UTA}{University of Texas, Arlington, TX, 76019, USA}
\newcommand{\Tufts}{Tufts University, Medford, MA, 02155, USA}
\newcommand{\VTech}{Center for Neutrino Physics, Virginia Tech, Blacksburg, VA, 24061, USA}
\newcommand{\Warwick}{University of Warwick, Coventry CV4 7AL, United Kingdom}
\newcommand{\Yale}{Wright Laboratory, Department of Physics, Yale University, New Haven, CT, 06520, USA}

\affiliation{\Bern}
\affiliation{\BNL}
\affiliation{\UCSB}
\affiliation{\Cambridge}
\affiliation{\CIEMAT}
\affiliation{\Chicago}
\affiliation{\Cincinnati}
\affiliation{\CSU}
\affiliation{\Columbia}
\affiliation{\Edinburgh}
\affiliation{\FNAL}
\affiliation{\Granada}
\affiliation{\Harvard}
\affiliation{\IIT}
\affiliation{\KSU}
\affiliation{\Lancaster}
\affiliation{\LANL}
\affiliation{\Manchester}
\affiliation{\MIT}
\affiliation{\Michigan}
\affiliation{\Minnesota}
\affiliation{\NMSU}
\affiliation{\Oxford}
\affiliation{\Pitt}
\affiliation{\Rutgers}
\affiliation{\SLAC}
\affiliation{\SDSMT}
\affiliation{\Maine}
\affiliation{\Syracuse}
\affiliation{\TelAviv}
\affiliation{\Tennessee}
\affiliation{\UTA}
\affiliation{\Tufts}
\affiliation{\VTech}
\affiliation{\Warwick}
\affiliation{\Yale}

\author{P.~Abratenko} \affiliation{\Tufts} 
\author{R.~An} \affiliation{\IIT}
\author{J.~Anthony} \affiliation{\Cambridge}
\author{L.~Arellano} \affiliation{\Manchester}
\author{J.~Asaadi} \affiliation{\UTA}
\author{A.~Ashkenazi}\affiliation{\TelAviv}
\author{S.~Balasubramanian}\affiliation{\FNAL}
\author{B.~Baller} \affiliation{\FNAL}
\author{C.~Barnes} \affiliation{\Michigan}
\author{G.~Barr} \affiliation{\Oxford}
% ...

\collaboration{The MicroBooNE Collaboration}
\thanks{microboone\_info@fnal.gov}\noaffiliation

\begin{abstract}

We present a measurement of the $\nu_e$-interaction rate in the MicroBooNE detector that addresses the observed MiniBooNE anomalous low-energy excess (LEE).
The approach taken isolates neutrino interactions consistent with the kinematics of charged-current quasi-elastic (CCQE) events. The topology of such signal events has a final state with 1 electron, 1 proton, and 0 mesons ($1e1p$).
Multiple novel techniques are employed to identify a $1e1p$ final state, including particle identification that use two methods of deep-learning-based image identification, and event isolation using a boosted decision-tree ensemble trained to recognize two-body scattering kinematics.
This analysis selects 25 $\nu_e$-candidate events in the reconstructed neutrino energy range of 200--1200\,MeV,  while $29.0 \pm 1.9_\text{(sys)} \pm 5.4_\text{(stat)}$ are predicted when using $\nu_\mu$ CCQE interactions as a constraint.
We use a simplified model to translate the MiniBooNE LEE observation into a prediction for a $\nu_e$ signal in MicroBooNE. A $\Delta \chi^2$ test statistic, based on the combined Neyman--Pearson $\chi^2$ formalism, is used to define frequentist confidence intervals for the LEE signal strength. Using this technique, in the case of no LEE signal, we expect this analysis to exclude a normalization factor of 0.75 (0.98) times the median MiniBooNE LEE signal strength  at 90\% ($2\sigma$) confidence level, while the MicroBooNE data yield an exclusion of 0.25 (0.38) times the median MiniBooNE LEE signal strength at 90\% ($2\sigma$) confidence level.

\end{abstract}

\maketitle

\section{Introduction}
\label{sec:new_intro}
We report the results of an analysis of data taken with the MicroBooNE liquid argon time projection chamber (LArTPC) that addresses an unexpected excess of electron-like events observed in the MiniBooNE detector~\cite{MiniBooNE:2018esg}.
The high-resolution particle-imaging within the 85-metric-ton active volume of the MicroBooNE LArTPC~\cite{MicroBooNE:2016pwy} allows for detailed reconstruction of particle production at the vertex that is not possible using the MiniBooNE Cherenkov detector. This allows a clean test of the hypothesis that the MiniBooNE excess events are due to $\nu_e$ charged-current quasi-elastic (CCQE) interactions ($\nu_e + n \rightarrow e^- + p$).

The MiniBooNE ``low-energy excess'' (LEE) is a 4.8$\sigma$ enhancement of events identified as $\nu_e$ interactions observed primarily between 200--500\,MeV in data taken from 2002--2019~\cite{MiniBooNE:2020pnu}. Although MiniBooNE interprets the events under a $\nu_e$ CCQE hypothesis, the event information is limited to observation of a single Cherenkov ring consistent with an electromagetic shower. 
The LEE can be interpreted in many ways, including as a signal for $\nu_\mu \rightarrow \nu_e$ oscillations with $\Delta m^2 \sim 1(\text{eV/c}^2)^2$~\cite{Diaz:2019fwt},  production of new particles with mass of ${\sim}100\,\text{MeV}/c^2$  that decay to photons~\cite{Magill:2018jla, Fujikawa:1980yx,Palazzo:2007gz,Shrock:1982sc,Dvornikov:2003js,Giunti:2014ixa,Brdar:2020quo}, or both~\cite{Vergani:2021tgc}.  Explanations not invoking new physics include distortions of the energy spectrum from mismodeled meson-exchange currents (MEC)~\cite{Ericson:2016yjn}, and single-photon decays of the $\Delta$ baryon resonance~\cite{Alvarez-Ruso:2015kua}.

Because of the wide range of explanations,  the MicroBooNE experiment has developed four distinct searches targeting the LEE signal:  
1) an exclusive search using the $1e1p$ CCQE channel (this paper); 
2) a semi-inclusive search for events with one electron and no pions in the final state~\cite{PeLEE};  
3) a fully inclusive search for events with one electron and any final state~\cite{WCLEE}; 
and 4) a photon-based-search with a signal focusing on the $\Delta$ radiative-decay hypothesis~\cite{MicroBooNE:2021zai}.  
The results of the first three searches, including this analysis, are considered together in~Ref.~\cite{jointMicroBooNEPRL}.

\begin{figure*}[tb]
\centering
\includegraphics[width=0.85\textwidth]{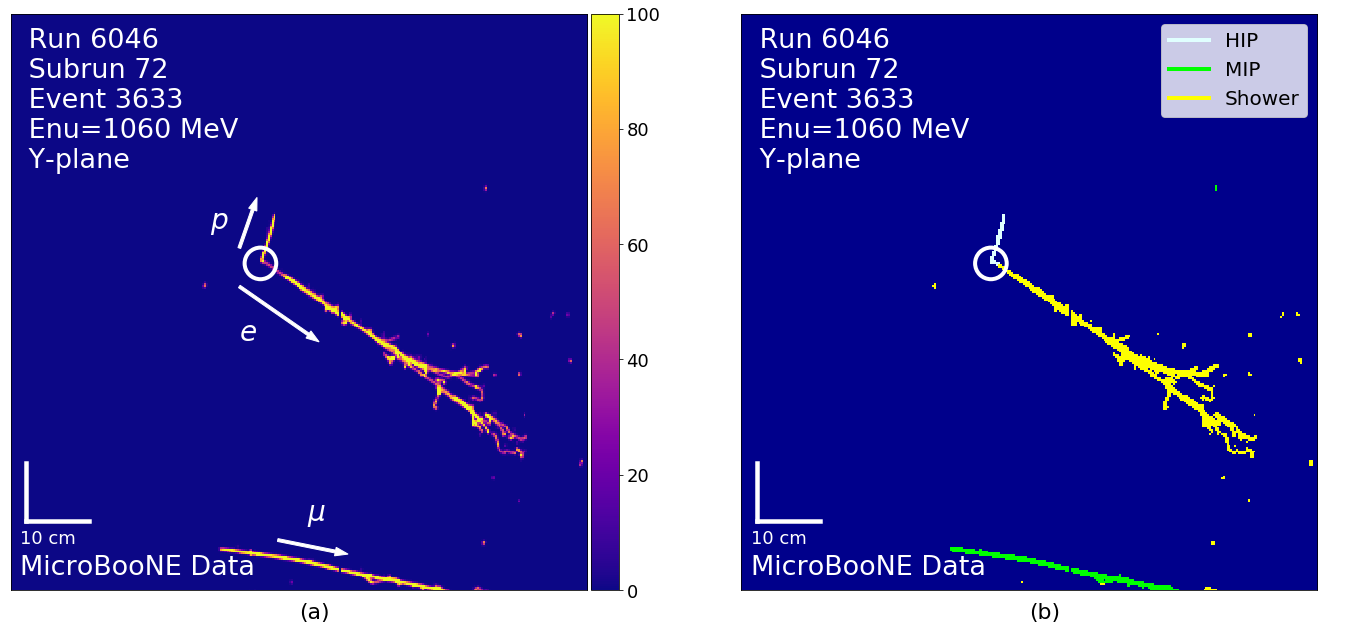}
\caption{A typical selected data event from this analysis, showing time tick vs wire number from the TPC. a), the pixel intensity images with annotation indicating the electron and proton. b), pixel labeling from the Deep-Learning-based semantic segmentation algorithm, discussed in Sec.~\ref{sec:reco}.}
\label{fig:eventexample}
\end{figure*}

\section{Synopsis of this Analysis}
\label{sec:intro} 
 
This paper reports the detailed results of the MicroBooNE search that uses the $1e1p$ CCQE exclusive channel and that is based upon the application of Deep Learning (DL) techniques. However, before describing this study in-depth, in this section we provide a synopsis of the analysis to guide the reader on this approach.
 
In Sec.~\ref{sec:uboone}, we describe the experiment.   
The MicroBooNE detector is located in the Booster Neutrino Beam (BNB) at Fermi National Accelerator Laboratory, 468.5\,m from the BNB beryllium target, 72.5~m upstream of the MiniBooNE detector. 
Collection of ionization electrons from charged particles passing through the argon in the LArTPC allows detailed reconstruction of the final state, including measurements of deposited energy and the particle identification.  
The detector provides detailed information on the events of interest: $\nu_e + n \rightarrow e^- + p$ interactions with minimal initial neutron momentum and no final state interactions. The LArTPC information is used to identify events having one electromagnetic shower 
and one proton track, forming a vertex with no gap between the two particles.  Fig.~\ref{fig:eventexample} (a),  shows an event display of a data event selected by this analysis, with the electron and proton annotated.   

As described in Sec.~\ref{sec:sig}, we use the unfolded median of the MiniBooNE LEE result to rescale the $\nu_e$ flux in order to simulate the signal. This populates neutrino energies from 200--500\,MeV, referred to as the ``LEE range,'' where  
CCQE interactions dominate.  Using visible energy and track angle relative to the beam direction ($z$-axis), many kinematic quantities can be reconstructed that will have specific correlations for well-reconstructed CCQE events, but not for most background events, allowing for a unique method of signal isolation. We use well-reconstructed events with kinematics that satisfy two-body scattering expectations as our operational definition for CCQE for this analysis.

In Sec.~\ref{sec:reco}, we present the reconstruction, with emphasis on how this analysis addresses application of DL to reconstruction of LArTPC data. As seen in Fig.~\ref{fig:eventexample}, the TPC data are represented as two-dimensional images, with wire number along the $x$ axis and drift time along the $y$ axis. Each bin represents a  ``pixel,'' where the intensity is the integrated reconstructed charge waveform over six time ticks (corresponding to a drift distance of $\approx0.33$\,cm ) after applying noise filtering~\cite{MicroBooNE:2017qiu} and signal processing~\cite{MicroBooNE:2018swd,MicroBooNE:2018vro}. 
The individual pixels of the image can be ``semantically segmented,'' or labeled, using a convolutional neural network (CNN)~\cite{MicroBooNE:2018kka,MicroBooNE:2020yze,MicroBooNE:2016dpb}.  Semantic segmentation is a well-known technique in the computer vision community. 
Fig.~\ref{fig:eventexample} (b) demonstrates the application of this algorithm, with pixels labeled as originating from minimum ionizing particles (MIP), highly ionizing particles (HIP), or as electromagnetic ``showers.''   This algorithm is applied as the first step of the event reconstruction. The results are passed into 
conventional algorithms for vertex-finding, track reconstruction~\cite{MicroBooNE:2020sar},  and shower clustering~\cite{MicroBooNE:2021nss}.  Lastly, we apply Multi-Particle IDentification (MPID)~\cite{MicroBooNE:2020hho}, a second DL algorithm that performs multiple-object classification on the whole intensity image.
This algorithm outputs scores indicating whether an image is consistent with containing one or more electrons, protons, muons, pions, or photons. 
MPID can be thought of as the complement of semantic segmentation--in this case, neither individual nor groups of pixels are labeled; instead holistic information on the contents of the image are reported.

In Sec.~\ref{sec:selects}, we explain signal and constraint-sample selection and background rejection.  We employ an ensemble of boosted decision trees (BDTs) that test for CCQE-consistent two-body-scattering kinematics.  The ensemble method trains multiple BDTs using the same input variables, but with different sets of training events, using the XGBoost~\cite{Chen_2016} gradient-boosting algorithm.  In principle, each BDT is identifying the same broad features characteristic of signal events, but in practice, the scores are not identical due to the individuality of each training sample. Taking the average score, referred to as the ``BDT score,'' as the selection variable reduces variance in the result. 
The ensemble method is a new approach to BDTs in particle physics, although it has been applied in other fields, including  medicine~\cite{8855897} and climate studies~\cite{Liu_2019} because it provides a very stable result. The combination of the BDT ensemble and other data selection criteria results in a 75\% $\nu_e$ CCQE purity in the ``analysis energy range'' of 200--1200\,MeV.

Although the CCQE interaction is a well-understood neutrino interaction~\cite{NuSTEC:2017hzk} in the analysis energy range, the uncertainties from the predicted cross section are, nevertheless, $\sim 15\%$~\cite{MicroBooNE:2021ldh,MicroBooNE:2021genie_tune}.  When combined with the $\nu_e$ flux uncertainties ($\sim 10\%$)~\cite{MiniBooNE:2008hfu, MicroBooNE:2019nio}, the  predicted rate is only known to the $\sim 20\%$ level, potentially obscuring a signal.   This problem was recognized by the MiniBooNE analyzers, who introduced the use of the $\nu_\mu$ data to constrain the $\nu_e$ rate~\cite{MiniBooNE:2007uho}.   This technique succeeds because $1\mu 1p$ CCQE  cross section systematic uncertainties are highly correlated to the $1e1p$ CCQE uncertainties, and the fluxes are connected through the meson/muon decay chains at production, assuming no $\nu_\mu$ disappearance due to oscillations. As a result, the most important control sample, or ``sideband,'' in this analysis is the $1\mu 1p$ CCQE event sample.

In Sec.~\ref{sec:sideBand}, we consider other dedicated data sets for studies of the reconstruction and selection.   A sample 
of neutral current $\pi^0$ (NC$\pi^0$) events in which the photon converts early, leaving 
no gap at the interaction vertex,
constrains the rate of $\pi^0$ backgrounds in the $1e1p$ CCQE analysis.    
Comparison of the rates of these events in data and in simulation demonstrates that the reconstruction efficiency is well understood. 
The data-to-simulation agreement of the observed rate of vertices formed by Michel electrons and muons in $\nu_\mu$ events gives further confidence that the reconstruction efficiency is understood, even at low electron energies. Lastly, the $\pi^0$ and Michel electron event samples were used to establish the conversion from the intensity of each shower pixel to energy in MeV (see Ref.~\cite{MicroBooNE:2021nss} and Sec.~\ref{sec:shower}).

In Sec.~\ref{sec:unc}, we review the systematic uncertainties associated with the analysis. The flux prediction and uncertainties are from Ref.~\cite{MiniBooNE:2008hfu} with minor changes~\cite{MicroBooNE:2019nio}.  The uncertainties on the tuned GENIE~\cite{Andreopoulos:2015wxa} model of the cross section are discussed in Ref.~\cite{MicroBooNE:2021ldh, MicroBooNE:2021genie_tune}. While all cross-section uncertainties are included, the discussion in this section focuses on those most relevant to this search.  MicroBooNE has developed a novel technique for determining detector uncertainties related to the TPC wire and electronics response~\cite{MicroBooNE:2021wiremod}.  To reduce statistical fluctuations in the Monte Carlo (MC) event samples used to construct the covariance matrix, this analysis employed a new approach---a kernel density estimator (KDE) smoothing algorithm~\cite{KDEcite}. The covariance matrix that includes the correlations in the allowed $1\sigma$ variations between the $1e1p$ and $1\mu 1p$ is presented in this section.

The steps to reach the final result are described in Sec.~\ref{sec:finalpred}.    
Within the analysis energy range (200--1200\,MeV), the data in the LEE range (200-500\,MeV) were sequestered from the analyzers until the final step of the analysis; consequently we describe these events as ``blinded.''   
The remaining CCQE $1e1p$ event samples were made available for study sequentially 
by first opening a ``far energy range'' of 700--1200\,MeV, 
and then a ``near energy range'' of 500--700\,MeV.
Prior to examining the LEE region, the $1\mu1p$ (one muon, one proton and no mesons) constrained prediction was established. 
The constraint leads to a small change in the prediction, but to a substantial reduction in systematic uncertainty. 

The statistical tests used to examine agreement between predictions and data are described in Sec.~\ref{sec:stat_test}.
In short, we compare to: 1) the constrained prediction based on the $1\mu1p$ sample, defined as background only (no LEE signal), and 2) the median LEE model described in Sec.~\ref{sec:sig}.
The primary test statistic used to quantify comparisons between the observation and the predictions is the Combined Neyman--Pearson $\chi^2$ ($\chi^2_\text{CNP}$)~\cite{Ji:2019yca}.

In Sec.~\ref{sec:res}, we explore this final result in detail.  This section provides data-to-simulation comparisons of various kinematic variables of $1e1p$ CCQE events, as well as an example of an event display from one of the selected $\nu_e$-candidates.   The rest of the event displays of the selected $\nu_e$-candidates are provided in App.~\ref{app:evtbyevt}.

\section{The MicroBooNE Experiment, Data Sets, and Simulation}
\label{sec:uboone}

The MicroBooNE neutrino flux~\cite{MiniBooNE:2008hfu,MicroBooNE:2019nio} is generated by the BNB by protons with kinetic energy of 8\,GeV impinging on a beryllium target, producing predominantly charged pions and kaons that decay in flight.
Protons are delivered in spills of $\sim 1.6$\,$\mu s$ duration consisting of 82 bunches of 2\,ns width each, with an intensity of $4\times10^{12}$ protons on target (POT).
This analysis makes use of $6.88\times 10^{20}$ protons that were delivered to 
a beryllium target between 2016 and 2018 in three runs, 
with $1.75\times 10^{20}$ POT in Run 1,  $2.70\times 10^{20}$ POT in Run 2 and $2.43\times 10^{20}$ POT in Run 3.
The beryllium target for meson production is located inside of the magnetic focusing horn.  For   
this data set, running was taken with magnet polarity that focuses positively charged mesons toward the detector. These 
subsequently decay to neutrinos in a 
50\,m steel decay pipe filled with air that terminates in a beam stop made of steel and concrete, followed by largely undisturbed earth.  
Within the analysis energy range, the neutrino flux is dominated by muon neutrinos (93.6\% $\nu_\mu$ and 5.9\% $\bar{\nu}_\mu$) with a small component of electron neutrinos (0.52\% $\nu_e$ and 0.05\% $\bar{\nu}_e$) referred to as intrinsic $\nu_e$. The mean  $\nu_\mu$ energy is $\sim 800$\,MeV. 

The beam has an ``intrinsic'' component of $\nu_e$---electron neutrinos produced by Standard Model processes in the beam line.  
These intrinsic $\nu_e$ form an irreducible background to this search.   The majority of the $\nu_e$ relevant to this analysis are produced through the decay chain of $\pi^+ \rightarrow \nu_\mu \mu^+$, followed by $\mu^+ \rightarrow e^+ \bar \nu_\mu \nu_e$.  
The remainder are intrinsic $\nu_e$ and $\bar \nu_e$ from three-body kaon decays. 
The MicroBooNE flux simulation is described in Ref.~\cite{MicroBooNE:2019nio}.

The detector is described in detail in~\cite{MicroBooNE:2016pwy}, and in this section we review those features salient to this analysis.  MicroBooNE employs a right-handed coordinate system to describe the detector, with $x$-axis toward the cathode, $y$-axis up, and $z$-axis downstream from the beam direction.  The angle $\theta$ is measured from the $z$-axis, and $\phi$ is measured from the $x$-$z$ plane.

The detector system consists of two subdetectors within an argon-filled volume:  a TPC for tracking and a light collection system. 
The neutrinos interact with the argon nuclei in the detector. The resulting charged particles traverse the liquid argon, producing scintillation light and ionization electrons. 

The light collection system is used to detect if activity occurs in the detector during the beam spill. 
Light is collected by 32 ``optical units'' in Run 1 and 31 optical units in Runs 2 and 3. 
Each optical unit consists of an 8-inch Hamamatsu R5912-02mod cryogenic photomultiplier tube (PMT) located behind an acrylic plate coated with tetraphenyl butadiene to shift the 128 nm scintillation photons to the visible range. 
The light is collected in samples, or ``time ticks,'' of 15.625\,ns each.  
A minimum threshold of 3.5 photoelectrons detected in six time tick ($\approx100$\,ns) is required to trigger and record events.
A further ``common optical filter'' threshold is applied, requiring $>20$ photoelectrons detected within any 6 consecutive ticks in the beam spill window, and $\le 20$ photoelectrons detected within 6 ticks in the 2\,$\mu s$ period prior to the beam window. This common optical filter is applied to each of the MicroBooNE LEE analyses~\cite{MicroBooNE:2021zai,WCLEE,PeLEE}, in order to reduce non-neutrino triggers.

Ionized electrons collected by the TPC are used for particle tracking, energy determination, and particle identification.   
Protons at MicroBooNE energies will be highly ionizing, muons will be minimally ionizing, and electrons will have radiative (``shower'') behavior, even at the lowest energies, since 
the critical energy of liquid argon is 39\,MeV~\cite{MicroBooNE:2019rgx}.
At low energies, electron tracks also can be distinguished from muon tracks due to the enhanced level of multiple Coulomb scattering observed.

The TPC detector has an active region of $2.6\times 2.3 \times 10.4$~m$^3$ containing about 85~metric tons of liquid argon. 
The TPC comprises three key elements: the anode wires, cathode plane, and field cage. In the coordinate system used in this analysis, the nominal drift direction is parallel to the x-axis and the beam direction is parallel to the z-axis.
The ionized electrons drift in an applied electric field to three wire planes that provide the charge read-out.  The wire angles are $\pm60^\circ$ from vertical for the $U$ and $V$ planes and vertical for the $Y$ plane; the wire spacing is 3\,mm.
The $U$ and $V$ planes detect signal via induction, while the $Y$ plane, which is farthest from the cathode, collects the charge.  The $Y$ plane generally has the largest signal, hence the best resolution when charge is converted to energy.   

The applied electric field is 273\,V/cm, leading to $\approx0.11$\,cm/$\mu$s electron drift velocity.  The data acquisition system triggers on each beam spill with a readout window of 4.8\,ms, during which 9600 time ticks are recorded.  Therefore, the drift distance spanned over six time ticks is about 0.33\,cm, which is roughly equivalent to the detector's 0.3\,cm wire pitch. This determines the size of a pixel when the data are converted to images for use by the semantic segmentation algorithm described in Sec.~\ref{sec:sssnet}.

Several issues limit the quality of the data.
The long wires of the LArTPC are subject to pickup noise. This is removed through noise filtering and signal processing, which extracts the real signal from the measured output, given the detector response.   
Some transient noise remains, especially in the $U$ plane--a feature that affects data selection criteria for this analysis.
The detector also suffers from $\sim10\%$ unresponsive wires. 
These unresponsive regions will be visible in the event displays presented in this paper (see example in Fig~\ref{fig:evd10}, middle row). The analysis makes use of the MicroBooNE ``Good Runs List'' which removes periods of poor detector response from noise, etc., from the data sample. The resulting good runs are used for all analyses.   Lastly, 20 to 30 cosmic rays enter the detector per TPC readout window because MicroBooNE is a near-surface detector.   Interference between cosmic rays and tracks from neutrino events can reduce efficiency of reconstruction.

In order to accurately describe the cosmic rays, noise, and unresponsive regions in the detector as a function of time in the simulation, the analysis overlays beam-off data taken throughout the three run periods onto simulated neutrino events.  
Events are simulated throughout the total volume of the liquid argon, allowing particles produced upon interactions with argon outside of the active volume to enter it, where they may be mistakenly reconstructed as a neutrino interaction. However, because of the highly specialized signature of this analysis and the excellent tools for cosmic background removal, the background rate from these events is found to be negligible. Therefore, we explicitly neglect external events in this analysis.

For generating simulated events, we employ GENIE~\cite{Andreopoulos:2009rq,Andreopoulos:2015wxa,GENIE:2021npt,Tena-Vidal:2021rpu} v3.00.06 and model set G18\textunderscore10a\textunderscore02\textunderscore11a as the primary model.  This generator features the Valencia CCQE and MEC (2p2h) models~\cite{ValenciaModel1} and the Local Fermi Gas nuclear model. These were found to give a good description of the MiniBooNE CCQE-like data~\cite{miniboone-ccqe}.  The generator also has an improved data-driven final state interaction model and a new tune to bubble chamber data for pion production.  To further improve the description, the collaboration undertook a tuning effort utilizing the T2K $\nu_\mu$ charged current, zero-pion data~\cite{t2k2016}, which is fully described in Ref.~\cite{MicroBooNE:2021genie_tune}. The propagation of particles is simulated using the Geant4 toolkit~\cite{Agostinelli:2002hh} V10.3.03c.

\section{The Analysis Goal and Approach}
\label{sec:sig}

The goal of this analysis is to test whether the MiniBooNE low-energy excess is due to an anomalously high rate of low-energy $\nu_e$ charged current (CC) interactions.  
This section describes the strategy used to test this hypothesis. If a signal is observed, then two alternative explanations are rejected:  1) the MiniBooNE LEE events are due to an unsimulated non-$\nu_e$-CC signal, for example an additional source of photons;  and
2) the events are due to sources of $\nu_e$-CC events included at an incorrect rate in the MiniBooNE simulation, for example MEC interactions.  

\subsection{The MicroBooNE simplified test model for the MiniBooNE LEE}
\label{sec:lee_model}

A phenomenological model for the LEE signal in MicroBooNE is needed for several purposes.   
The most important is to define the signal region to be blinded.    
A model also provides a benchmark for the analysis, in order to understand the potential signal-to-background ratio as the analysis develops.   
However, obtaining a good model for the MiniBooNE LEE signal in MicroBooNE is not straightforward.   
To date, there is no consensus on a theory-based prediction for the MiniBooNE result that can be used as a model.  
The MiniBooNE and MicroBooNE detectors, as well as the LEE analysis approaches, are sufficiently different that one cannot simply scale the MiniBooNE LEE for detector mass and POT to obtain a prediction.  Therefore, MicroBooNE has developed a purely-phenomenological, simplified model for the LEE.

\begin{figure}[t]
    \centerline{\includegraphics[width=\linewidth]{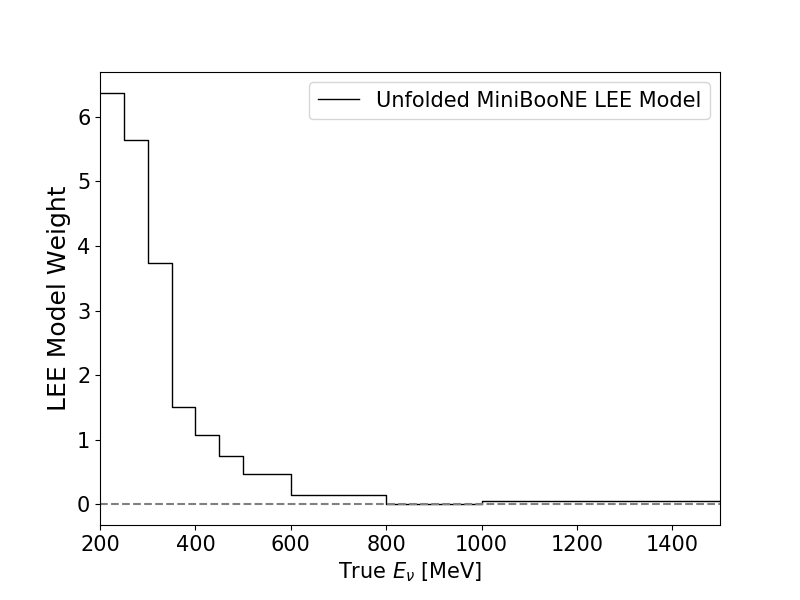}}
    \caption{MicroBooNE LEE model weights, based on unfolding the MiniBooNE electron-like excess assuming a $\nu_e$ hypothesis. These weights are applied to the beam-intrinsic CC $\nu_e$ prediction in order to obtain the MiniBooNE median LEE model prediction used in this analysis, which is added to the background contributions. In this analysis, LEE model weights above a true neutrino energy of $800$\,MeV are set explicitly to zero.}
    \label{fig:MBModel}
\end{figure}

The MicroBooNE LEE signal prediction is obtained by unfolding MiniBooNE simulated and measured neutrino spectra with reconstructed neutrino energy $>$200\,MeV~\cite{MiniBooNE:2018esg}. This produces a ``true'' underlying distribution, where the difference between the data and simulation is attributed to an excess of $\nu_e$ in the beam.
The revised electron neutrino flux is then applied to $\nu_e$ events in the MicroBooNE simulation at truth-level using the weights shown in Fig.~\ref{fig:MBModel}. These events are generated according to the $\nu_e$ interaction kinematics encoded in the GENIE event generator.
This LEE signal contribution is added to the beam-intrinsic CC $\nu_e$ and misidentified background prediction from the standard simulation to obtain the total predicted event rate under this model.
All three MicroBooNE $\nu_e$-based LEE searches use this model, which is called the ``MiniBooNE median LEE model''.   
The ``LEE signal strength,'' $x_\text{LEE}$, describes the normalization of the signal relative to the median model, where $x_\text{LEE} = 0$ represents the no-LEE hypothesis and $x_\text{LEE} = 1$ represents the median signal.

There are two important limitations of this model that must be noted.  
First, uncertainties are not propagated from MiniBooNE to the MicroBooNE prediction. 
However, we use the signal strength parameter to represent the MiniBooNE uncertainties in a naive way. Given that the significance of the MiniBooNE excess is $4.8\sigma$ ($12.2\sigma$) with (without) the experiment's systematic uncertainties, we estimate a $1\sigma$ confidence interval on the LEE signal strength of $1 \pm 0.21$ ($1 \pm 0.08$). %
Second, the model was initially developed based on the MiniBooNE 2018 data set that does not include information below 200\,MeV, and the excess rate in that range is taken to be zero.    
This assumption is not consistent with the MiniBooNE results released in 2020~\cite{MiniBooNE:2020pnu}, which presented a non-zero excess in the 150--200\,MeV region; however, because the MicroBooNE analysis was well underway, the model was not changed to reflect this change.
As a result, there can be spill-out of events due to reconstruction resolution into the $<200$\,MeV region, but there is no spill-in from energies below 200\,MeV. As a result, this should be considered a simplified model.

Given this simplified model, the LEE prediction is used to define the LEE region to be 200--500\,MeV in reconstructed neutrino energy, and to quantitatively evaluate the comparison between the data observation and the prediction with and without this signal model. %

\subsection{CCQE interactions in MicroBooNE}

As seen in Fig.~\ref{fig:MBModel}, the simplified LEE model described above predicts a significant excess in the MicroBooNE data set for energy $<500$\,MeV.
At ``truth level,'' before detector and reconstruction effects are considered, CC $\nu_e$ interactions in this energy range are 77\% CCQE interactions, with 17\% MEC and 5\% $\Delta$ resonance contributions also present. 
Thus, an exclusive analysis that considers only CCQE events, which are defined as:
\begin{equation}
    \nu_\ell + n \rightarrow \ell^- + p, \label{CCQEint}
\end{equation}
where the lepton $\ell$ is an electron in this case,  is relatively efficient for the LEE signal. This channel has several other benefits.
First, among the interactions in this energy range, the CCQE interaction has the lowest uncertainties due to cross section modeling. Second, in the MicroBooNE detector the neutrino CCQE interaction products---one lepton and one proton---can, in principle, be accurately reconstructed. In practice, the reconstructed kinematics of the CCQE event may suffer substantial smearing from the initial neutron momentum, which is unknown, and from final state interactions as the proton exits the nucleus.
These effects are taken into account in the analysis. 

\subsection{Using two-body kinematics to isolate a well-reconstructed CCQE signal}
\label{sec:twobodyccqe}

To achieve the goal of isolating CCQE events, we make use of the fact that, in principle, the CCQE signal is a two-body interaction, with fully constrained kinematics if the incoming and outgoing 4-vectors are known.  Even when realistic effects are introduced, 
as shown by the GENIE cross section model, a significant subset of CCQE interactions are expected to exhibit nearly ideal two-body scattering characteristics. For these events, energy-momentum conservation constraints allow the neutrino energy to be determined completely from the final-state lepton energy and angle, the final-state proton energy and scattering angle, or a combination of the final-state lepton and proton measurements.
Non-CCQE interactions will typically fail energy- and momentum-conservation tests when the event is reconstructed as a CCQE event.  
To achieve the goal of accurate neutrino energy reconstruction, we define a well-reconstructed CCQE signal in simulation as one in which the reconstructed neutrino energy agrees with the true neutrino energy within 20\%, which is approximately the neutrino energy resolution of the event reconstruction.

As a first illustration of the power of this method, consider the case of the kinematics of the outgoing proton.   
If the struck neutron is at rest, then two-body kinematics requires the proton to move forward in the lab frame, for all neutrino energies.   On the other hand, misreconstructed events, such as $\Delta \rightarrow \pi^0 p$, where one photon from the $\pi^0$ is not reconstructed and the other is mistaken for the electron, often have backward-going protons.
Rarely, CCQE events with significant final state interactions can also lead to a backward-going proton. As this analysis focuses on reconstructing clean CCQE evens without such final state interactions, two-body kinematic requirements help reduce contamination from these events.

As a second illustration of the method, note that for a CCQE interaction with a neutron at rest, the neutrino, lepton, and proton momentum vectors lie in a single plane.   
Thus, if the goal is to isolate events consistent with the kinematics of CCQE interactions, then events with high transverse momentum ($p_T$) can be rejected.  
The $p_T$ is a particularly useful variable when isolating CCQE $1\mu 1p$ events to use as a constraint.  
The multiple Coulomb scattering of low-energy electrons causes this to be a weaker discriminator for CCQE $1e1p$ scatters, but it is, nonetheless, useful for removing unwanted events.

A third example relies upon reconstruction of the neutrino 4-vector.   
Because the MicroBooNE detector is located far from the target,  we assume that the neutrinos enter the detector parallel to the $z$ axis.   Given this information, the neutrino energy can be reconstructed in three ways:
\begin{widetext}
\begin{eqnarray}
E_\nu^{\rm range}&=& \textnormal{K}_{p} + \textnormal{K}_{\ell} + M_{\ell} + M_{p} - (M_{n} - B),  \label{eq:erange} \\
E_\nu^{QE-p}&=& \left(\dfrac{1}{2}\right) \/\dfrac{2\cdot(M_{n}-B)\cdot E_{p}-((M_n-B)^{2}+M_{p}^{2}-M_{\ell}^{2})}{(M_{n}-B)-E_{p}+\sqrt{(E_{p}^{2}-M_{p}^{2})}\cdot \cos\theta_{p}}, \label{enuqep}\\
E_\nu^{QE-\ell}&=&\left(\dfrac{1}{2} \right)\/ \dfrac{2\cdot(M_{n}-B)\cdot E_{\ell}-((M_n-B)^{2}+M_{\ell}^{2}-M_{p}^{2})}{(M_{n}-B)-E_{\ell}+\sqrt{(E_{\ell}^{2}-M_{\ell}^{2})}\cdot \cos\theta_{\ell}}, \label{eq:enuqelepton}
\end{eqnarray}
\end{widetext}
where $K$ is kinetic energy determined from the track length or charge clustered into the electromagnetic shower; $\theta$ is measured with respect to the beam axis; $M$ is mass; $p$ is proton in nucleus; $n$ is neutron in nucleus; and $B$ is the average binding energy, assumed to be $40$\,MeV~\cite{Ankowski:2005wi}.
In this analysis, the reconstructed neutrino energy is defined as $E_\nu \equiv E_\nu^{\rm range}$. In the case of a well-reconstructed CCQE event that is a scatter from a neutron at rest,  $E_\nu^{QE-\ell}$ and $E_\nu^{QE-p}$ will be in good agreement with $E_\nu^{\rm range}$.   
Thus, comparison of the results of these three variables represents a third example of a powerful use of two-body kinematics to reduce backgrounds.

Given the capability to fully reconstruct the event, a fourth method of identifying the CCQE interactions of interest is to form the common variables used in scattering analyses, including $Q^2$ and the Bjorken scaling variable, $x_{Bj}$.   Calculation of these variables depends upon the lepton mass.    If the lepton is misidentified, leading to an incorrect lepton mass used in the calculation, then
the variables will reconstruct with unphysical values.  Also, CC resonance and deep inelastic events in the sample tend to reconstruct with very small values of $x_{Bj}$ compared to quasi-elastic events.

As discussed in Secs.~\ref{sec:1e1p} and~\ref{sec:1m1p} below, the primary requirement on expected correlations between variables reconstructed under the CCQE hypothesis is implemented through the BDT ensemble.
The above discussion has focused on the case where the target neutron is at rest, but in the analysis we cannot make such a strict requirement.   When implementing the CCQE kinematic conditions in the BDT ensemble as discussed below, events for the training sample are required to have reconstructed neutrino energy within 20\% of the true neutrino energy.    
Because the final outcome of the analysis is presented as a function of $E_\nu$, it is worth noting here that the BDT learns the correlation between the kinematic variables during training in order to categorize events, but it does not learn the relative rate of events as a function of any variable, and hence will not distort the outcome. 
This is further supported by the good agreement observed between data and simulation in the reconstructed neutrino energy distribution before and after a requirement on the BDT ensemble score, indicating that the BDT ensemble efficiency is well-modeled. 

\subsection{Blinded Analysis Procedure}
\label{subsec:blindnessprocedure}

We adopted a ``blind'' analysis approach where different portions of the data were made available to analyzers only after certain conditions were met.
The exception is a small sample of data with an integrated exposure  of approximately $5\times10^{19}$ POT, which was provided at early stages in order  to validate the analysis.
Otherwise, reconstruction and selection criteria were developed primarily on simulated data.

Data samples outside the LEE range were made available only after the reconstruction and the selection criteria were frozen, documented, and reviewed by the collaboration. For this analysis the first available data included event samples targeting $\nu_\mu$ interactions and a high-energy $\nu_e$ sample whose events were above a reconstructed energy of 700\,MeV. The data were compared to the prediction to look for evidence of major issues in the implementation of the analysis.

In addition, the analysis chain was tested using simulated data samples, created with and without an LEE excess, as well as variations of systematic effects. Blind analysis of each simulated data set produced results in agreement with the expectation.

After these first stages, a second data sample was made available that included $\nu_e$ candidates between the reconstructed energies of 500--700\,MeV. 
This served as a final check of the reconstruction and selection at energies close to the defined signal region.
After a final review, the remainder of the data samples containing low-energy $\nu_e$ candidates were unblinded.

\section{Reconstruction Chain}
\label{sec:reco}

This section describes the steps in the reconstruction chain.   These provide the information used for event selection, which is described in Sec.~\ref{sec:selects}.

\subsection{Preparation of Input}

Prior to the reconstruction, events are subject to the common optical filter, good run selections, noise filtering, and signal processing described in Sec.~\ref{sec:sig}, as is done for all of the MicroBooNE analyses introduced in Sec.~\ref{sec:intro}. Events with in-time light are neutrino candidates.  These events are converted to the images of the type seen on Fig.~\ref{fig:eventexample} (a) for use in this analysis.  In addition, we define the pixel intensity as the integrated reconstructed charge waveforms over six time ticks and measure it in Pixel Intensity Units (PIU). Thus, the effective size of each pixel is 3.3 mm along the y-axis and 3.0 mm along the x-axis of the plot. The wire images have an intensity threshold applied to retain only the major topological features. As the distribution of pixel intensity originating from MIPs peaks at $\sim 40$\,PIU, a lower threshold of 10\,PIU is assigned, rejecting regimes where the detector response is less optimal, while keeping most of the expected signal. Pixels with intensity lower than 10\,PIU are assigned a value of zero.

\subsection{SparseSSNet}
\label{sec:sssnet}

The reconstruction chain begins with semantic segmentation--the use of a CNN that determines a classification label for each pixel in the image.
 We employed a U-Net-style network with residual sparse convolutional layers. The implementation of the network follows a sparse 
algorithm, which improves the resource consumption by an order of magnitude and improves the accuracy of the results.
We refer to the network as ``\sssnet.''
A full description of this network can be found in~\cite{MicroBooNE:2020yze}, 
therefore we only supply a brief discussion here.

The output of \sssnet\ is a normalized probability vector 
(also referred to as ``scores'') for 5 different classes: 1) HIP; 2) MIP; 3) electromagnetic activity; 4) delta ray; and 5) Michel electron. 
These categories are mapped into two classes:  
\begin{itemize}
\item track $\equiv$ HIP or MIP, 
\item shower $\equiv$ electromagnetic activity or delta ray or Michel electrons. 
\end{itemize}
The five-category pixel labeling remains available and is useful for interpreting event displays, as seen in Fig.~\ref{fig:eventexample} (b).

A pixel is labeled as ``shower'' if its shower score, $p_{shower}$, is  $\geq 0.5$ and ``track" otherwise, where the scores are constructed such that $p_{shower} + p_{track} = 1$. 
There is an emphasis in the training on low-energy particles ($\sim15\%$ of the training set), 
so as to allow better identification of events in the signal energy range. 

The class accuracy of \sssnet\ is defined as the fraction of correctly predicted pixel labels out of all truth label pixels from that class. As an example, using a Y-plane test sample, the track pixel accuracy is 99.2\% and the shower pixel accuracy is 99.6\%. This very high accuracy provides an essential foundation for the next steps in the analysis.

\subsection{Cosmic Ray Tagging}
\label{sec:cosmic}

Events are then run through the Wire-Cell charge-light matching algorithm~\cite{WCJINST,MicroBooNE:2021zul},  
used jointly in this analysis and the Wire-Cell inclusive LEE search~\cite{WCLEE}, to map charge clusters to flashes of scintillation light observed by the PMTs.
Charge clusters mapped to flashes outside the beam spill are tagged as cosmic in origin.
These pixels are masked off so that they are not used in the analysis reconstruction algorithms that follow.

\subsection{Vertex Finding}
\label{sec:vtx}

The vertex-finding algorithm, described in detail in Ref.~\cite{MicroBooNE:2020sar}, is run prior to particle reconstruction.  
The first step in the vertex algorithm searches for the characteristic ``vee'' shape of a $1 \ell 1p$ event in the image (as can be seen in Fig.~\ref{fig:eventexample}), which may be due to a  ``track-track'' pair or a ``shower-track'' pair, as identified by clustering pixels from \sssnet.   The search is performed in each wire plane.
The algorithm requires a vee with opening angle $>10^\circ$ in at least one wire-plane image, which results in clearly separated leptons and protons.

The track-track vertex-finding algorithm masks out non-track pixels as labeled by \sssnet, and then follows the remaining sets of connected pixels in search of the vee topology.  
Kinks in the two dimensional (2D) images are identified using defect detection of 2D convex hulls ~\cite{rVicThesis}.  
The three-dimensional (3D) vertex positions are found by matching the 2D seeds across the wire planes.   
The shower-track algorithm reduces the initial number of 2D vertex seeds by requiring that the kink be consistent with a shower-track meeting point pixel, as labeled by \sssnet.   
This generally prevents vertices from being assigned to branches of the electromagetic showers.  
An event may have multiple 3D vertices.   
These may be real---for example, a $1 \mu 1p$ interaction may have a Michel electron attached as a second vertex---or they may be spurious.  
The latter case usually arises near event vertices with multiple protons, which leads to vertices found at similar, but not identical points.  
In each selection presented, the best vertex in a given event is chosen from all reconstructed vertices if it passes certain criteria specific to each selection sample.
If multiple vertices still pass the criteria, a best vertex is chosen for each selection using various criteria described below. 

\subsection{Track Energy and Angle Reconstruction}
\label{sec:track}

Track reconstruction ~\cite{MicroBooNE:2020sar} begins in 3D at a given vertex and follows ionization trails in all three wire images at once. A cluster of charge forming an ionization trail from the vertex is referred to as a ``prong.''   While a prong is a collection of continuous charge, it need not be a single line of charge.  The prong may comprise connected branches, as will be the case for most electromagnetic showers.  Each prong is assumed to come from a single initiating particle.  Thus, in the event selection described below, the selection criterion of 2-prongs requires that the event have two and only two particles exiting the vertex. 

The 2-prongs cases are consistent with $1e1p$ or $1\mu1p$ events.
To associate the prong with the \sssnet-identified pixels, the 3D prong is projected onto the 2D images and the pixels are then associated to the prong.      
The fraction of shower pixels for each of the two prongs is  calculated, and the prong with greatest shower pixel fraction is identified.  
If this prong has more than 20\% shower-labeled pixels, the event is tagged for the $1e1p$ reconstruction chain, and the other prong is identified as the proton. Otherwise, the event enters the $1\mu 1p$ reconstruction chain, where we identify the shorter prong as the proton. The prongs described here are not adequate to reconstruct the shower-like particle kinematics. A dedicated algorithm is used to reconstruct the energy and angle of shower-like prongs as described in Sec.~\ref{sec:shower}.

Track-like prongs next go through a dedicated track reconstruction which takes as input the full pixel intensity images for each of the three planes and the vertex 3D position. 
Then, 3D points (i.e., points in the detector coordinate frame) are added to the track using an iterative stochastic search in the vicinity of previously chosen points, starting from the vertex. 
In order to find the minimal set of 3D points that best describes the track, a regularization is performed.
Once the track is reconstructed, the energy is calculated based on the length using the known stopping power of muons and protons in LAr~\cite{pdgMuonAr,NistAr}. The energy resolution for muons (protons) is 3.4\% (2.5\%). The $\theta $ resolution for muons (protons) is $3^\circ$ ($3.6^\circ$), and the $\phi$ resolution is $3.3^\circ$ ($6.4^\circ$).

\subsection{Shower Energy and Angle Reconstruction}
\label{sec:shower}

Shower-like prongs, as identified by  \sssnet, are reconstructed by an algorithm described in~\cite{MicroBooNE:2021nss}. 
The first step of the reconstruction is to mask the image to only use the \sssnet\ shower pixels with a shower score of $>$0.5 and intensity $>$10\,PIU.  
A template in the shape of a triangle is placed with the apex at the reconstructed vertex position. 
The triangle is then optimized to find the configuration for which the triangle contains the largest number of shower pixels with non-zero intensity. The direction, gap size, opening angle, and length parameters are optimized sequentially. The full range of allowed values for each parameter is shown in Table~\ref{tab:showerreco_params}. The ranges of the parameters were chosen by investigating a simulated sample of $\nu_e$ events and observing the true values of these parameters in simulation. More details on the triangle-fitting and an example are provided in Ref.~\cite{MicroBooNE:2021nss}.

\begin{table}[htb]
    \centering {
       \caption{ Allowed range of parameters for the shower reconstruction algorithm.   
    \label{tab:showerreco_params}}
    \resizebox{\linewidth}{!}{

    \begin{tabular}{lcc}
    \hline \hline
    & Minimum value & Maximum value \\
    \hline
    Direction        & $0^\circ$ & $360^\circ$\\
    Gap Size         & 0\,cm & 17\,cm \\
    Opening angle    & $17^\circ$ & $75^\circ$ \\
    Length           & 3\,cm & 35\,cm \\
    \hline \hline
    \end{tabular}}
    }
\end{table}

The next step is to calculate the energy of the shower. The intensity of each shower pixel enclosed in the $Y$-view triangle is summed. 
This total is denoted by $Q_{sh}$. 
$Q_{sh}$ must then be converted to energy units (MeV). 
To do so, the $Q_{sh}$ in a sample of simulation events was compared to the generated electron energy in the events. 
A $Q_{sh}$-to-E conversion is determined by fitting a linear function to the distribution of generated energy of the electrons vs. $Q_{sh}$ as discussed in more detail in~\cite{MicroBooNE:2021nss}, resulting in:

\begin{equation}
\label{eq:orig_calib}
     ~\text{E}~\text{[MeV]} = (0.0126 \pm 0.0001)  \times Q_{sh}~\text{},
\end{equation} 
where the error here comes from the statistical error on the simulated events used for the fit. The next step of shower reconstruction is to obtain the 3D direction, The 2D showers are first compared between planes for pixel overlap in time. The overlap fraction is defined as the fraction of shower pixels in the collection plane shower that overlap in time with shower pixels from a 2D shower in another plane, where the $U$ and $V$ planes are considered separately, as described in ~\cite{MicroBooNE:2021nss}. 
If the overlap fraction is $> 0.5$ in either plane, then overlapping pixels are used to calculate a cluster of 3D shower points. If both the U and V planes meet this criterion, then the pixels from the plane with the highest overlap fraction are used.  
The direction is found by using the calculated geometric center of the point cluster and the event vertex.

\subsection{Multi Particle Identification}
\label{sec:MPID}

MPID, described in detail in~\cite{MicroBooNE:2020hho}, is a neural network that identifies whether a given particle species is produced in an interaction. The input to the network is a 512~$\times$~512 (1.5\,m~$\times$~1.5\,m) cropped image centered around a candidate vertex. The network consist of ten convolutional layers followed by two fully connected ones. The MPID algorithm classifies the probability of finding specific particles (muons, charged pions, protons, electrons and photons) in the image. Each category is considered individually and assigned a score between 0 and 1 (higher scores correspond to higher probabilities). This algorithm operates separately from the reconstruction algorithms described above, except for use of the vertex points, which improves the robustness of the analysis.


MPID provides two types of scores for each category: image score and interaction score.     
The image score indicates the probability that a given type of particle is anywhere within a the image; while the interaction score indicates the probability that a given type of particle is attached to the vertex. The scores from MPID are used as additional data selection criteria (see Sec.~\ref{sec:1e1p},~\ref{sec:1m1p}, and~\ref{sec:pi0recomass})

\begin{figure}
    \centering
    \includegraphics[width=0.45\textwidth]{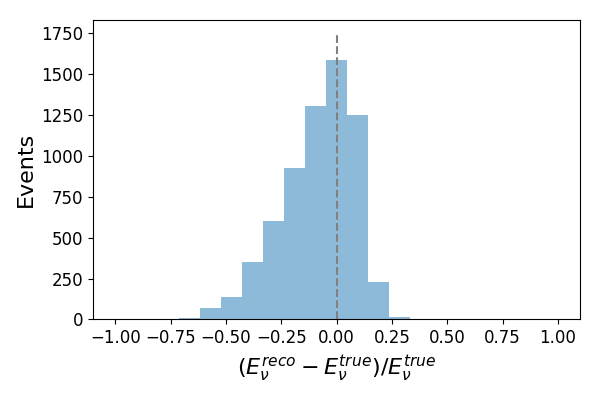}
    \caption{The fractional error distribution of the neutrino energy reconstruction for simulated $\nu_e$ CCQE events that pass $1e1p$ selection criteria and reconstruct with $200<E_\nu<1200$\,MeV.}
    \label{fig:nuenergy_frac}
\end{figure}

\subsection{$\pi^0$ Mass Reconstruction}
\label{sec:pi0recomass}

An important potential background to $\nu_e$ CCQE events comes from neutral current (NC) events with a $\pi^0$, where an early conversion of one photon produces a $1e1p$-like vertex with no apparent gap.  
To remove these events, a $\pi^0$ search algorithm is applied to the $1e1p$-identified images.
To allow capture of both decay photons from $\pi^0$ events, the electromagnetic shower-finding algorithm described above is extended. The pixels found in the first shower are temporarily masked out. 
If the sum of shower pixel intensity remaining is greater than 5000\,PIU, then the algorithm is run a second time. This requirement is fairly loose due to leftover cosmic background and meant to speed up processing. In a simulation sample of CC $\pi^0$ events, nearly all events with a contained second decay photon and a well reconstructed first photon pass this selection criterion. In the second shower search, the maximum length is increased to 60\,cm and the maximum gap is increased to 90\,cm. 

Next, simple shower quality data selection criteria are applied to search for events with two 3D reconstructed showers. The selected events must have: 1) a reconstructed vertex within the fiducial volume; 2) two collection plane showers, each with reconstructed energy greater than 35\,MeV; 3) showers in the collection plane that have overlap fraction in another plane of greater than 0.5; 
and 4) unique shower matching across planes.
If a collection ($Y$) plane shower matches with showers in both the induction ($U$ and $V$) planes, then the one with the highest overlap fraction is chosen. 

Given the reconstructed 4-vectors of the photon candidates for the selected events, the invariant mass, which is potentially that of a $\pi^0$, is calculated using the following equation:
\begin{equation}
\label{eq:pi0mass}
M_{\pi^0} = \sqrt{4\sin^2\left(\frac{\theta}{2}\right)(E_1)(E_2)},
\end{equation}
where $E_1$ and $E_2$ are energies of leading and subleading photons, and $\theta$ is the opening angle between the two showers. 
The resulting value $M_{\pi^0}$ is used as a test variable for $\pi^0$ identification in the $\nu_e$ event selection, as described in the next section.

\section{Selection of the Signal and Constraint Samples}
\label{sec:selects}

This section describes the selection criteria for the $1e1p$ and $1\mu 1p$ CCQE events that form the basis of the final analysis. The $1e1p$ CCQE events are the sample that potentially demonstrates an LEE signal. The $1\mu 1p$ CCQE sample is used to constrain the MC prediction of the $1e1p$ rate.

\subsection{Signal Selection: $1e1p$ CCQE}
\label{sec:1e1p}
\subsubsection{Overall data selection criteria for the $1e1p$ sample}

The data selection criteria applied to select the $1e1p$ CCQE sample fall into three broad categories:  1)  basic data selection criteria;  2) BDT score selection criterion, and 
3) $\pi^0$ rejection criteria.

The basic selection criteria are applied to all 2-prong vertices identified as $1e1p$ using the 20\% shower pixel fraction requirement (see Sec.~\ref{sec:track}). First, accepted vertices  must be reconstructed within the defined fiducial volume, 
$>10$\,cm from active volume edges and outside of the region of unresponsive wires from $z=700$ to 740\,cm. Second, a containment selection criterion is required. We define the distance of a prong from the edge of the detector as the minimal distance from all the prong's 3D points to an edge. If the both prongs have a similar distance, we require that the distance of both prongs is $>5$\,cm, otherwise we require that the minimal distance of both prongs is $>15$\,cm (one radiation length). Third, $1e1p$ events are rejected if the event enters the inefficient region of the $U$ plane. 
Although the analysis uses the shower energy determined only from the $Y$ view plane, to assure good 3D reconstruction and shower identification, the fourth basic selection criteria requires the energy in the $Y$ plane to be consistent with the energy in the other two views. This is accomplished by requiring
\begin{equation}
\frac{\sqrt{(E_e^U-E_e^V)^2 + (E_e^U-E_e^Y)^2 + (E_e^V-E_e^Y)^2}}{E_e^Y} < 2,
\end{equation}
where $E_e^{[U,V,Y]}$ denote the reconstructed electron candidate energy on the $U$, $V$, and $Y$ plane, respectively.
Energy selection criteria are placed on the reconstructed particles:   neutrino energy $200<E_\nu<1200$\,MeV; electron kinetic energy $K_e>35$\,MeV;  and proton kinetic energy, $K_p>50$\,MeV.  Three further kinematic requirements remove misreconstructed events and backgrounds: the opening angle between the electron and the proton must be $> 0.5$\,rad; 
the proton must be forward going, and the reconstructed velocity must have a physically allowed boost from the lab frame to the neutron rest frame (reconstructed $|\beta|$ between 0 and 1).

\begin{figure*}
    \centering
    \subfloat[]{
    \includegraphics[width=0.45\textwidth]{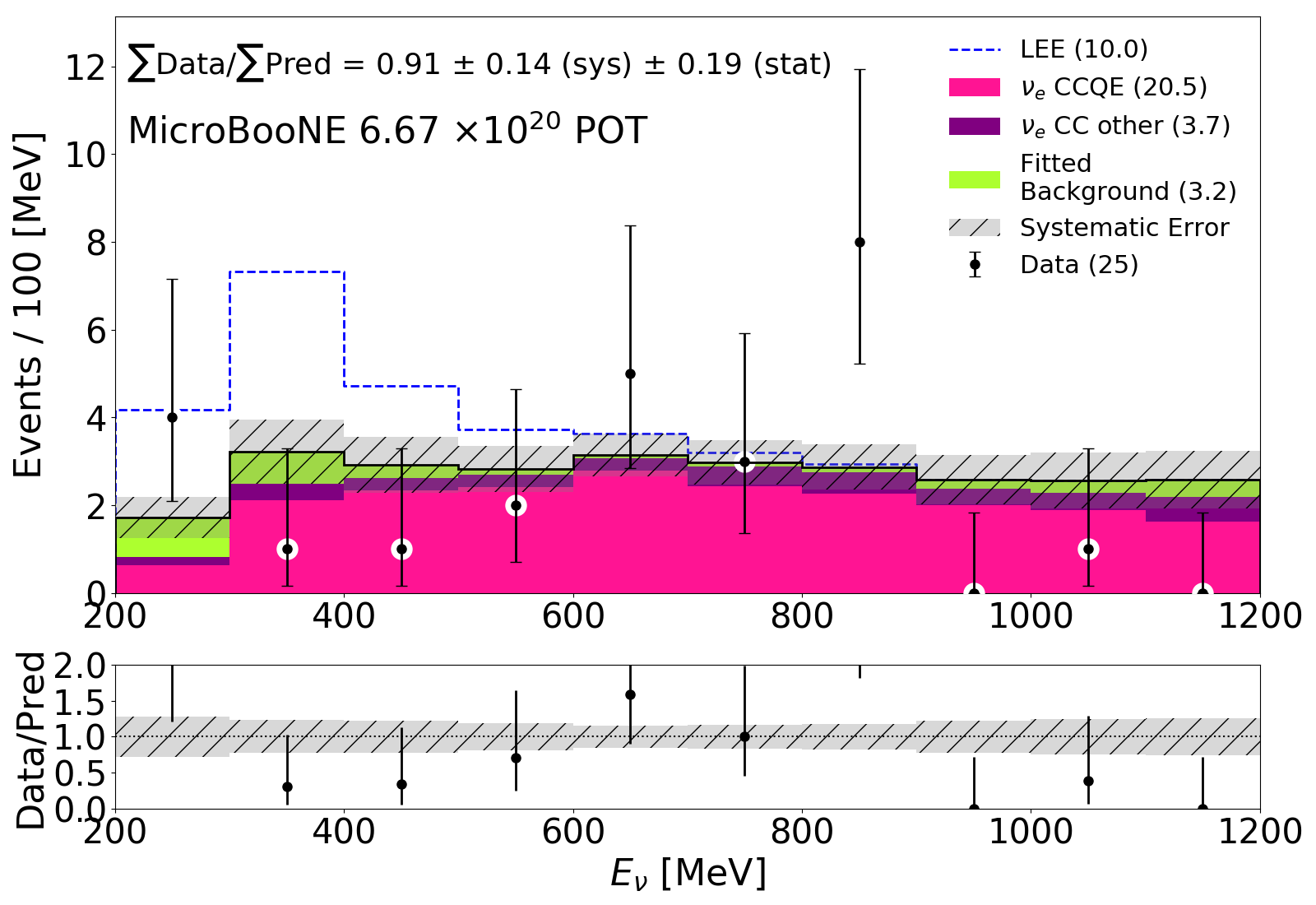}}
    \subfloat[]{
    \includegraphics[width=0.45\textwidth]{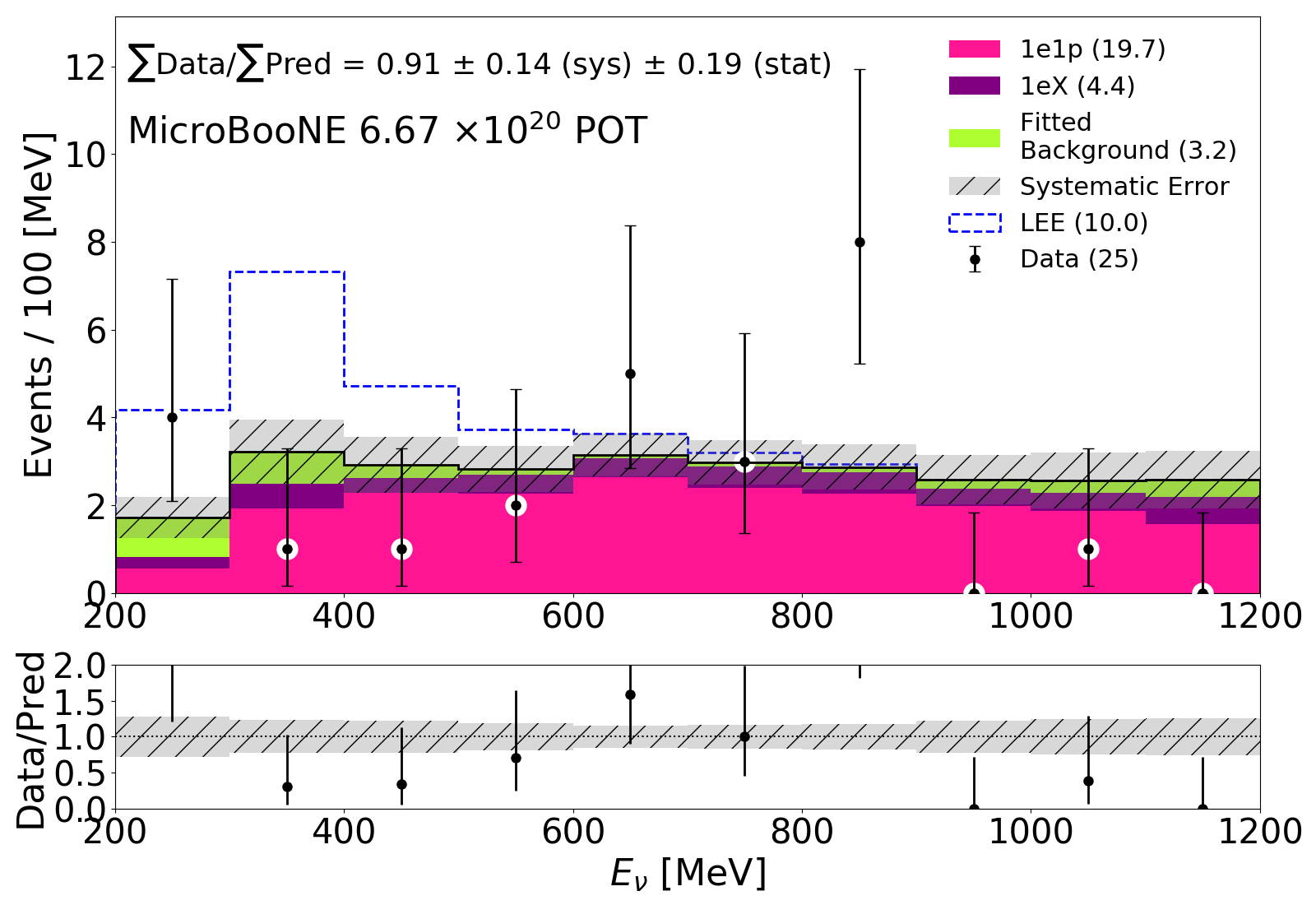}}
    \\
    \caption{$E_\nu$ distribution comparing data (black points) to the unconstrained prediction (stacked histogram) in the $200<E_\nu<1200$\,MeV region. 
    The green contribution represents the prediction from the background fit described in Sec.~\ref{sec:numubkgfit}and the dashed blue line gives the prediction of the unfolded median MiniBooNE excess model described in Sec.~\ref{sec:lee_model}. The $\chi^2_\text{CNP}$/dof is 23.02/10, corresponding to $p_{\rm val}$ = 0.024. The prediction is presented in terms of both interaction type (a) and topology (b). The numbers in parentheses indicate the integrated number of events over the range shown.
    }
    \label{fig:Enu_stacked}
\end{figure*}

\begin{figure*}
    \centering
    \includegraphics[width=0.5\textwidth]{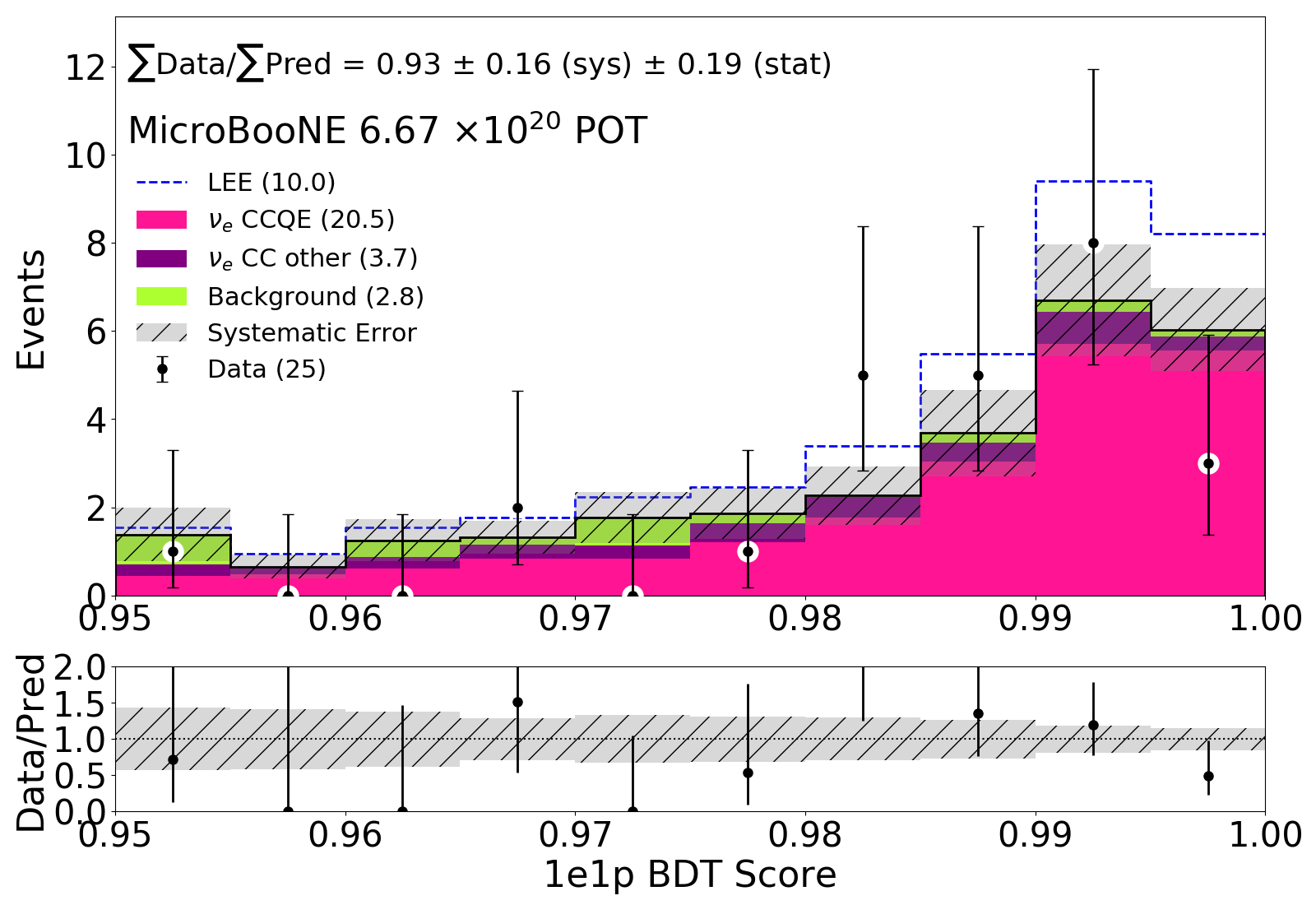}
    \caption{$1e1p$ BDT ensemble average score distribution comparing data (black points) to the unconstrained prediction (stacked histogram) in the $[0.95,1.0]$ region. The green background contribution comes directly from our simulation sample and the dashed blue line gives the prediction of the unfolded median MiniBooNE excess model described in Sec.~\ref{sec:lee_model}. The $\chi^2_\text{CNP}$/dof is 11.06/10, corresponding to $p_{\rm val}$ = 0.43.}
    \label{fig:unconstr_1e1pBDT}
\end{figure*}

Following the basic selection, this analysis uses an ensemble of BDTs, as described in the next section.   An ensemble is trained for each of the three runs. Each BDT categorizes events on the basis of 19 kinematic measurements.   These variables are related to transverse momentum, energy, angle, and combinations that test the correlations between these variables. One such variable is the consistency between the three different neutrino energy definitions given in Eqs.~\ref{eq:erange}--\ref{eq:enuqelepton}.  Four measurements related to ionization charge are also included: total charge within 5\,cm of the vertex, shower pixel fraction in the electron prong, shower pixel fraction in the proton prong, and the ratio of the number of shower pixels in the image to the number of shower pixels connected to the electron prong. The full list of training variables is provided in~\cite{Supp}.   The BDT is trained using simulated $\nu_e$ CCQE $1e1p$ events. For these training events, we require the reconstructed vertex to fall within 5\,cm of the true vertex and the reconstructed $E_\nu$ to fall within 20\% of the true $E_\nu$. The BDT is trained against $\nu_\mu$ interactions (including those with a $\pi^0$ in the final state) and cosmic muon interactions. BDT scores are normalized to the range $[0,1]$, where higher values indicate a higher probability for the event to come from a true $\nu_e$ CCQE $1e1p$ interaction. The analysis requires the average BDT score within the ensemble to be~$>0.95$, which was optimized for sensitivity to the LEE model described in Sec.~\ref{sec:lee_model}.

\begin{figure*}
    \centering
    \subfloat[]{
    \includegraphics[width=0.45\textwidth]{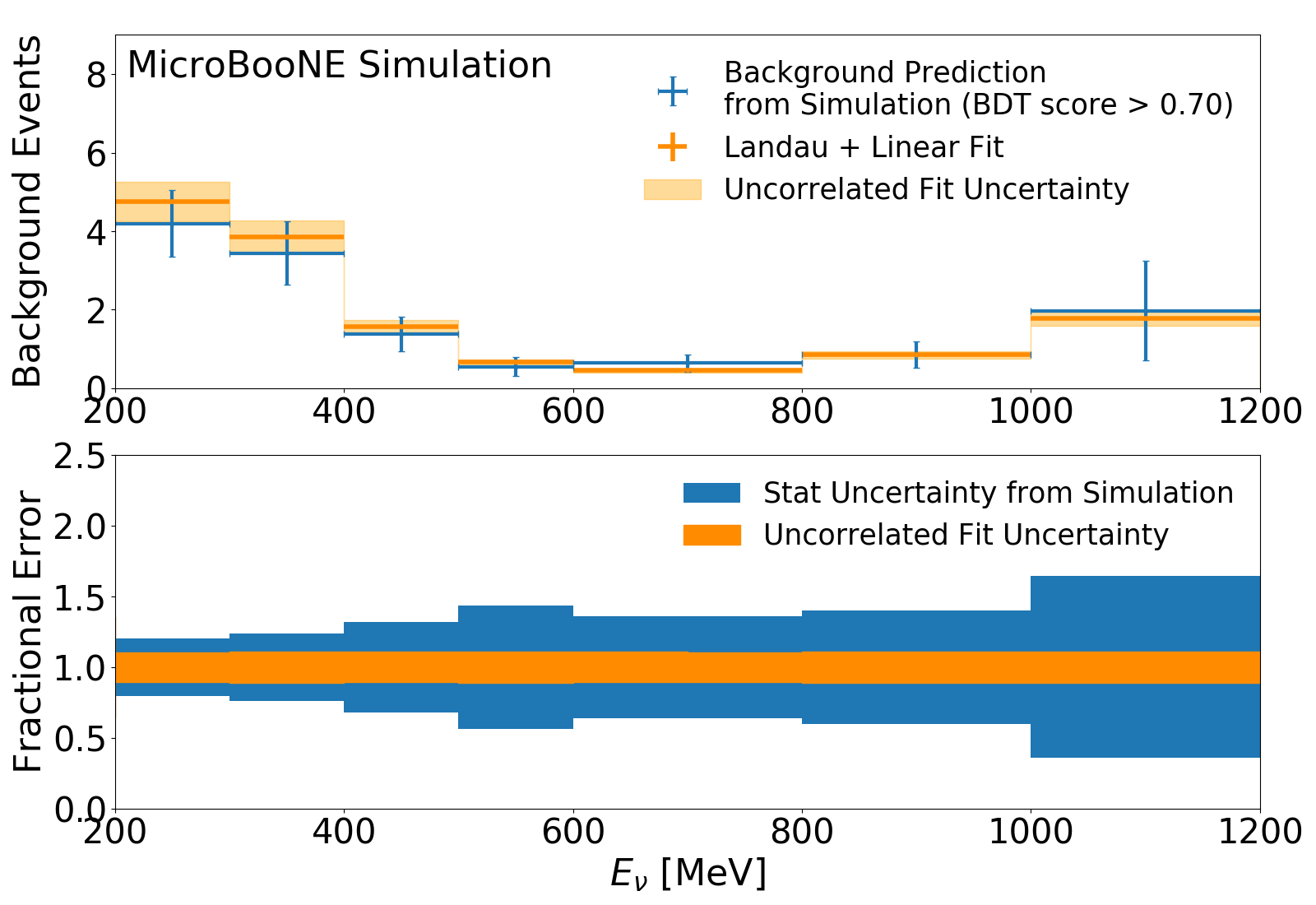}}
    \subfloat[]{
    \includegraphics[width=0.45\textwidth]{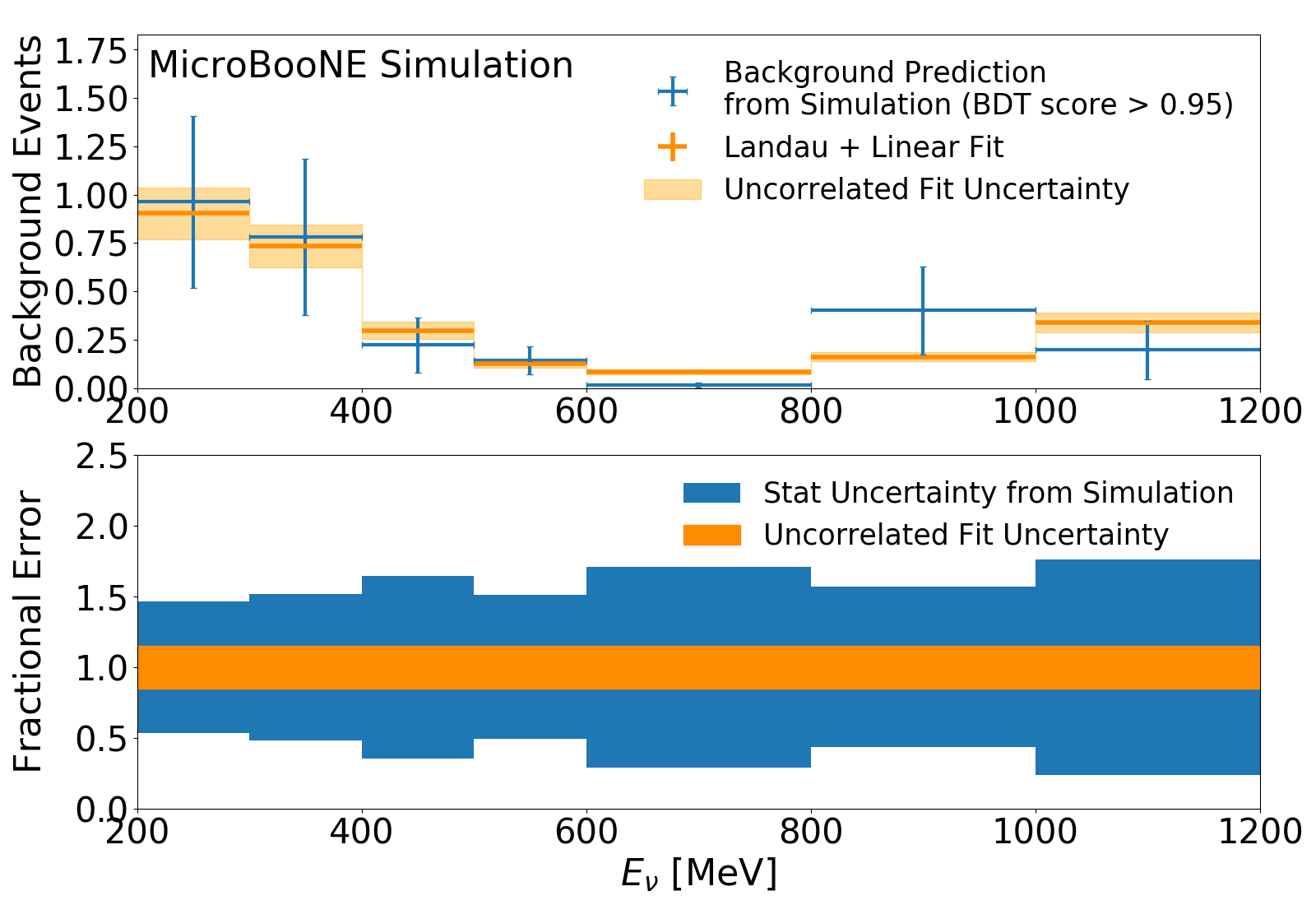}}
    \\
\caption{The fit to the $\nu_\mu$ background distribution to the $1e1p$ analysis. The shape fit is performed at a loose BDT score cutoff of 0.7 (a) and scaled to the signal cutoff of 0.95 (b). Blue points represent the prediction from the simulation, with error bars representing the Gaussian approximation of the statistical error (quadrature sum of event weights). The orange line and corresponding shaded region represent prediction and uncertainty, respectively, coming from the Landau+linear fit.} 
\label{fig:bkgfitresult}
\end{figure*}

The last category of data selection criteria is related to $\pi^0$ rejection.   The reconstructed $\pi^0$ mass test variable, $M_{\pi^0}$, must be $<50$\,MeV/$c^2$.   Note that if a second shower is not identified, then the $\pi^0$ mass test variable is set to zero. We also employ two cuts using the MPID network: (1) the MPID $\gamma/e$ image score ratio must be $<2$, and (2) the MPID muon interaction score must be $<$ 0.2 if $E_e> 100$\,MeV.  This addresses background from misreconstructed $\nu_\mu$ CC$\pi^0$ events.  
The final step of the selection addresses events with more than one identified vertex passing all the other criteria. In this case, only the vertex with the highest BDT score is considered.

The result of this selection is a highly pure ($75\%$) sample of $\nu_e$ CCQE events with good $E_\nu$ resolution. These events are an irreducible background in our test of the LEE model described in Sec.~\ref{sec:lee_model}.   
Because we aim for an exclusive, highly pure sample, the overall $\nu_e$ CCQE selection efficiency for events in the active volume across a true neutrino energy range of 200--1200\,MeV is~$\sim 6.6\%$. This efficiency peaks at a true neutrino energy of~$\sim 400$\,MeV, which increases the sensitivity of this analysis to the MiniBooNE median LEE model. 
Figure~\ref{fig:nuenergy_frac} gives the fractional error distribution of the reconstructed neutrino energy of selected $\nu_e$ CCQE events from the simulation sample. The standard deviation of this distribution is $16.5$\%.

The predicted neutrino energy spectrum is displayed in Fig.~\ref{fig:Enu_stacked} (a) as a stacked histogram where the events are partitioned by interaction type.
Alternatively, in Fig.~\ref{fig:Enu_stacked} (b), the data are compared to prediction partitioned by true topological labels. The distribution of $1e1p$ BDT ensemble average scores above the analysis cutoff of $0.95$ is shown Fig.~\ref{fig:unconstr_1e1pBDT}.
In all plots comparing data to prediction throughout this work, including Figs.~\ref{fig:Enu_stacked}~and~\ref{fig:unconstr_1e1pBDT}, the numbers in parentheses indicate the integrated number of events over the range shown.
In Fig.~\ref{fig:Enu_stacked}, the green contribution representing the background prediction comes from the Landau+linear fit described in Sec.~\ref{sec:numubkgfit}.   The systematic uncertainty will be explained in Sec.~\ref{sec:unc}.
The remainder of this paper presents information allowing interpretation of these figures, followed by discussion of these results in detail in Sec.~\ref{sec:res}.

\subsubsection{Boosted Decision Tree Ensemble Studies and Results}

The $1e1p$ BDT ensemble is a suite of 20 BDTs for each Run period, resulting in a total of 60 BDTs.
Each individual BDT is trained on half of the simulation samples for the given run, while the other half is reserved for evaluation. This splitting is done at the event level and is unique for each individual BDT in the ensemble, thus reducing the importance of the arbitrary training set draw and allowing the BDT to better map the entire simulation set. As stated previously, the signal likelihood used in this analysis is taken as the average score of each BDT in the ensemble. When calculating this average for a given event, we omit any BDT which included the event in its training sample. One of the advantages of this method for our specific analysis is the ability to retain all available simulated background events when calculating predicted event rates. As each BDT in the ensemble is trained on a random 50\% subset of the simulation sample, it is highly unlikely for a simulated event to show up in the training sample of every BDT in the ensemble ($P = 2^{-20}$). Therefore, it is possible to calculate the average BDT score for every event in our simulation sample using only BDTs that were not trained on the event in question. Because this analysis aims to achieve a high-purity $\nu_e$ CCQE $1e1p$ sample, it is crucial to maximize the available simulated background events. Additionally, ensemble-based BDT classification methods reduce the variance of the signal likelihood variable compared to the output from a single BDT, as has been shown in studies using public classification-based data sets~\cite{ganjisaffar2011bagging}.

We have performed a number of studies to ensure the robustness of the BDT ensemble used in this analysis. They show, among other things, that the BDT ensemble generalizes well between different Run periods (i.e., the change in efficiency is much smaller than the systematic and statistical error on our prediction) and behaves similarly on both data and simulation (where training events are removed for the latter). Further discussion on the BDT ensemble is given in~\cite{Supp}.%

\subsubsection{$\nu_\mu$ Background Prediction for the $1e1p$ sample}
\label{sec:numubkgfit}
The tools developed for this analysis produce a highly pure $\nu_e$ CCQE sample, resulting in a low-statistics simulation sample for assessing the $\nu_\mu$ background to the $1e1p$ signal. In order to obtain a more robust prediction of this background, we have elected to leverage information on the energy distribution of $\nu_\mu$ background events at a loose BDT score cutoff of 0.7 and extrapolate this to our signal cutoff of 0.95. This is accomplished by fitting a parameterized Landau+linear probability density function (PDF) to the $\nu_\mu$ background energy distribution at the loose cutoff and scaling this prediction to the signal cutoff. This shape is motivated empirically by the observation of a rise in the background rate toward the lowest energies ($\lesssim 500$\,MeV) and a smaller rise toward the highest energies ($\gtrsim 800$\,MeV), both for the loose BDT score cutoff and the signal one, as shown by the blue points in Fig.~\ref{fig:bkgfitresult}. The Landau portion of the fit is additionally motivated by the observation that a majority of $\nu_\mu$ background events contain a $\pi^0$ in the final state, for which one of the decay photons is misinterpreted as an electron shower. The reconstructed neutrino energy of these events is governed predominately by the energy of this photon, which has a tail out to higher energies caused by pions with high momentum in the lab frame. This tail is a characteristic feature of the Landau function. The output of the fit and resulting error are used in place of the raw prediction and statistical error for the $\nu_\mu$ backgrounds in this analysis. We note here that the simulated $\nu_\mu$ events with a $\pi^0$ in the final state are weighted according to the observation in our dedicated $\pi^0$ sample. The method for calculating these weights is described in Sec.~\ref{sec:pi0weights}.

The predicted background spectrum in each reconstructed neutrino energy bin is generated by integrating a Landau+linear PDF within that bin (where we use the Moyal approximation of the Landau function~\cite{moyalapprox}). The Landau+linear fit is carried out using only shape information at a loose BDT score cutoff of 0.7. In order to get the overall normalization, we fit the BDT score distribution of the $\nu_\mu$ backgrounds to a linear PDF, which we can integrate to get the total expected number of background events for a given BDT score cutoff. The resulting shape+normalization fit for the $\nu_\mu$ background distribution at the loose BDT cutoff of 0.7 and signal cutoff of 0.95 are shown in Fig.~\ref{fig:bkgfitresult}. One can see that the fit agrees with the raw prediction within statistical error in both cases, and that the error on the fit is generally reduced compared to the simulation statistical error. 

The errors on the fits in Fig.~\ref{fig:bkgfitresult} are obtained by simulating pseudo-experiments according to the covariance matrix of the fit parameters. This is accomplished efficiently using Cholesky decomposition~\cite{cholesky}. The overall normalization error on the background rate at a given BDT score cutoff is calculated using the error on the linear fit to the BDT score distribution of background events. For the Landau+linear fit, we generate a shape-only covariance matrix for the $\nu_\mu$ backgrounds in the nominal reconstructed neutrino energy bins. The overall normalization error has been added as a fully-correlated contribution to each bin of the shape-only $E_\nu$ covariance matrix. The uncorrelated uncertainty from the fit, given by the square root of the diagonal entries of the full (shape + normalization) covariance matrix, is indicated by the orange band in Fig.~\ref{fig:bkgfitresult}. This covariance matrix replaces the nominal statistical uncertainty on the simulated background, given by the quadrature sum of weights in a given $E_\nu$ bin, which is indicated by the shaded blue band in Fig.~\ref{fig:bkgfitresult}.

\begin{figure}
    \centering
    \includegraphics[width=0.48\textwidth]{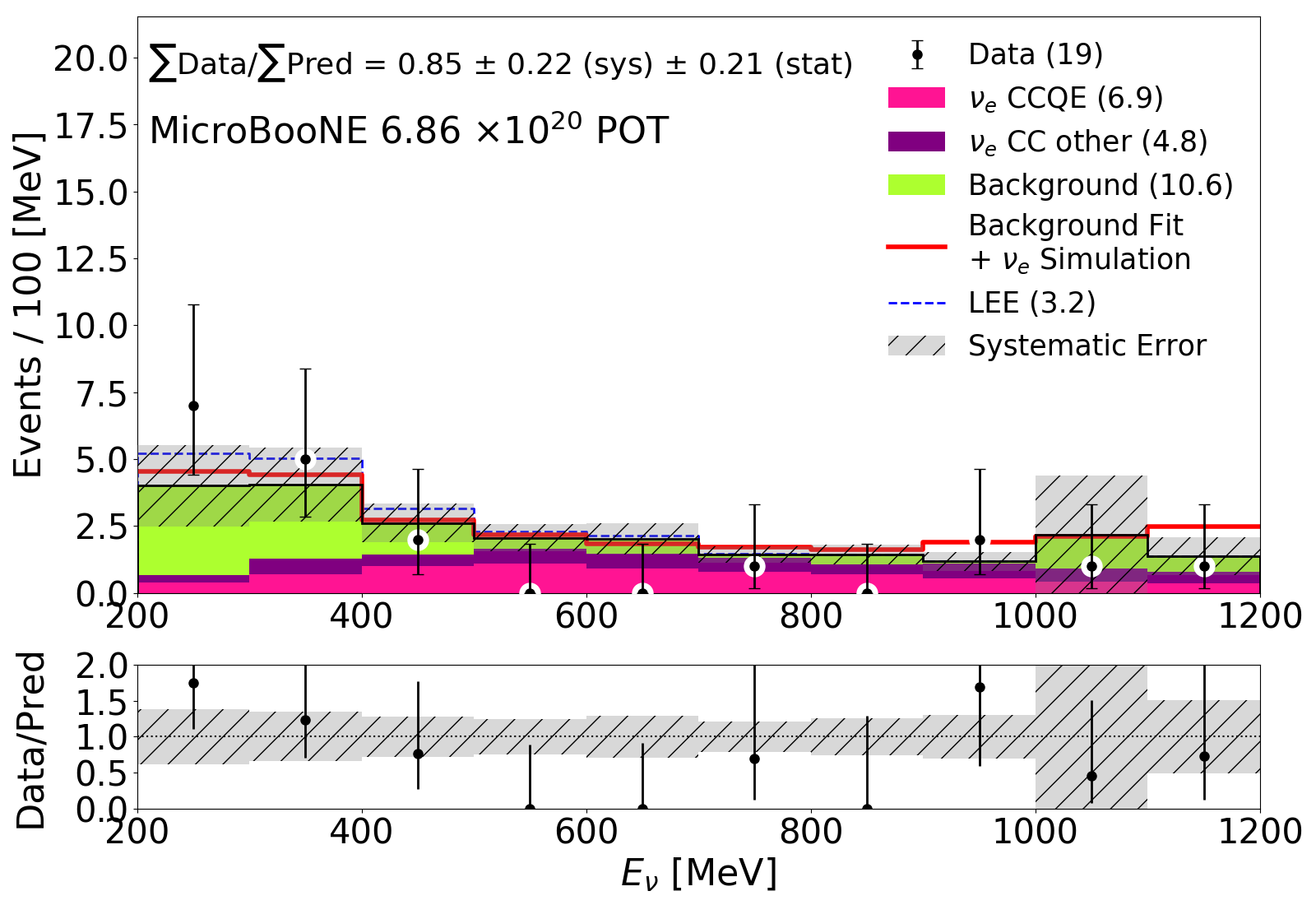}
    \caption{The data and MC prediction for events with a BDT score inside $[0.7,0.95]$. The $\chi^2_\text{CNP}$/dof for this distribution is 11.35/10, corresponding to $p_{\rm val} = 0.38$. This indicates good agreement between observed data and prediction. The prediction incorporating the Landau+linear background fit is shown by the red line.} 
    \label{fig:midBDTplot}
\end{figure}

The performance of the fit is evaluated on data by examining events slightly below the signal region, with a BDT score in $[0.7,0.95]$. These events are shown in Fig.~\ref{fig:midBDTplot}. The raw MC prediction is given by the stacked histogram, while the prediction incorporating the $\nu_\mu$ background fit on top of the simulated $\nu_e$ prediction is given by the red line. The data agree well with both predictions, indicating the consistency of the fit method presented here in predicting the $\nu_\mu$ background distribution. Further discussion on the backgrounds to this analysis can be found in~\cite{Supp}.

\subsection{Constraint Sample:  $1\mu1p$ CCQE}
\label{sec:1m1p}

To constrain the $1e1p$ prediction, to be discussed in Sec.~\ref{sec:constr}, we isolate a $1\mu1p$ CCQE sample.
The $1\mu1p$ CCQE signal selection criteria overlap with the $1e1p$ case in order to maximize the correlations between the two channels, which allows us to achieve a more powerful constraint.
The principal differences in this selection arise from lower multiple Coulomb scattering and improved energy measurement of the muon. The $1\mu 1p$  sample is differentiated from the $1e1p$ sample by applying the shower pixel fraction requirement of $f_{sh}<20\%$ on the lepton prong, which is the inverse of the $1e1p$ shower pixel fraction requirement.    This ensures the selections are orthogonal, therefore no single event can be selected by both the $1e1p$ and the $1\mu 1p$ selections.

The basic selection criteria are the same as the $1e1p$, except that the inefficient $U$-plane region is included in the fiducial volume, and there is no shower energy consistency requirement.  The input to the BDTs are also very similar, with the addition of a transverse $\phi$ angle variable in order to capture information on planarity of the event. The shower charge variables are not utilized, as we no longer expect a shower to be one of the two prongs in the final state.  Finally instead of the MPID network cuts used for the 1e1p selection, the $1\mu1p$ selection requires the MPID proton score be greater than 0.9. This serves to better reject background neutrino events from the $1\mu1p$ signal.  The training data are similarly changed. Now the signal definition is muon neutrino CCQE events with one muon and one proton in the final state, and the background being trained against consists of the following: neutrino events where the reconstructed vertex is further than 5cm from the true simulated vertex, neutrino events where the reconstructed energy is more than 20\% different from the true simulated energy, non-CCQE neutrino events, and off-beam sample cosmic muon background events.  Ensembles of 10 BDTs are trained for each Run.  

The selection criterion on the ensemble average BDT score was optimized to produce the highest sensitivity to the median LEE model after applying the constraint, and was found to be 0.5. In events with multiple reconstructed vertices, the vertex with the highest BDT score is selected.  Lastly, this sample does not require $\pi^0$ identification selection criteria, but it does require a criterion of proton MPID $>0.9$ (interaction score) across all three planes. This is applied only on events with reconstructed neutrino energy $E_\nu< 400$\,MeV, resulting in removal of $1\mu 1p$ events that are misreconstructed due to a crossing cosmic-ray muon.

\begin{figure}
    \centering
    \includegraphics[width=0.45\textwidth]{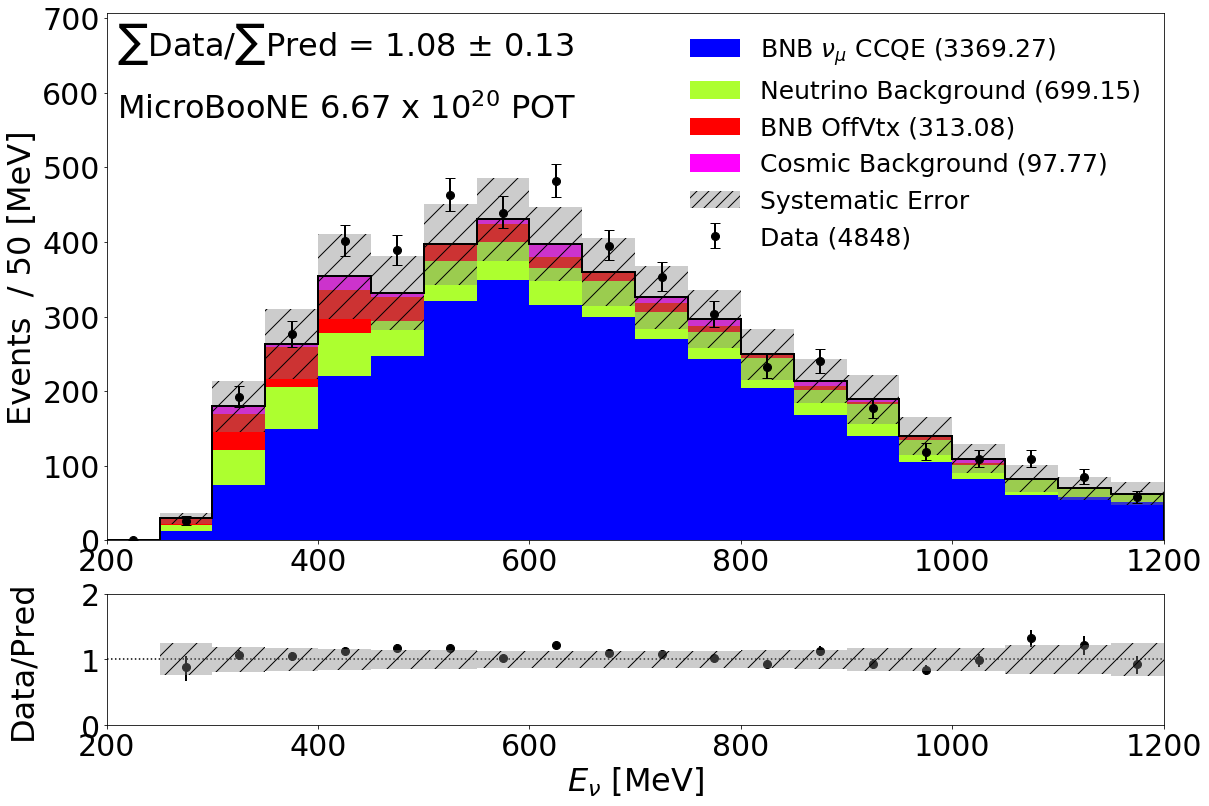}
    \caption{The reconstructed neutrino energy spectrum of the data and simulation for events passing the $1\mu 1p$ selection, calculated using the track lengths of the proton and muon. The $\chi_\text{CNP}^2/19(dof) = 1.314$ with a p-value of 0.162.}
    \label{fig:Enu_1m1p}
\end{figure}

The energy spectrum of the selected data/simulation events that are used as a constraint in the final analysis is shown in Fig.~\ref{fig:Enu_1m1p}.  The p-values quoted for this section are calculated analytically.
The selected simulation events are separated into several categories. The most frequently selected event is the signal category: $\nu_\mu$ CCQE. The next-most frequent category is our neutrino background. This category encompasses all neutrino interactions with interaction types other than the CCQE signal. The off-vertex category contains any events where the reconstructed interaction vertex is greater than 5 cm from the simulated neutrino vertex. This label takes priority over the first two. Namely if a $\nu_\mu$ CCQE interaction were to occur, but the reconstructed vertex was too far from the simulated vertex, the event would be categorized as off-vertex, not $\nu_\mu$ CCQE. Our final category is cosmic background. This portion of our prediction comes from the off-beam sample of events containing cosmic muons. The resulting signal purity of the sample is 77.3\%; this yields the signal efficiency of 4.3\% relative to all the $\nu_\mu$ CCQE interactions occurring within the active volume of the detector. The data agree with simulation with a p-value of 0.162.

\begin{figure}
    \centering
    \includegraphics[width=0.45\textwidth]{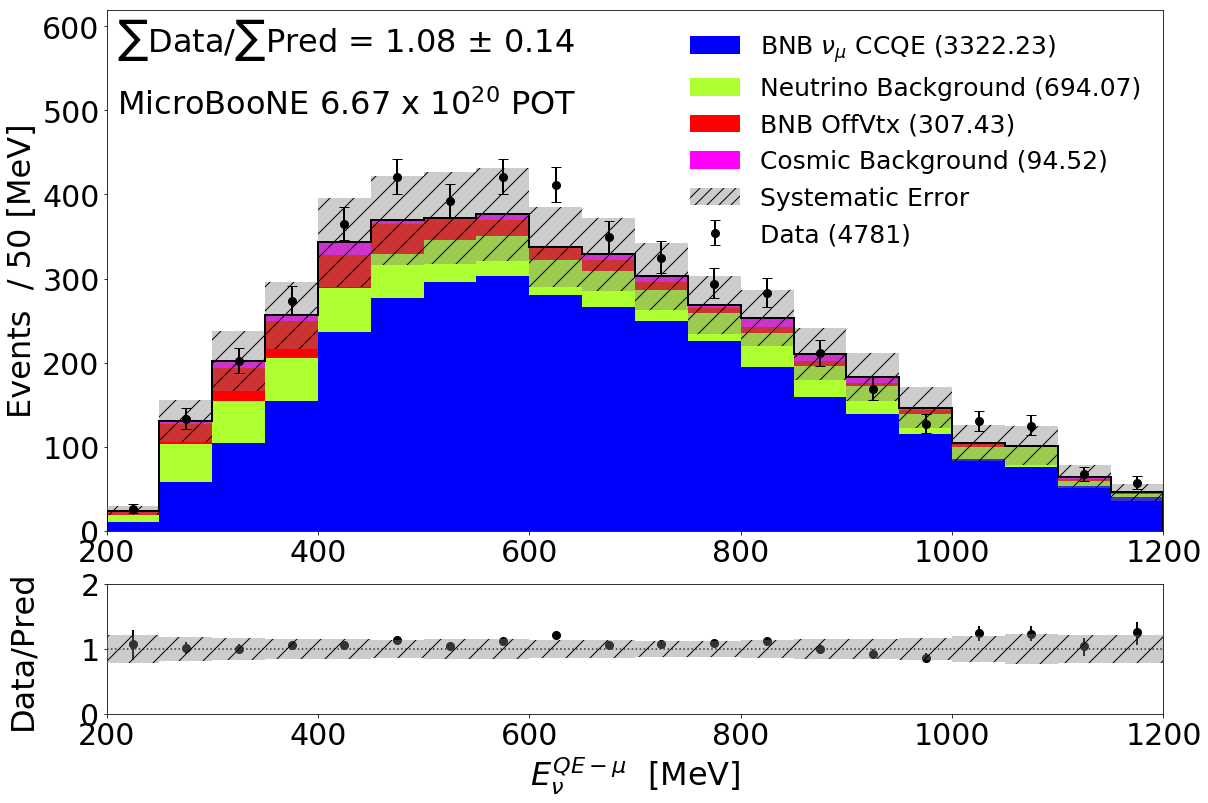}
    \caption{For comparison to Fig.~\ref{fig:Enu_1m1p}, the reconstructed neutrino charged current quasi-elastic energy spectrum of the data and prediction for events passing the $1\mu 1p$ selection.   This is calculated using Eq.~\ref{eq:enuqelepton}. The $\chi_\text{CNP}^2/14(dof) = 0.950$ with a p-value of 0.522.}
    \label{fig:EnuQE_muon}
\end{figure}

As this analysis targets events that are kinematically consistent with CCQE interactions, we can cross check the selection by looking at the data to prediction agreement of events using the neutrino charged current quasi-elastic energy calculated based on the lepton energy (see Eq.~\ref{eq:enuqelepton}).  This is the energy distribution that has been used in past experiments, including MiniBooNE.  The neutrino energy calculated via this method is shown in Fig.~\ref{fig:EnuQE_muon}.  The result is very similar to the spectrum shown in Fig.~\ref{fig:Enu_1m1p}, which is used in the final analysis, in agreement with the goal of obtaining a sample of events with two-body kinematics consistent with CCQE scattering.

\section{Other Electromagnetic Samples for Dedicated Studies}
\label{sec:sideBand}

The reconstruction was tuned and verified through comparison to two event types that contain electromagnetic showers in the energy range of the analysis: 1) $\nu_\mu$ $\pi^0$ production and 2) muon decays producing Michel electrons.    This section describes how these samples were isolated and studied.

\subsection{The $\pi^0$ sample}

Events with a $\pi^0$ in the final state make the largest contribution to the background from $\nu_\mu$ interactions contributing to the $1e1p$ CCQE event sample.  To better understand this background, a sample of $\pi^0$ events has been isolated and analyzed. The most common source of such events in our selection is due to CC and NC production of the $\Delta$ baryon that subsequently decays to a $\pi^0$ and a proton or neutron. These studies rely upon the two-particle vertex reconstruction discussed in Sec~\ref{sec:vtx}, resulting in the bulk of the $\pi^0$ events having one of two topologies.  The first, and most prevalent one, is from CC$\pi^0$, where the scattered muon and the proton from the $\Delta$ decay form the vertex, and with two disconnected electromagnetic showers from a $\pi^0$ decay. This can form a background if one electromagnetic shower overlaps the muon, and the other is undetected.
The second is from NC$\pi^0$, where one photon converts within the 0.3\,cm wire spacing, forming a vertex with the proton, while the second photon is displaced from the vertex.  This forms a background if the second photon is not detected.  

The selection of the dedicated sample focuses especially on these topologies described above. The shower quality selection criteria described in Sec.~\ref{sec:pi0recomass} are applied, as well as selection criteria to remove misreconstructed events and backgrounds. A $\Delta$ mass test variable is constructed for use in these cuts by using the kinematics of the two reconstructed showers as well as the proton prong discussed in Sec. \ref{sec:track}. In the context of this selection, a background event is any event without a $\pi^0$ in the final state. These selection cuts are:
1) the reconstructed $\pi^0$ mass test variable is less than 400\,MeV/$c^2$;
2) the reconstructed energy of the leading photon is greater than 80\,MeV;
3) the total pixel intensity (both track and shower) within 2\,cm of the vertex is greater than 250 pixel intensity units;
4) the leading shower reconstructed angle with respect to the beam direction is less than 1.5\,rad;
5) the angle between the two photons is less than 2.5\,rad;
and 6) the reconstructed $\Delta$ mass test variable is between 1000--1400\,MeV/$c^2$.

In addition, to remove the LEE-range $1e1p$ sample, to maintain blindness, a requirement of a $1e1p$ BDT score $<$ 0.7 was applied. Finally, in events with more than one identified vertex that passes all the other criteria, only the vertex with the highest leading shower energy is considered.

\begin{figure}[t]
\centering
\includegraphics[width=0.45\textwidth]{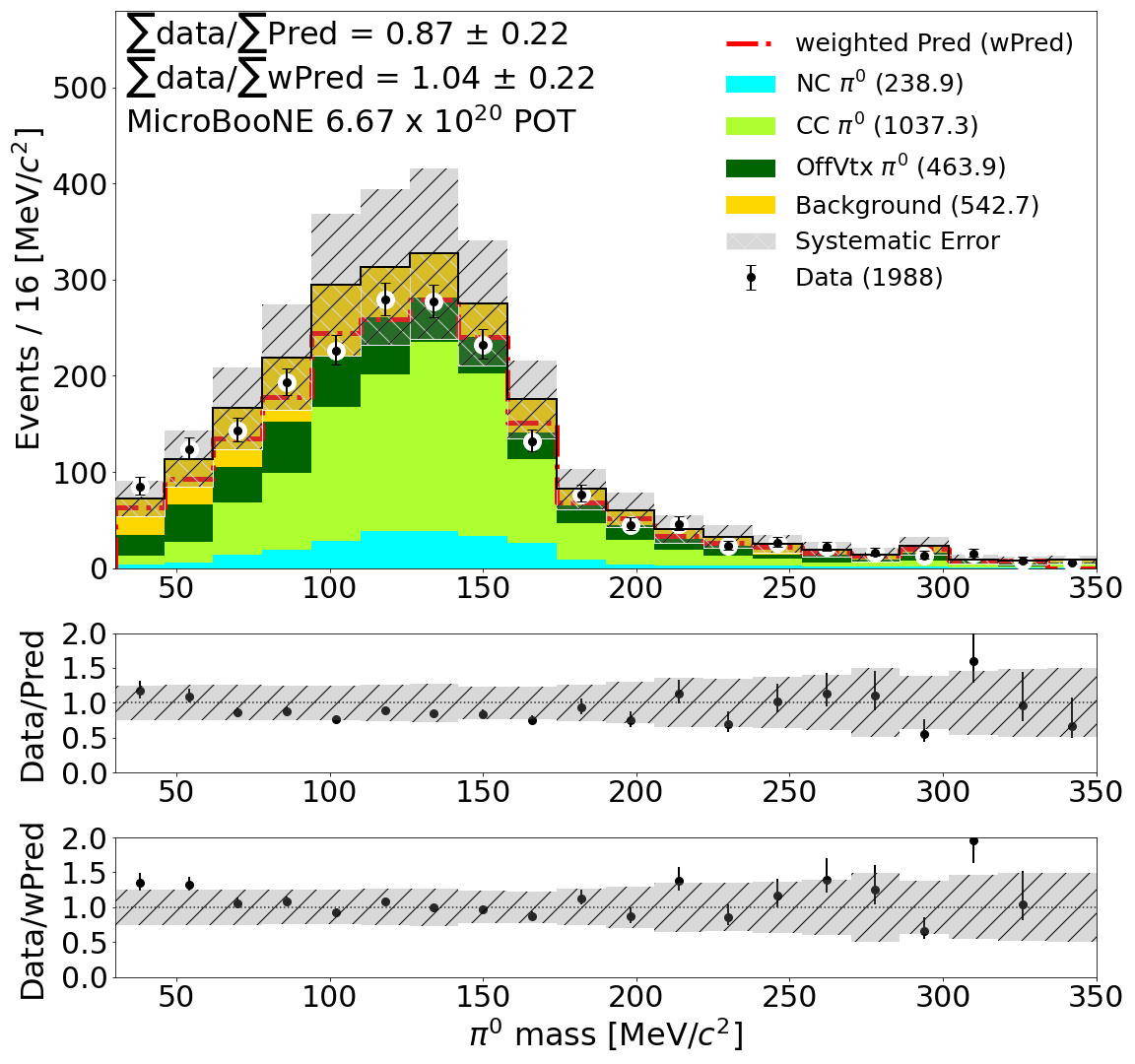}
\caption{The $\pi^0$ mass test variable for events passing the $\pi^0$ selection criteria. The colored stacked histograms represent the standard simulation (MC) prediction. The red dashed line is the total weighted (wMC) prediction. The data points are shown by the black points. The lower two panels show the data/MC ratio and data/wMC ratio respectively. The $\chi_\text{CNP}^2/20(dof) = 0.709$ with a p-value of 0.821 for the MC prediction. The $\chi_\text{CNP}^2/20(dof) = 0.778$ with a p-value of 0.744 for the wMC prediction. }
\label{fig:pi0mass}
\end{figure}

The resulting invariant mass peak calculated using Eq.~\ref{eq:pi0mass} is shown in Fig.~\ref{fig:pi0mass}. In addition to the standard  simulation, the $\pi^0$ weighted distribution is also shown. The weights will be discussed in the next section.

\subsection{Use of the $\pi^0$ sample to constrain the $\pi^0$ background rate}
\label{sec:pi0weights}

An overall deficit of the data-to-prediction ratio is observed in Fig.~\ref{fig:pi0mass}.  While the ratio is unity within systematic uncertainties, we correct it to better estimate the $\pi^0$ background to the 1e1p signal. We apply a re-weighting scheme based on the true $\pi^0$ momentum of the event that is similar to the process used by MiniBooNE~\cite{ref:MinibooneReweight}.  

\begin{figure*}[t]
    \centering
    \subfloat[CC Sample]{
    \begin{tabular}{@{}c@{}}
    \includegraphics[width=.48\textwidth]{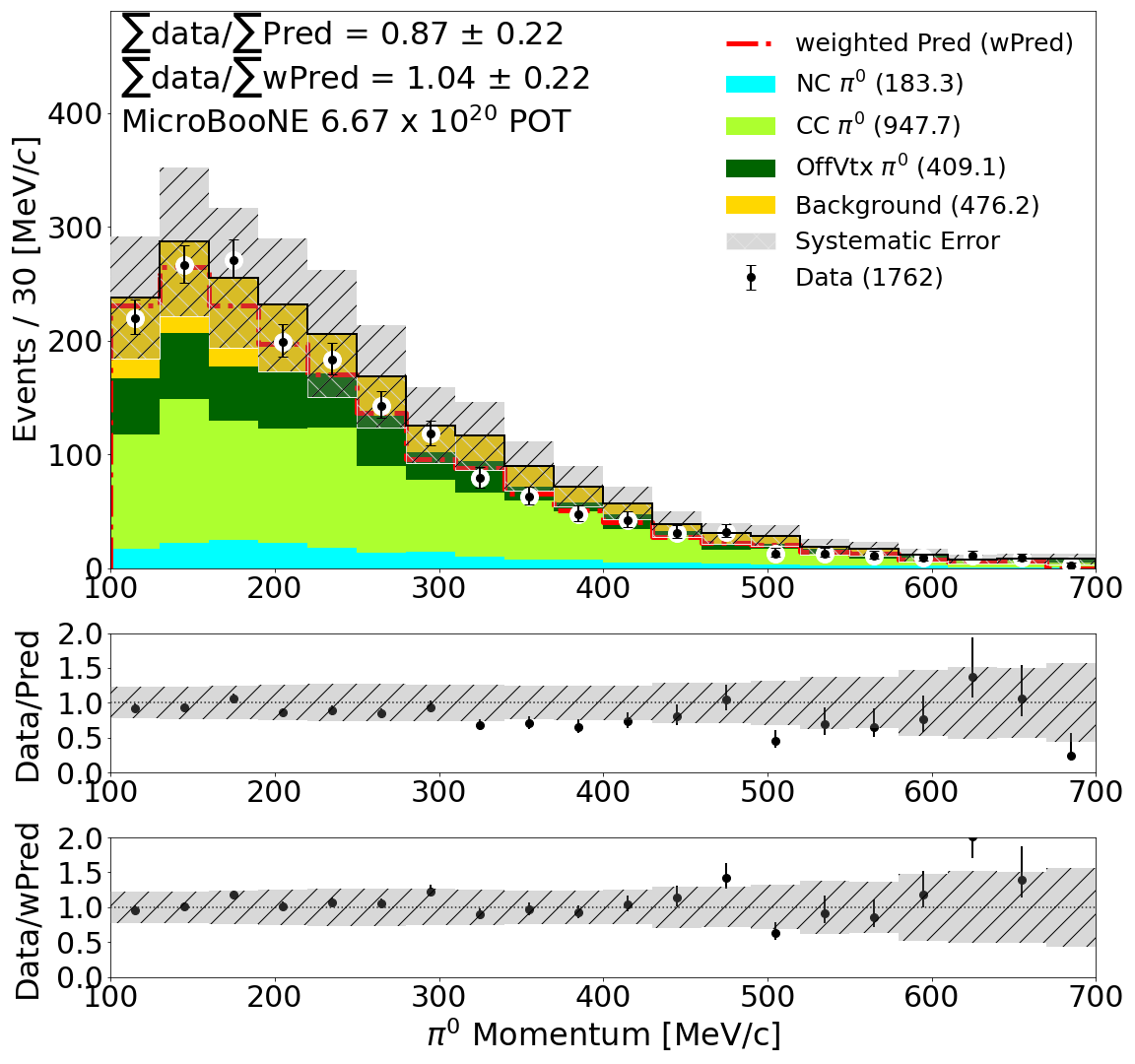}
    \end{tabular}}
    \subfloat[NC Sample]{
    \begin{tabular}{@{}c@{}}
    \includegraphics[width=.48\textwidth]{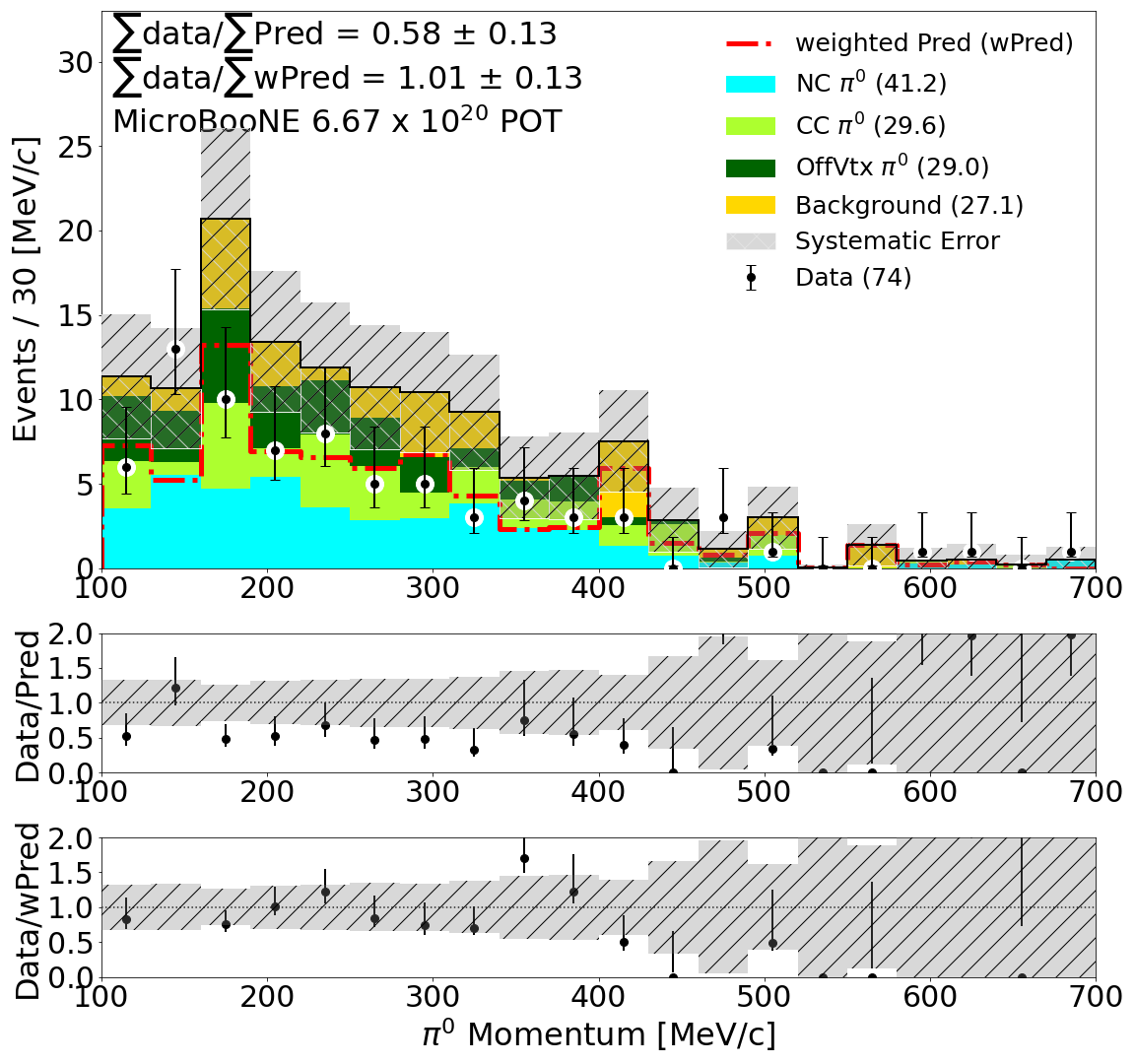}
    \end{tabular}}
    \\
\caption{Reconstructed $\pi^0$ momenta for the CC (a) and NC (b) samples. The colored stacked histograms represent the standard simulation (MC) prediction. The red dashed line is the total weighted (wMC) prediction. The data points are shown by the black points. The lower two panels of each figure show the data/MC ratio and data/wMC ratio respectively. In the CC sample, the $\chi_\text{CNP}^2/20(dof) = 0.619$ with a p-value of 0.902 for the MC prediction and the $\chi_\text{CNP}^2/20(dof) = 0.405$ with a p-value of 0.991 for the wMC prediction. In the NC sample, the $\chi_\text{CNP}^2/20(dof) = 0.555$ with a p-value of 0.944 for the MC prediction and the $\chi_\text{CNP}^2/20(dof) = 0.490$ with a p-value of 0.972 for the wMC prediction.
}
\label{fig:pi0momentum}
\end{figure*}

The first step is to divide the sample into sets that enhance CC and NC events using the MPID muon interaction score. This allows for different weights based on event type. A selection criterion value of 0.05 MPID muon interaction score was chosen to retain a purity of both the CC ($>0.05$) and NC ($<0.05$) samples above 60\%.  The simulated sample is further broken into two categories: ``good'' and ``background.'' Good simulated events for the purpose of this re-weighting study include all selected events that contain a true $\pi^0$ in the final state. Fig.~\ref{fig:pi0momentum} shows the reconstructed $\pi^0$ momentum for the CC and NC samples.

The ratio fit for reweighting is:
\begin{equation}
    R = \frac{ \textrm{(Data--background)}:\textrm{unfolded reco momentum}}{ \textrm{(Good simulated Events):true momentum}}.
\end{equation}
The numerator represents ``good" data events, with the reconstructed distribution unfolded back to truth. A polynomial fit is calculated for each  distribution. These fits give the reweighting formulae as a function of $p$, the true $\pi^0$ momentum in (MeV/$c$):
\begin{equation}
    \text{CC Weight} = (3.4\times{10^{-6}})p^2-(0.0039)p +1.784,
\end{equation}
\begin{equation}
    \text{NC Weight} = (1.89\times{10^{-6}})p^2-(0.0032)p +1.442.
\end{equation}
The weight formula is applied to every MC simulation event that contains a true $\pi^0$. The result on the $\pi^0$ selection is seen in Fig.~\ref{fig:pi0momentum}. This plot shows improved agreement between data and simulation for both the CC and NC samples. These weight formulas are further applied to selected 1e1p and 1$\mu$1p MC simulation events which contain a true $\pi^0$ in the final state. In the $1e1p$ analysis, the application of these weights reduces the total predicted non-$\nu_e$ CC background rate across 200--1200\,MeV from 3.6 events to 2.8 events. The Landau+linear background fit procedure described in Sec.~\ref{sec:numubkgfit} then increases the total prediction to 3.2 events. An additional use of the $\pi^0$ selection to investigate $1\gamma 1p$ events is described in  Ref.~\cite{Supp}.

\subsection{The sample of Michel electrons from $\nu_\mu$ charged-current events }
\label{sec:michel}

In order to assess the performance of the shower reconstruction algorithm at low energies, we have identified a sample of Michel electron events. This is possible using our reconstruction tools because stopped muons that decay to a Michel electron form a two-prong topology in the detector. The selection criteria used to define this sample are as follows; 1) two prongs are reconstructed at the vertex, the longer (shorter) track being the candidate muon (Michel); 2) the long-prong track-length is $>$ 100\,cm;
3) the short-prong track-length is $<$ 30\,cm; 
4) the long track consists of $< 20\%$ shower pixels;
5) the short track consists of $> 80\%$ shower pixels; and
6) $\phi_\mu <$ 0.5\,rad (where $\phi_\mu$ is the angle between the muon track and the $x$-$z$ plane ). 
Lastly, in events with more than one identified vertex that passes all the other criteria, we keep only the first vertex found by the vertex algorithm~\cite{MicroBooNE:2020sar}. This approach is chosen because it does not bias the Michel energy spectrum.

 The purpose of this sample is to compare the shower reconstruction between data and simulation.  Therefore it is important to reject contributions from cosmic overlay, which are{ \it in situ} measurements rather than simulation, while isolating the simulated Michel electrons coming from muons created in charged-current (CC) $\nu_\mu$ interactions. The final selection criterion on $\phi_\mu$ above is designed to remove Michel electrons from downward-going cosmic ray muons to optimize the purity of Michel electrons from these CC $\nu_\mu$ interactions. The data set used for this sample is the $5 \times 10^{19}$~POT subset described in Sec.~\ref{subsec:blindnessprocedure}.
 
\begin{figure}
     \centering
     \includegraphics[width=0.5\textwidth]{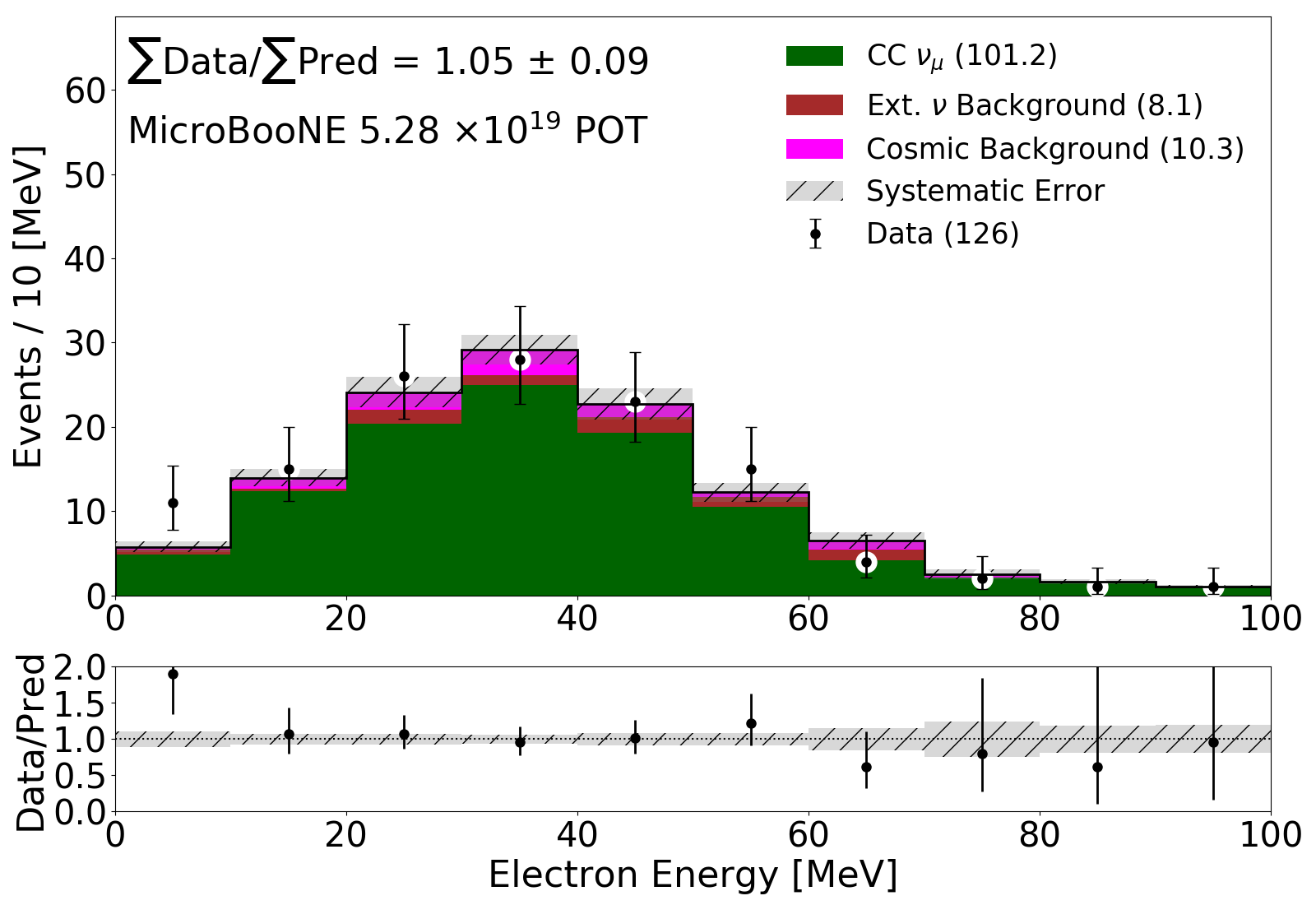}
     \caption{The Michel electron energy spectrum. The data observation is given by the black points and the prediction is given by the stacked histogram. The systematic error here corresponds only to the statistical uncertainty on the simulated events. The $\chi^2_\text{CNP}$/dof for this distribution is 5.9/10.}
     \label{fig:michelshowerE}
 \end{figure}

 The electron energy distribution for this Michel electron sample is shown in Fig.~\ref{fig:michelshowerE}. Though we explicitly neglect external events in the other samples isolated for this analysis, we do find a non-negligible contribution to this Michel sample consisting of muons from external $\nu_\mu$ interactions that enter the detector, stop, and subsequently decay. Therefore, we do not ignore these events when estimating the Michel prediction—they are represented by the brown region (labeled ``Ext. $\nu$ Background'') in Fig~\ref{fig:michelshowerE}.
 The data/prediction agreement indicates that the shower algorithm is well modeled by the simulation down to low energies. This agreement has been explored in more quantitative detail in~\cite{MicroBooNE:2021nss} and will be discussed further in the next section.
 
\subsection{ Pixel intensity-to-energy conversion from the $\pi^0$ and Michel electron samples}

A detailed study validating the total pixel intensity ($Q_{sh}$)-to-energy conversions for showers using the $\pi^0$ and Michel electron samples is presented in~\cite{MicroBooNE:2021nss} and summarized here. The $Q_{sh}$-to-energy conversion factor is derived from simulation, so this study validates the conversion factor and the agreement seen between data and simulation. 
For both the $\pi^0$ and Michel electron sample, the conversion factor that gives the best fit to a known physical value is found for the data and simulation samples. For the $\pi^0$ fit, the $\pi^0$ rest mass of 135\,MeV/$c^2$ is used, while for the Michel electron fit, the Michel energy spectrum cut-off at 52.8\,MeV is used.

As discussed in Ref.~\cite{MicroBooNE:2021nss}, the best fit $Q_\mathrm{sh}$-to-energy value agrees between data and simulation for the two samples within $3\%$. Further, the data and simulation best fit values agree with the simulation-derived $Q_\mathrm{sh}$-to-energy conversion value within $6\%$. While the best fit $Q_\mathrm{sh}$-to-energy values for each of the two samples do not exactly match, the difference seen between the values is sufficiently low given the $E_{\nu}$ resolution of $16.5\%$.

\section{Uncertainties}
\label{sec:unc}

\begin{figure*}
    \centering
    \subfloat[$1e1p$ uncertainties]{\includegraphics[width=0.95\columnwidth]{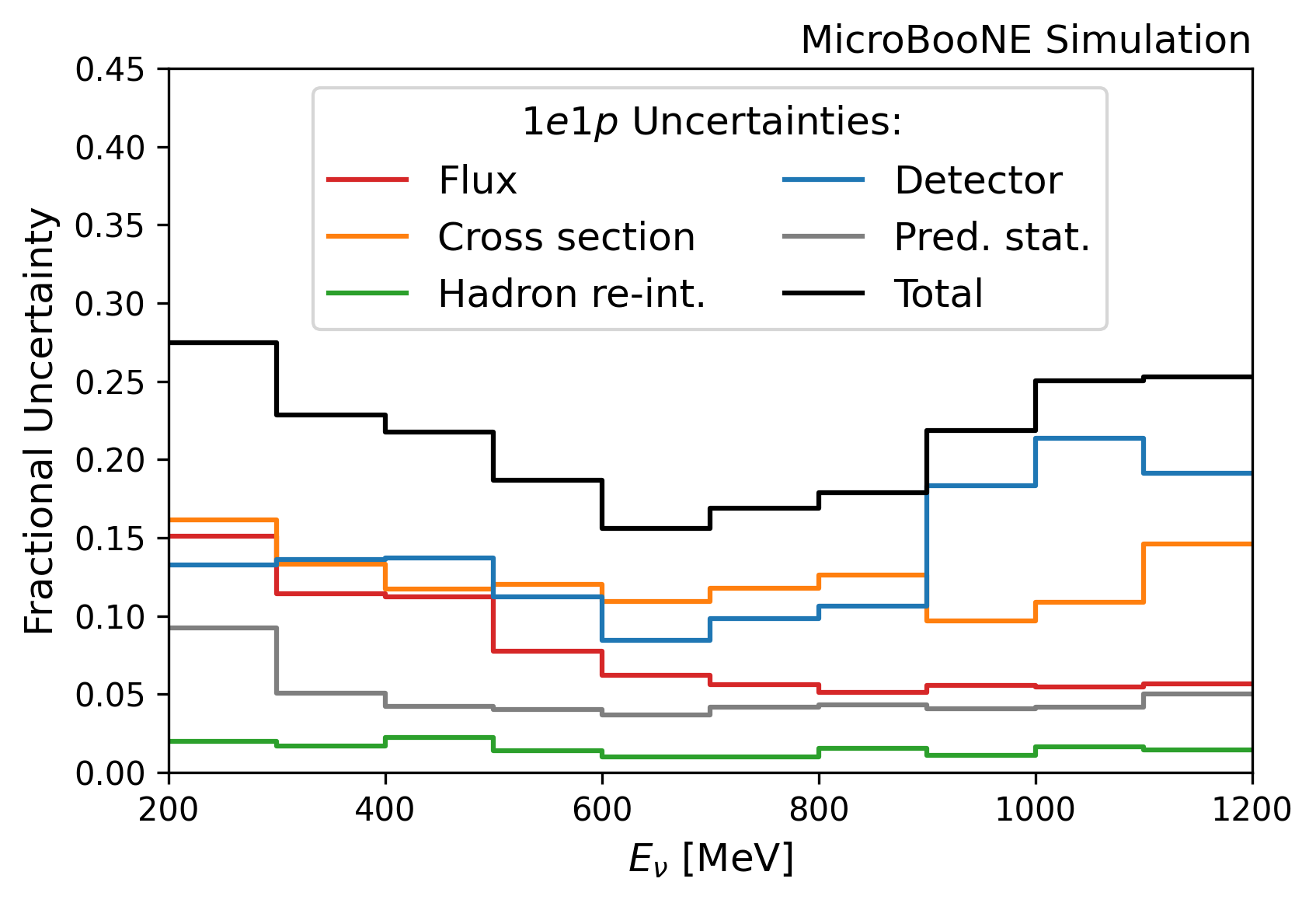}}
    $\quad$
    \subfloat[$1\mu 1p$ uncertainties]{\includegraphics[width=0.95\columnwidth]{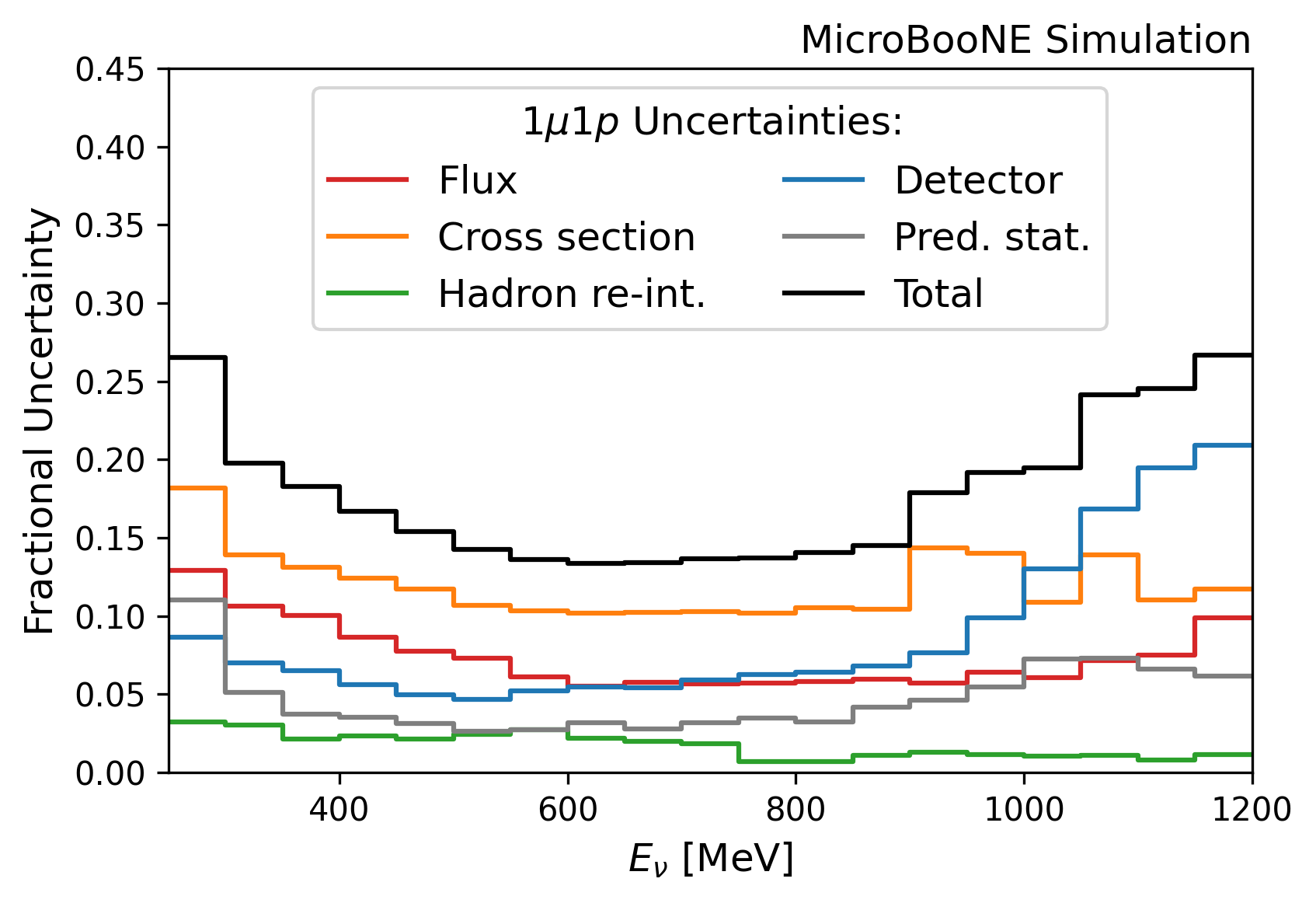}}
    \caption{Fractional uncertainties on the prediction for the $1e1p$ (a) and $1\mu 1p$ (b) event selections as a function of reconstructed neutrino energy.}
    \label{fig:1l1p_unc}
\end{figure*}

Uncertainties on the prediction in this analysis arise from five sources: the beam flux prediction, the neutrino--nucleus interaction model, the hadron re-interaction model, the detector simulation, and the finite statistics in the samples used to form the prediction. The fractional uncertainties due to each of these sources of systematic error for the $1e1p$ and $1\mu 1p$ predicted event spectra are shown in Fig.~\ref{fig:1l1p_unc}.

Two distinct methods are used to assess the systematic uncertainties.
Where possible, the uncertainties are evaluated using event reweighting that is applied to the primary simulation samples. For each systematic parameter variation, events are assigned a weight based on their truth-level information. The uncertainty is determined from the resulting changes in the reconstructed spectrum. This method is used for the flux, neutrino interaction, and hadron re-interaction uncertainties.
On the other hand, the detector-related systematic uncertainties are evaluated using a set of samples in which the detector simulation has been varied. Comparisons between these modified samples and the nominal simulation are used to estimate the uncertainty on the prediction.
Additionally, we note that the uncertainties on the $\nu_\mu$ backgrounds to the $1e1p$ selection are assessed based on the simulation events that pass the selection. The fractional uncertainty calculated using those events is applied to the background prediction derived from the procedure described in Sec.~\ref{sec:numubkgfit}.

\begin{figure*}
    \centering
    \subfloat[Fractional covariance matrix]{\includegraphics[height=0.75\columnwidth]{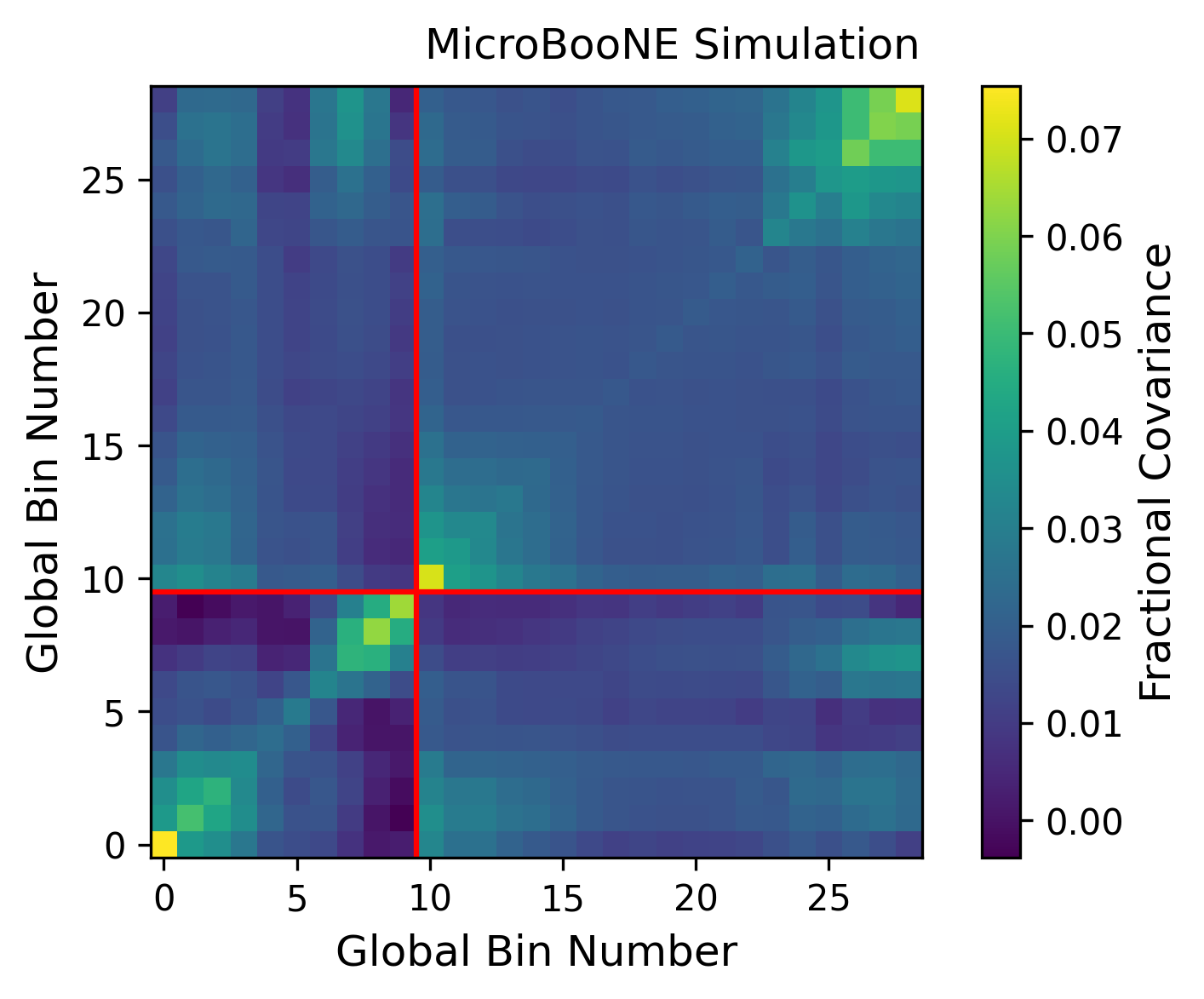}}
    $\quad$
    \subfloat[Correlation matrix]{\includegraphics[height=0.75\columnwidth]{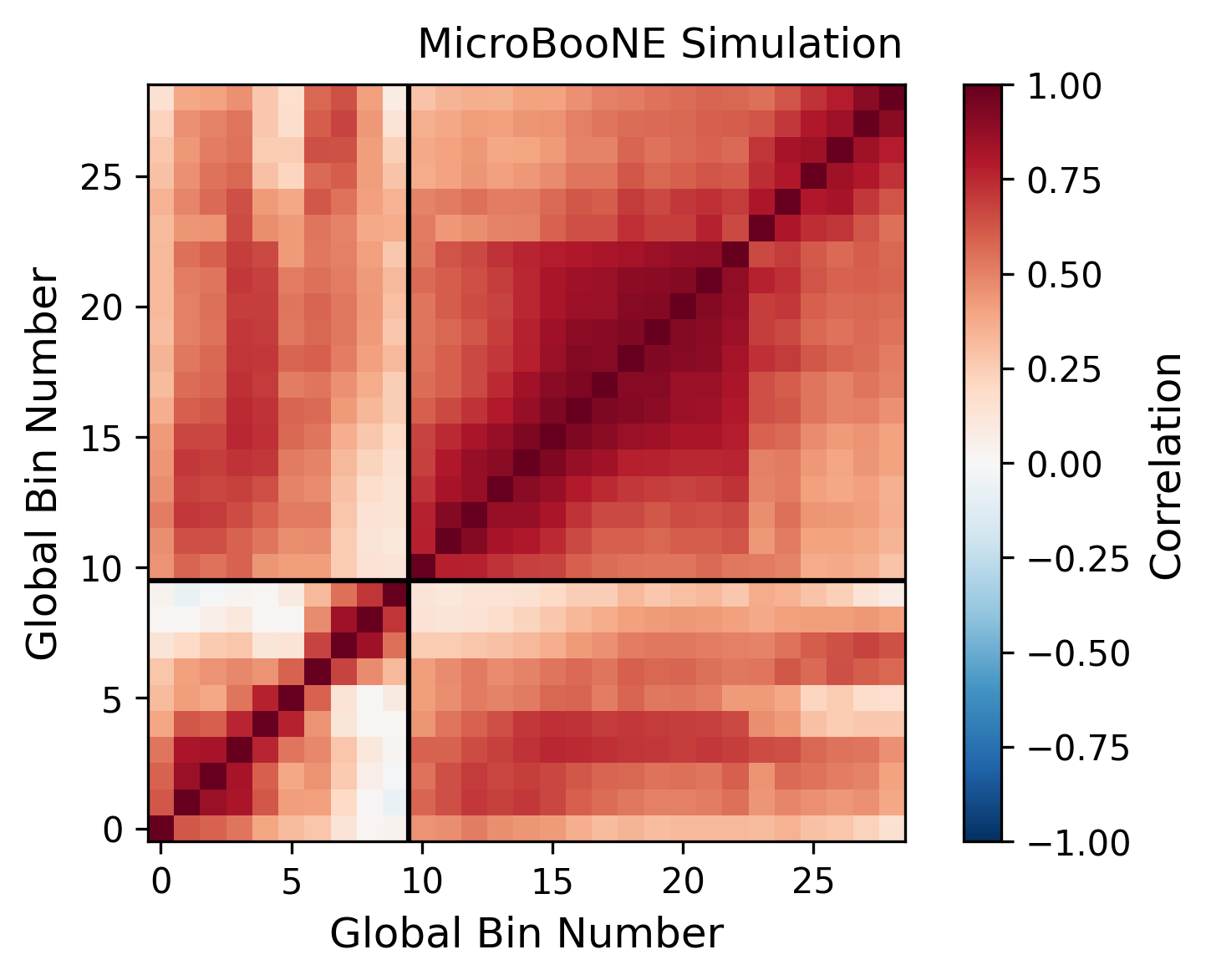}}
    \caption{Total fractional covariance (a) and correlation (b) matrices for the $1e1p$ and $1\mu 1p$ selections. The $1e1p$ events are on the lower-left and are binned from 200--1200\,MeV in 100\,MeV bins. The $1\mu 1p$ events are on the upper-right and are binned from 250--1200\,MeV in 50\,MeV bins. The solid bold lines indicate the boundary between the $1e1p$ and $1\mu 1p$ selections.}
    \label{fig:total_covar}
\end{figure*}

The uncertainties are incorporated into this analysis using the covariance matrix formalism. A covariance matrix encodes the variance ($\sigma^2$) of the contents of each histogram bin in diagonal entries, and the covariance between the contents of pairs of histogram bins in off-diagonal entries. The off-diagonals therefore provide information about correlations between the analysis bins.
Having such information is particularly important for applying the constraint from the $1\mu 1p$ selection to the $1e1p$ prediction, as this relies on the correlations between the selected events in the $1e1p$ and $1\mu 1p$ histograms.
However, given the fitting procedure used to obtain the prediction for the $\nu_\mu$ backgrounds to the $1e1p$ selection that was described in Section~\ref{sec:numubkgfit}, we do not use the correlations for that contribution to the spectrum; their contribution to the off-diagonals is removed from the covariance matrix.
The total covariance matrix is the sum of the covariance matrices due to each source of uncertainty.
The joint fractional covariance and correlation matrices for the $1e1p$ and $1\mu 1p$ selections as a function of the reconstructed neutrino energy is shown in Fig.~\ref{fig:total_covar}. Given the covariance matrix $M$ with entries $M_{ij}$ and a spectrum with bin contents $N_i$, the fractional covariance matrix is given by $F_{ij} = M_{ij}/(N_i N_j)$ and the correlation matrix is given by $\rho_{ij} = M_{ij} / \sqrt{M_{ii} M_{jj}}$.
Additional covariance matrices are provided in Ref.~\cite{Supp}.

\subsection{Flux Uncertainties}
\label{subsec:flux_sys}

The systematic uncertainties on the flux prediction can be grouped into three main sources: hadron production in the target, secondary hadron interactions, and the properties of the magnetic focusing horn.
Flux uncertainties are evaluated by reweighting events according to the properties of the parent hadron, including the hadron species and its direction and momentum after leaving the focusing horn. The MicroBooNE flux reweighting method is based on the techniques previously developed by the MiniBooNE collaboration~\cite{MiniBooNE:2008hfu} and adapted to the MicroBooNE detector location~\cite{MicroBooNE:2019nio}.
The 2\% uncertainty on the number of protons delivered is neglected, as such normalization effects are effectively constrained by the $1\mu 1p$ data observation via the procedure described in Sec.~\ref{sec:constr}.

The largest flux uncertainty for this analysis is from $\pi^+$ hadron production, which is the main source of neutrinos in the BNB below 1\,GeV.
The flux effects are generally highly correlated between the $1e1p$ and $1\mu 1p$ selections because the neutrinos in both are coming from the same population of parent hadrons.

\subsection{Neutrino Interaction Uncertainties}
\label{subsec:xsec_sys}

The neutrino interaction model includes a wide variety of reweightable parameters that can be broken down into two broad categories: parameters associated with each interaction mode, and parameters for final state interactions (FSI) that affect all neutrino interaction modes.
Events are reweighted according to the properties of the neutrino's interaction with the nucleus and the interactions of the resulting particles as they exit the nucleus. These parameters and their uncertainties are described in detail in Ref. ~\cite{MicroBooNE:2021genie_tune}.
We also consider uncertainties related to differences in the CCQE cross sections for electron and muon neutrinos due to second-class currents~\cite{Day:2012}.
Some inconsistencies were identified in the GENIE v3.0.6 reweighting code used to evaluate FSI-related systematic uncertainties, but the effect of these were found to have a negligible impact on the overall analysis sensitivity and so have been neglected \cite{MicroBooNE:2021genie_tune}.

As this analysis targets events that are kinematically consistent with CCQE interactions, the neutrino interaction model parameters that dominate our cross section systematic uncertainties are primarily those related to the CCQE interaction mode and, secondarily, those related to FSI.
The FSI parameters are important because they can change the fraction of true CCQE interactions that will have a $1\ell 1p$ (one lepton, one proton, and no mesons) final-state topology with kinematics consistent with two-body scattering, and, therefore, the rate at which CCQE interactions will be selected in this analysis.  
Similarly, FSI can allow non-CCQE events to appear CCQE-like.
These effects are common to the $1e1p$ and $1\mu 1p$ selections, so the FSI systematic uncertainties are well correlated between the two channels.
The MEC contribution has been minimized by the two-body scattering kinematic consistency requirement. As a result the MEC uncertainties are relatively small.

\subsection{Hadron Re-Interaction Uncertainties}
\label{subsec:g4_sys}

The systematic uncertainties on hadron re-interactions consider variations in the hadron--argon interaction cross sections for protons, $\pi^+$, and $\pi^-$.  
Hadron re-interaction systematic uncertainties are evaluated by reweighting events according to the Geant4 truth information that describes the trajectories of the hadrons after they leave the argon nucleus~\cite{geant4reweight:2021}.
These uncertainties are small in this analysis. %

\subsection{Detector Response Uncertainties}
\label{subsec:detsys}

The systematic uncertainties related to the detector response are evaluated using a set of simulation samples in which the detector model has been varied in some way.
We consider several types of variations: modifications in the amplitude and width of signals on the wire waveforms as a function of $x$ position, $(y,z)$ position, and detector angles $\theta_{XZ}$ and $\theta_{YZ}$ of the local particle trajectory~\cite{MicroBooNE:2021wiremod}; variations in the electron--ion recombination parameters; an alternative electric field map in the TPC; and variations in the light yield, the light attenuation, and the Rayleigh scattering length.
Comparisons between these modified samples and the nominal simulation are used to estimate the uncertainty on the prediction.

Producing the detector variation samples is computationally expensive, and as a result they suffer from limited statistics. Thus the comparisons between the modified samples and the nominal simulation contain some statistical jitter. 

As the underlying true spectra (e.g., the neutrino energy spectrum) are expected to be smooth, we exploit a KDE algorithm~\cite{KDEcite} to estimate most contributions to our detector variation predictions. For a given detector variation sample, each event is assigned a kernel function with some width, and the final spectrum for this variation is obtained by summing all kernels. This allows us to reduce the effect of the statistical fluctuations in the spectrum obtained from each detector variation sample without making any other assumptions about the impact of the variation on our analysis variables.

The kernel used in this analysis is an Epanechnikov kernel, which is optimal in a mean square error sense~\cite{epach}, and the bandwith used is the Sheather-Jones bandwidth~\cite{SJ}. The KDE-smoothed spectra are used in the computation of the covariance matrix, including both the on- and off-diagonal entries.


The only case where we do not use KDE smoothing for evaluating the detector response uncertainties is the misidentified $\nu_\mu$ backgrounds to the $1e1p$ prediction. The limited statistics in the detector variation sample combined with the high purity of the $1e1p$ selection leave us with too few events to obtain robust results from the KDE algorithm.
We instead assess a 20\% systematic uncertainty on these backgrounds, based on the magnitude of the differences observed in the detector variation samples within the limited statistics available. This uncertainty is treated as uncorrelated between the analysis bins.
We note that this will have a small effect compared to the $\mathcal{O}(100\%)$ Poisson statistical errors on the number of muon neutrino and neutral current events in each $1e1p$ analysis bin.

\subsection{Uncertainties due to Finite Statistics}
\label{subsec:mc_stat}

The predicted spectra are also subject to statistical uncertainties due to finite statistics in the samples used to form the prediction, including both simulated samples and off-beam data.
For the muon neutrino and neutral current contributions to the $1e1p$ prediction, the statistical errors are determined as part of the background fitting procedure described in Sec.~\ref{sec:numubkgfit}.
For all other contributions to the predicted spectra, the statistical uncertainty in each bin is calculated as
\begin{equation}
    \sigma_\text{pred stat} = \sqrt{\sum_i w_i^2}
\end{equation}
where the sum runs over all of the events in the bin and $w_i$ is the weight of each event, including the scaling factor needed to match the data POT, the cross section model tune weight, and the $\pi^0$ weights where applicable.
This uncertainty is added to the covariance matrix.

\section{Final Predictions and Statistical Method}
\label{sec:finalpred}

This section describes the method for applying the $1\mu 1p$ constraint to the $1e1p$ prediction using the selections and uncertainties described above.   
This section also discusses the statistical tests that we use to quantitatively compare our observation to the constrained prediction without and with the LEE signal.

\subsection{Constrained Predictions}
\label{sec:constr}

A key component of this analysis is the use of well-measured data sets to constrain the predicted $1e1p$ spectrum and its uncertainties. The concepts used here were originally developed for the MiniBooNE analysis~\cite{MiniBooNE:2007uho}. This procedure uses the $1\mu1p$ data sample described in Sec.~\ref{sec:1m1p}.
We implement the constraint using the conditional covariance method. In this procedure, the $1e1p$ and $1\mu 1p$ spectra are treated as coming from a multivariate normal distribution defined by their predictions and the associated joint covariance matrix shown in Fig.~\ref{fig:total_covar}. We then condition on the $1\mu 1p$ data observation. The result is an updated prediction for the $1e1p$ selection and an associated updated covariance matrix, which incorporates the information provided by the data observation in the $1\mu 1p$ compared to its prediction.

\begin{figure}
    \centering
    \includegraphics[width=0.5\textwidth]{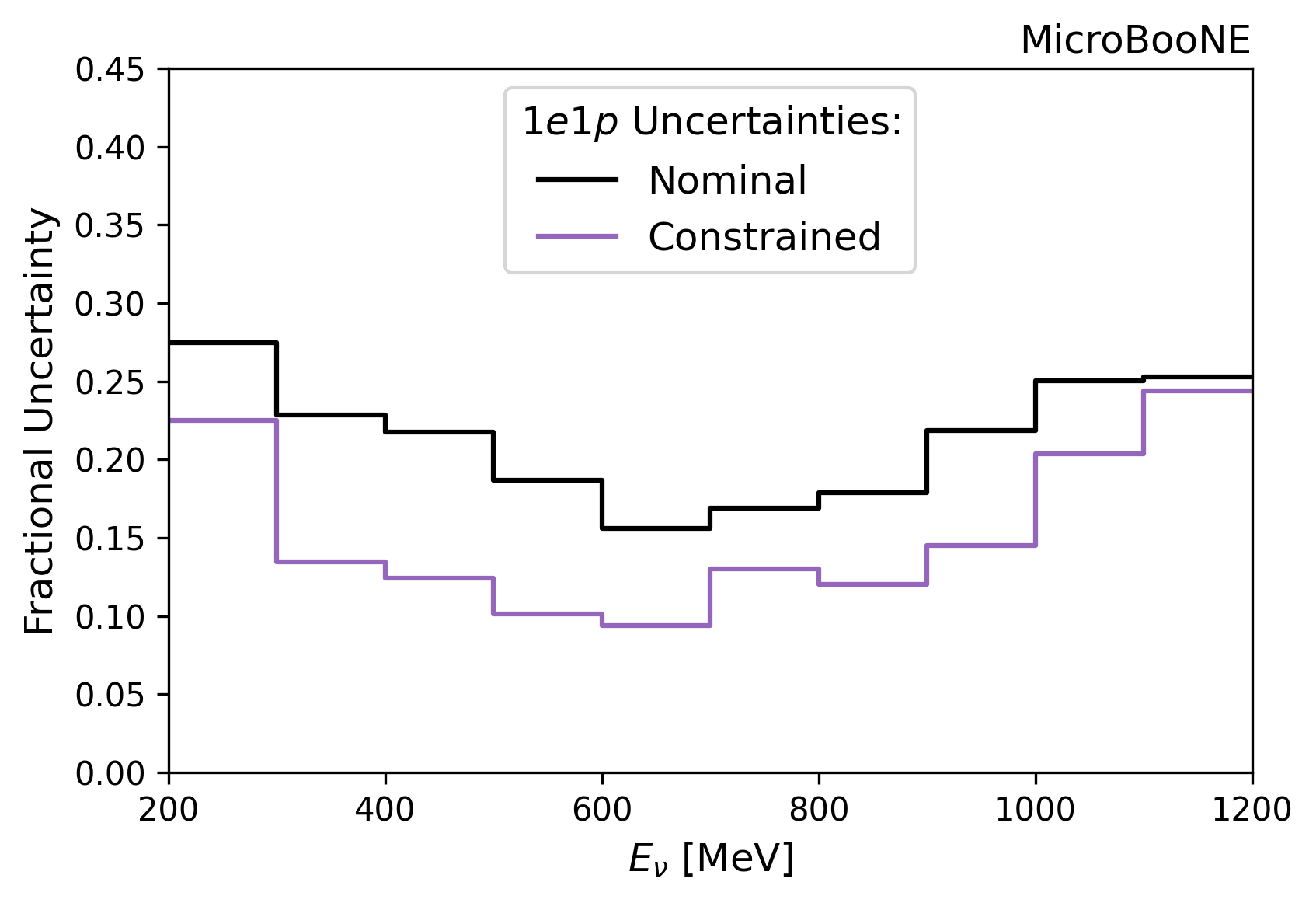}
    \caption{Fractional systematic uncertainties for the $1e1p$ prediction as a function of reconstructed neutrino energy, before and after the $1\mu 1p$ constraint. 
    The black curve is the same as the total systematic uncertainty shown in Fig.~\ref{fig:1l1p_unc}(a).}
    \label{fig:constr_unc}
\end{figure}

The uncertainties on the prediction decrease due to the information that is gained from the $1\mu 1p$ observation. The decrease in the fractional uncertainties is shown in Fig.~\ref{fig:constr_unc}.
After applying the constraint from the $1\mu 1p$ observation, the uncertainties on the $1e1p$ are dominated by the data statistical uncertainties.
The final constrained $1e1p$ prediction is shown in Fig.~\ref{fig:1e1p_constrained}, and is further discussed in Sec.~\ref{sec:res}.
We note here that the normalization of the $\nu_e$ contribution to the prediction increases somewhat compared to the unconstrained prediction shown previously in Fig.~\ref{fig:Enu_stacked}, consistent with the slight excess observed in the $1\mu 1p$ data.

\subsection{Statistical Tests}
\label{sec:stat_test}

We test the agreement of the $1e1p$ observation with two benchmark hypotheses: the prediction consisting of the irreducible $\nu_e$ backgrounds and the misidentified $\nu_\mu$ backgrounds, called $H_0$; and prediction containing the median MiniBooNE LEE prediction described in Sec.~\ref{sec:lee_model} stacked on top of the expected backgrounds, called $H_1$.
To quantify the agreement of the $1e1p$ data with these hypotheses, we utilize three sets of statistical tests, which are described below.
The results of these tests are presented in Sec.~\ref{sec:res}.

The first set of statistical tests check the goodness-of-fit between the observed $1e1p$ data and the predictions, considered both with and without the $1\mu 1p$ constraint described in Sec.~\ref{sec:constr}. This test is performed separately for each hypothesis. It is also performed for both the LEE range (200--500\,MeV) and the full analysis range (200--1200\,MeV).
A goodness-of-fit test quantifies the level of agreement between the observation and a given prediction within the uncertainties without reference to an alternative prediction.
For each goodness-of-fit test, we use the $\chi^2_\text{CNP}$ test statistic~\cite{Ji:2019yca} and apply a frequentist method to determine the corresponding $p$-value.

The frequentist studies are performed using $10^5$ pseudo-experiments that are generated assuming that the hypothesis being tested is the underlying truth.
In each pseudo-experiment, we sample from the multivariate normal distribution defined by the hypothesis's mean prediction and the associated covariance matrix. In each bin, we then sample from a Poisson distribution defined by the rate parameter (mean) that is equal to the value sampled from the multivariate normal distribution. This gives a pseudo-observation that has integer numbers of events in each bin and incorporates both systematic and statistical uncertainties.
The results of these pseudo-experiments allow us to build up a probability distribution for the test statistic under the given hypothesis. The value of the test statistic in data is compared to this probability distribution in order to determine the $p$-value.
The frequentist studies are carried out in the SBNfit software package, which provides support for such calculations~\cite{sbnfit}.

The second set of statistical tests is two-hypothesis tests which provide another method for determining compatibility between the observed $1e1p$ data and the $H_0$ and $H_1$ hypotheses.
For these tests, we use the $1e1p$ predictions for the full analysis range incorporating the $1\mu 1p$ constraint.
The test statistic is $\Delta \chi^2 = \chi^2_{H_0} - \chi^2_{H_1}$, where each $\chi^2$ is computed using the $\chi^2_\text{CNP}$ formalism. This $\Delta \chi^2$ is as an approximation of the log-likelihood ratio.
For these tests, we are primarily interested in the probability that each hypothesis would give a $\Delta \chi^2$ value greater than (less than) the data observation in the case where the data observation falls above (below) the median of the $\Delta \chi^2$ for that hypothesis.
These probabilities are calculated using the same frequentist method described above.
The corresponding significance is computed in the one-sided manner. 

We can evaluate the analysis sensitivity in terms of the two-hypothesis tests.
We define our sensitivity to exclude the $H_0$ hypothesis if the $H_1$ hypothesis is true based on the probability that a $\Delta \chi^2$ from the $H_0$ pseudo-experiments is greater than the median $\Delta \chi^2$ of the $H_1$ pseudo-experiments. This probability is 0.003, which translates to a significance of $2.7\sigma$.
Similarly, we define our sensitivity to exclude $H_1$ if $H_0$ is true based on the probability that a $\Delta \chi^2$ from the $H_1$ pseudo-experiments is less than the median $\Delta \chi^2$ of the $H_0$ pseudo-experiments. This probability is 0.017, which translates to a significance of $2.1\sigma$.

The third statistical test that we consider is a fit for the LEE signal strength parameter, $x_\text{LEE}$, introduced in Sec.~\ref{sec:lee_model}. 
For this fit, we use the $1e1p$ predictions for the full analysis range that incorporate the $1\mu 1p$ constraint.
We fit by minimizing the $\chi^2_\text{CNP}$, where both the prediction and the associated covariance matrix are updated for each value of $x_\text{LEE}$.
We determine the confidence interval on the LEE signal strength using the Feldman--Cousins procedure~\cite{Feldman:1998}.

We can also evaluate the analysis sensitivity in terms of the confidence intervals that would be obtained if the observation was the expectation under a given hypothesis. In this case, we define the sensitivity based on the $\Delta \chi^2$ for the Asimov data set of the hypothesis being considered.
For $H_0$ ($x_\text{LEE} = 0$), the expected upper limit on the LEE signal strength is 0.75 (0.98) at the 90\% ($2\sigma$, or $\sim$95\%) confidence level.
For $H_1$ ($x_\text{LEE} = 1$), the expected confidence interval for the LEE signal strength is [0.53, 1.66] ([0.28, 2.67]) at the $1\sigma$ (2$\sigma$) confidence level. The expected significance to rule out $x_\text{LEE} = 0$ using this method is $2.8\sigma$.

\section{Results}
\label{sec:res}

A total of 25 $1e1p$ events are selected, while a total of $29.0 \pm 1.9_\text{(sys)} \pm 5.4_\text{(stat)}$ ($27.4 \pm 3.8_\text{(sys)} \pm 5.2_\text{(stat)}$) events are predicted for the analysis range (200--1200\,MeV) with (without) the $1\mu 1p$ constraint.
Using the goodness-of-fit test described in Sec.~\ref{sec:stat_test}, we find the $p$-value for the comparison between the $1e1p$ observation and the expected background prediction to be 0.014 (0.024) for the  prediction with (without) the $1\mu 1p$ constraint; this indicates $2.5\sigma$ ($2.3\sigma$) tension with the prediction.  
This and other goodness-of-fit test results are summarized in Table~\ref{tab:gof}.

\begin{figure}
    \centering
    \includegraphics[width=\linewidth]{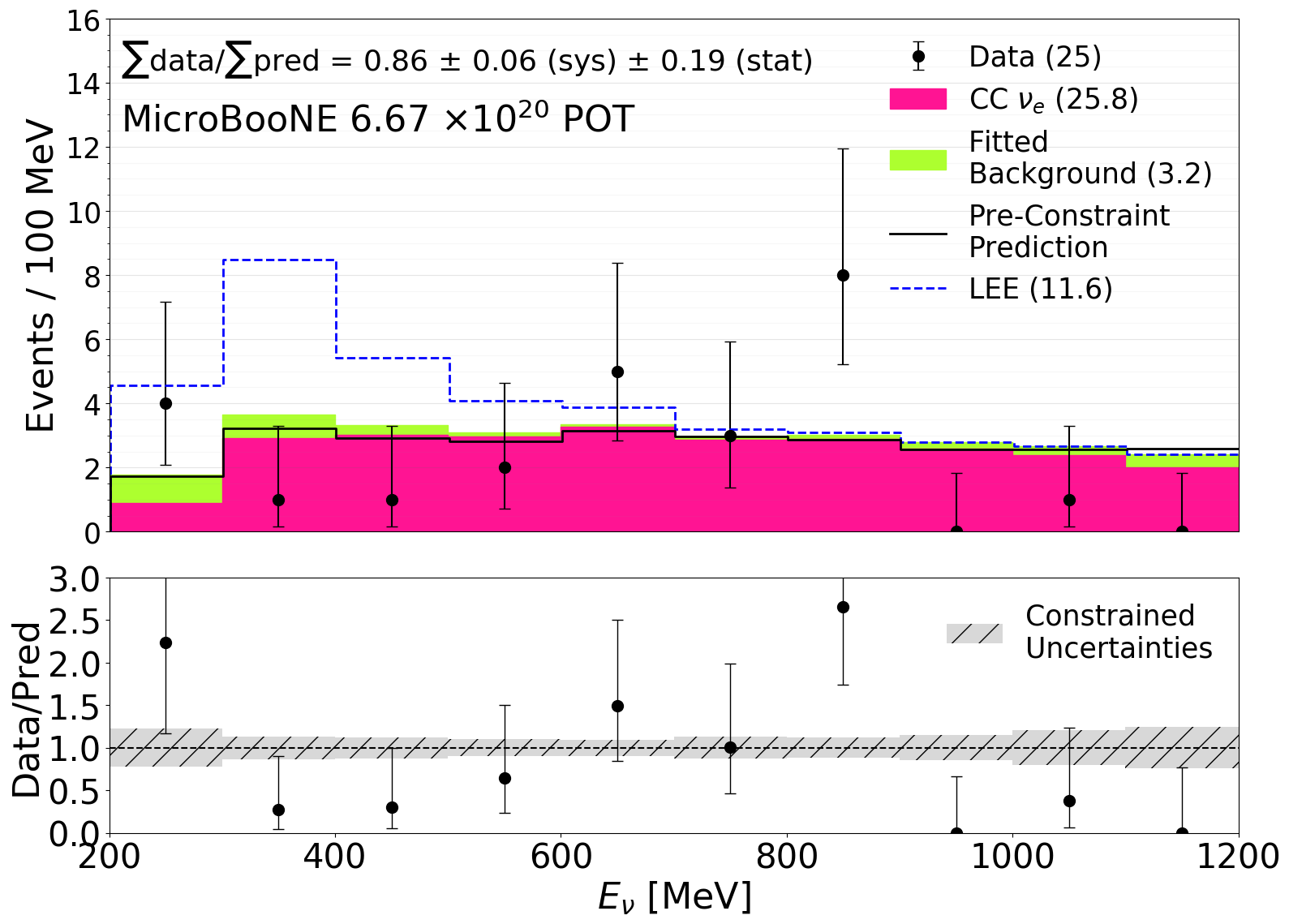}
    \caption{The final constrained prediction for $\nu_e$ signal and $\nu_\mu$ background events compared with the observed data events in the analysis range ($200<E_\nu<1200$\,MeV).  
    The final prediction before the constraint is shown by the black line, and the final constrained systematic errors are shown by the gray band. The $\chi^2_\text{CNP}$/dof is 25.28/10, corresponding to $p_{\rm val}$ = 0.014.}
    \label{fig:1e1p_constrained}
\end{figure}

A primary result of this analysis is the selected $\nu_e$ energy distribution compared to the constrained prediction.
The observed data are compared to the constrained prediction in Fig.~\ref{fig:1e1p_constrained}, showing also the constrained systematic errors and the original pre-constraint prediction. The reconstructed $E_\nu$ distribution for the observed data compared to the unconstrained prediction was shown in Sec.~\ref{sec:1e1p}, partitioned by interaction type in Fig.~\ref{fig:Enu_stacked} (a) and by topology in Fig.~\ref{fig:Enu_stacked} (b).

\begin{figure}
    \centering
    \includegraphics[width=0.5\textwidth]{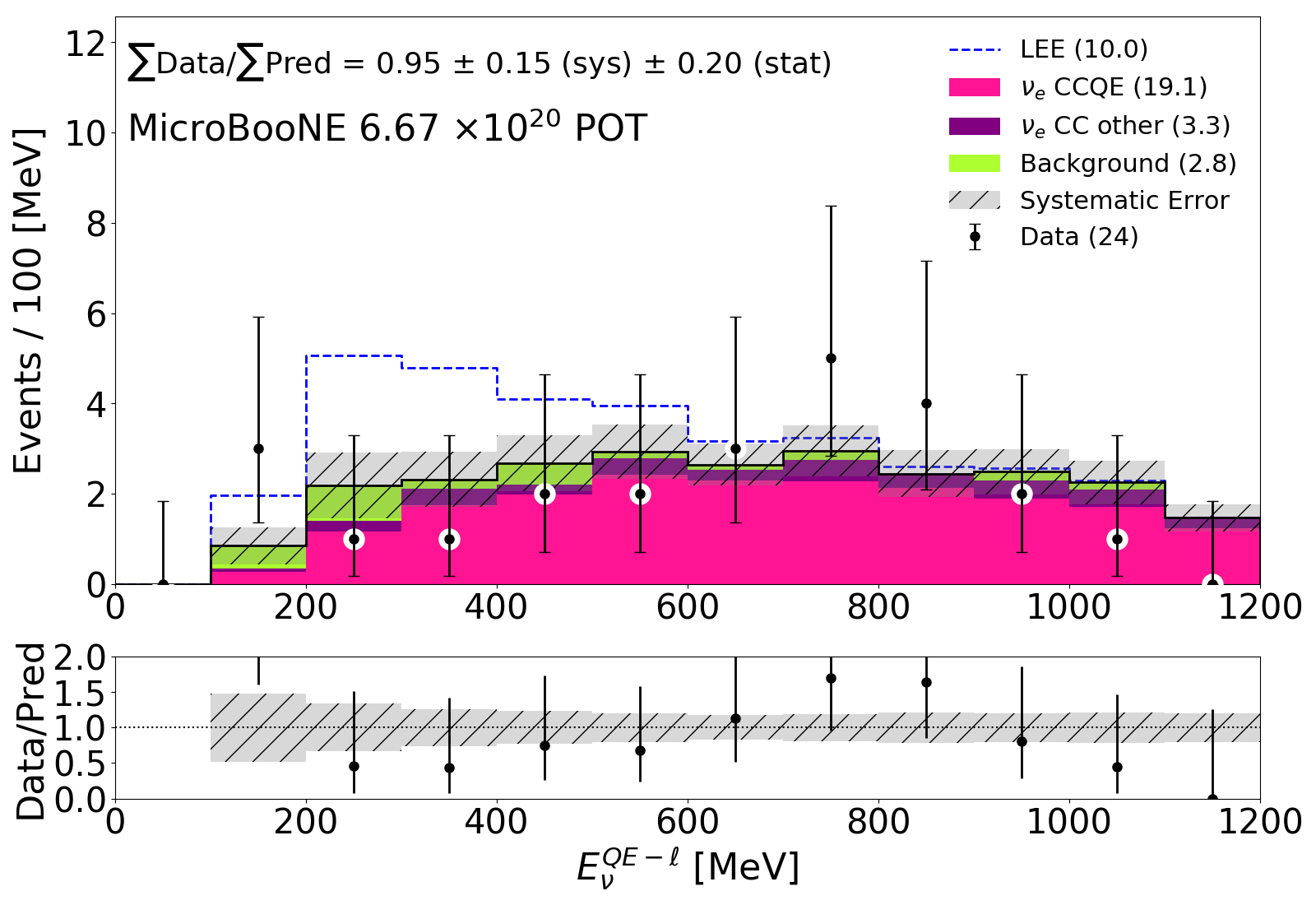}
    \caption{Comparison between data (black points) and simulation in the $E_\nu^{QE-\ell}$ distribution. The stacked histograms give the prediction directly from our simulation sample and the dashed line gives the prediction of the unfolded median MiniBooNE excess model described in Sec.~\ref{sec:lee_model}. The $\chi^2_\text{CNP}$/dof is 11.46/10, corresponding to $p_{\rm val}$ = 0.52.}
    \label{fig:unconstr_EnuQEe}
\end{figure}

\begin{figure}
    \centering
    \includegraphics[width=0.5\textwidth]{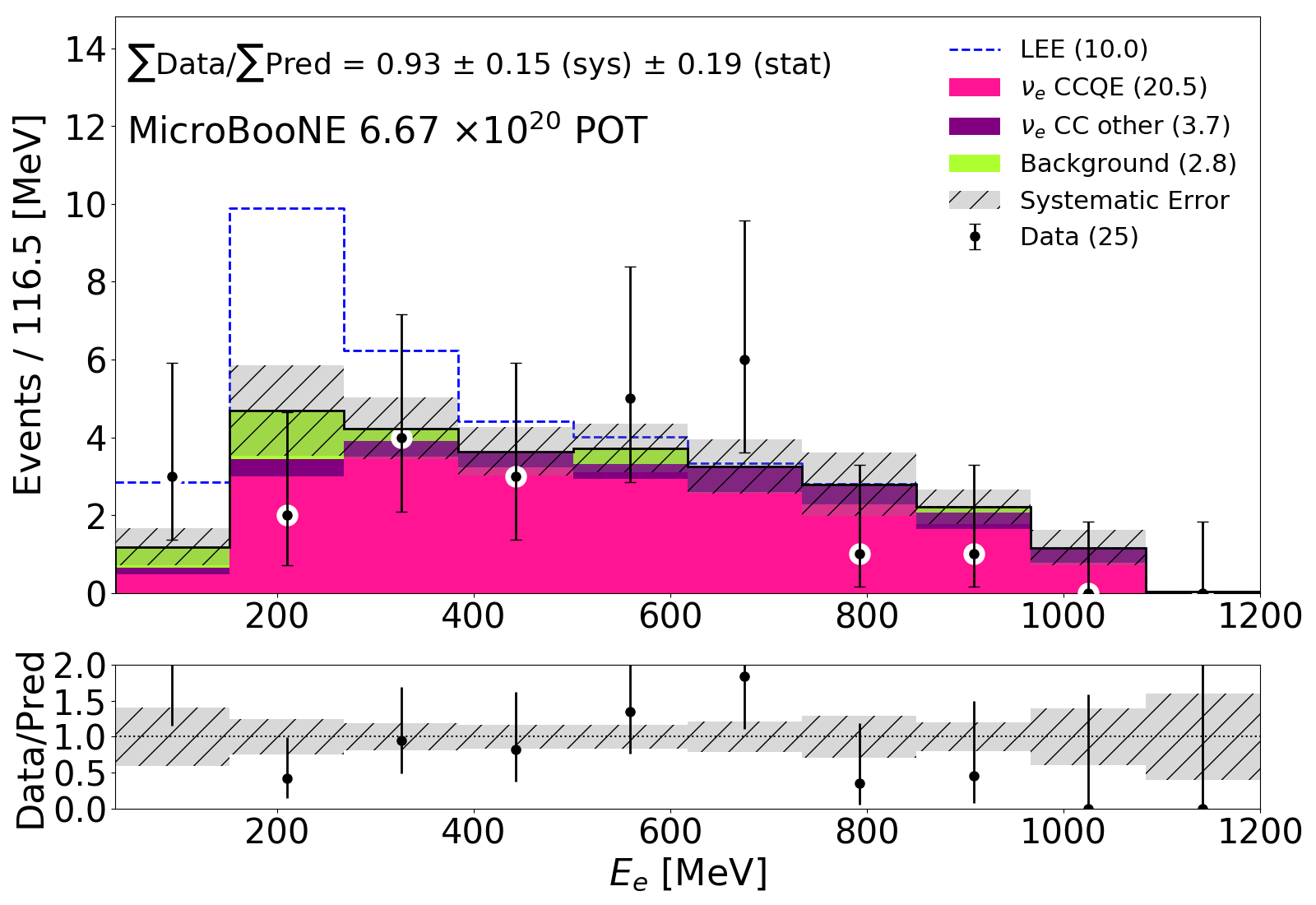}
    \caption{Comparison between data (black points) and simulation for the $E_{e^-}$ distribution. The stacked histograms give the prediction directly from our simulation sample and the dashed line gives the prediction of the unfolded median MiniBooNE excess model described in Sec.~\ref{sec:lee_model}. The $\chi^2_\text{CNP}$/dof is 10.68/10, corresponding to $p_{\rm val}$ = 0.42. }
    \label{fig:unconstr_Ee}
\end{figure}

\begin{figure}
    \centering
    \includegraphics[width=0.5\textwidth]{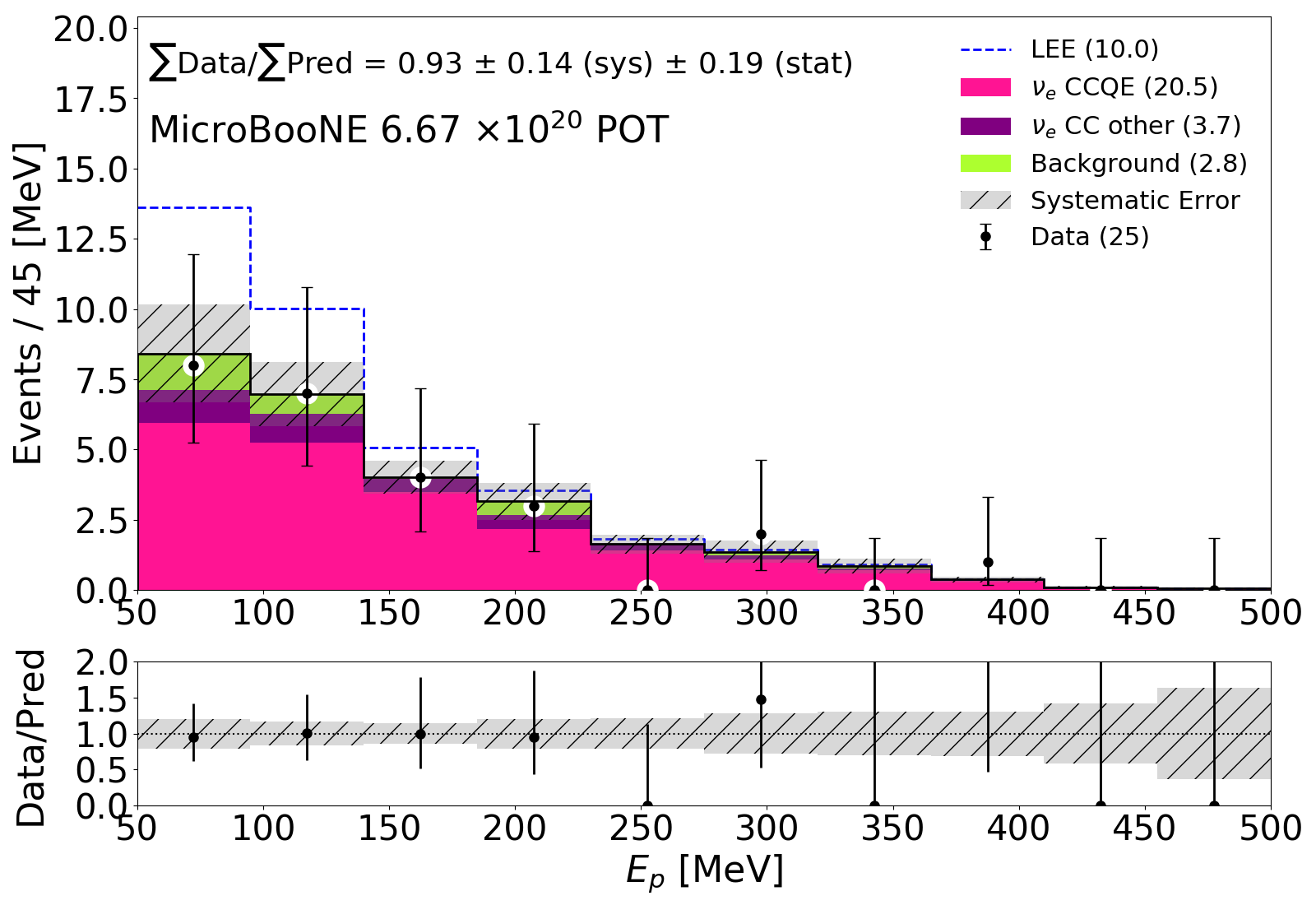}
    \caption{Comparison between data (black points) and simulation (stacked histograms) for the $E_p$ distribution. 
    The stacked histograms are as described in Fig.~18.
    The $\chi^2_\text{CNP}$/dof is 5.77/10, corresponding to $p_{\rm val}$ = 0.80.}
    \label{fig:unconstr_Ep}
\end{figure}

\begin{figure}
    \centering
    \includegraphics[width=0.5\textwidth]{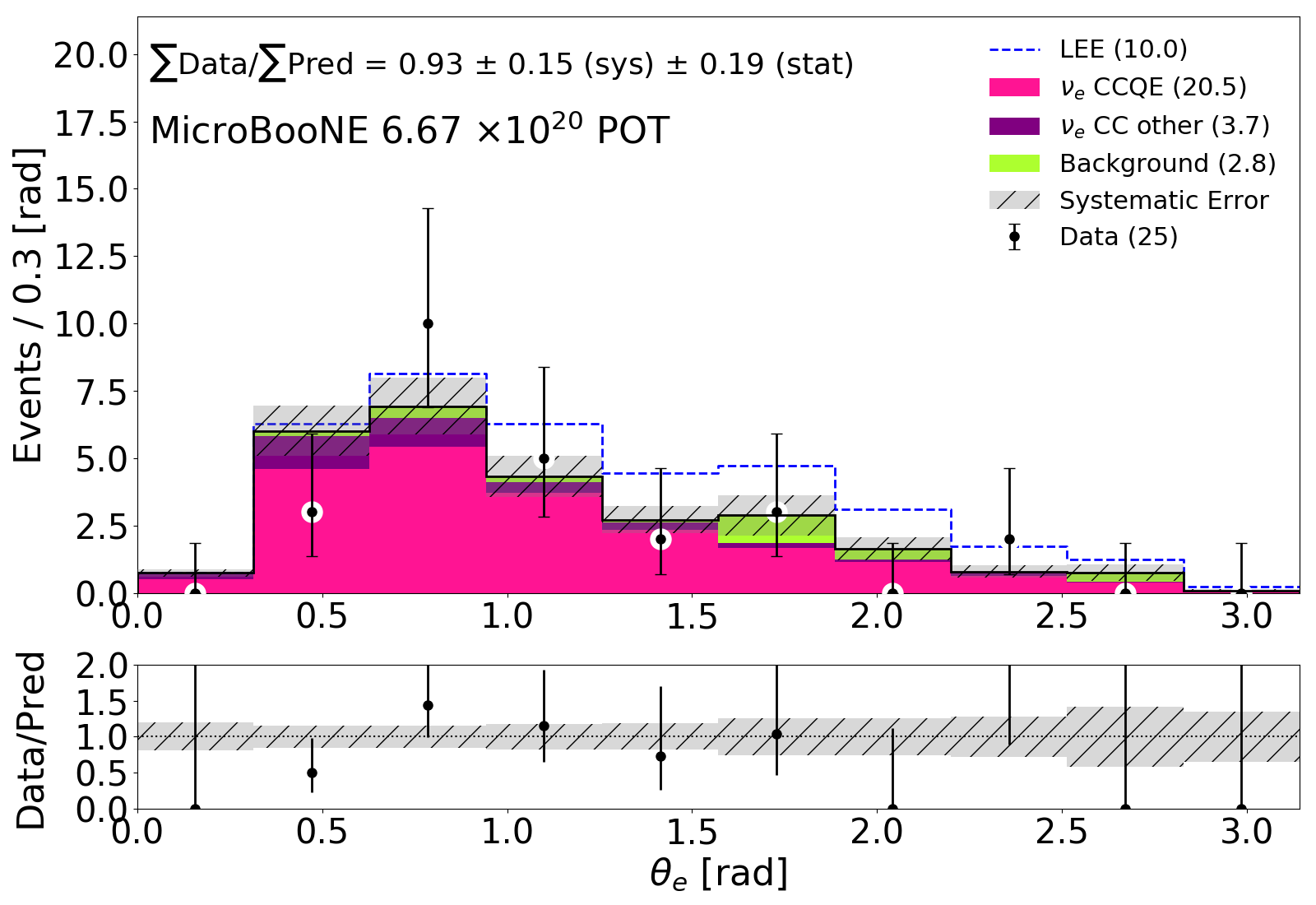}
    \caption{Comparison between data (black points) and simulation (stacked histograms) for the $\theta_{e^-}$ distribution. 
    The stacked histograms are as described in Fig.~18.   The distribution is not peaked at small angle due to the opening angle requirement for vertex-finding described in Sec.~\ref{sec:vtx}.
    The $\chi^2_\text{CNP}$/dof is 10.19/10, corresponding to $p_{\rm val}$ = 0.48.}
    \label{fig:unconstr_Thetae}
\end{figure}

There is some tension between the data and the constrained background predictions in the $E_\nu$ spectrum, more so than in other kinematic variables as seen in Figs.~\ref{fig:unconstr_EnuQEe}--\ref{fig:unconstr_Thetae}.  In particular, the data are lower than the prediction at low energy, which leads to a constraint on the median MiniBooNE LEE, presented below, that is somewhat better than the predicted sensitivity. We note that similar features are seen when data are presented as a function of $E_\nu^{QE-\ell}$, as shown in Fig.~\ref{fig:unconstr_EnuQEe}; thus the features are not tied to the proton reconstruction.  As described in Sec.~\ref{sec:michel}, the study of Michel electrons shows good agreement between data and prediction within statistical uncertainty, so the deficit at low energy is unlikely to arise from a systematic issue with low-energy electron reconstruction.

In Figs.~\ref{fig:unconstr_EnuQEe}--\ref{fig:unconstr_Thetae}, we show distributions for $E_\nu^{QE-\ell}$, $E_{e^-}$, $E_{p}$, and $\theta_{e^-}$. The set of observed distributions are in agreement with prediction as indicated by the $\chi^2_\text{CNP}$/dof (degrees of freedom) values and corresponding $p$-values. Note that the $p$-values are calculated using the same frequentist method as the goodness-of-fit tests described in Sec.~\ref{sec:stat_test}.
If the the low p-value in the Enu distribution was due to a systematic issue in the energy reconstruction, poor agreement would also be expected in these other related distributions; however, that has not been seen.
Note that the predicted distributions for Figs.~\ref{fig:unconstr_EnuQEe}--\ref{fig:unconstr_Thetae} do not apply the background fitting procedure described in Sec.~\ref{sec:numubkgfit} nor the constraint procedure described in Sec.~\ref{sec:constr}.
Both procedures were developed specifically for the $E_\nu$ distribution and not extended for these additional variables. However, the quantitative measures of the data-to-simulation agreement are dominated by the data statistical error and therefore the application of similar methods would be expected to have only modest effects.
Note also that Fig.~\ref{fig:unconstr_EnuQEe} shows only 24 of the 25 data events in our final selection; this is because the remaining event reconstructs with $E_\nu^{QE-\ell} = 1200.2$\,MeV, placing it just above the upper plotting bound of $1200$\,MeV.
A more extensive collection of plots comparing data to simulation in other reconstructed variables can be found in~\cite{Supp}.

\begin{figure*}
    \centering
    \includegraphics[width=0.92\textwidth]{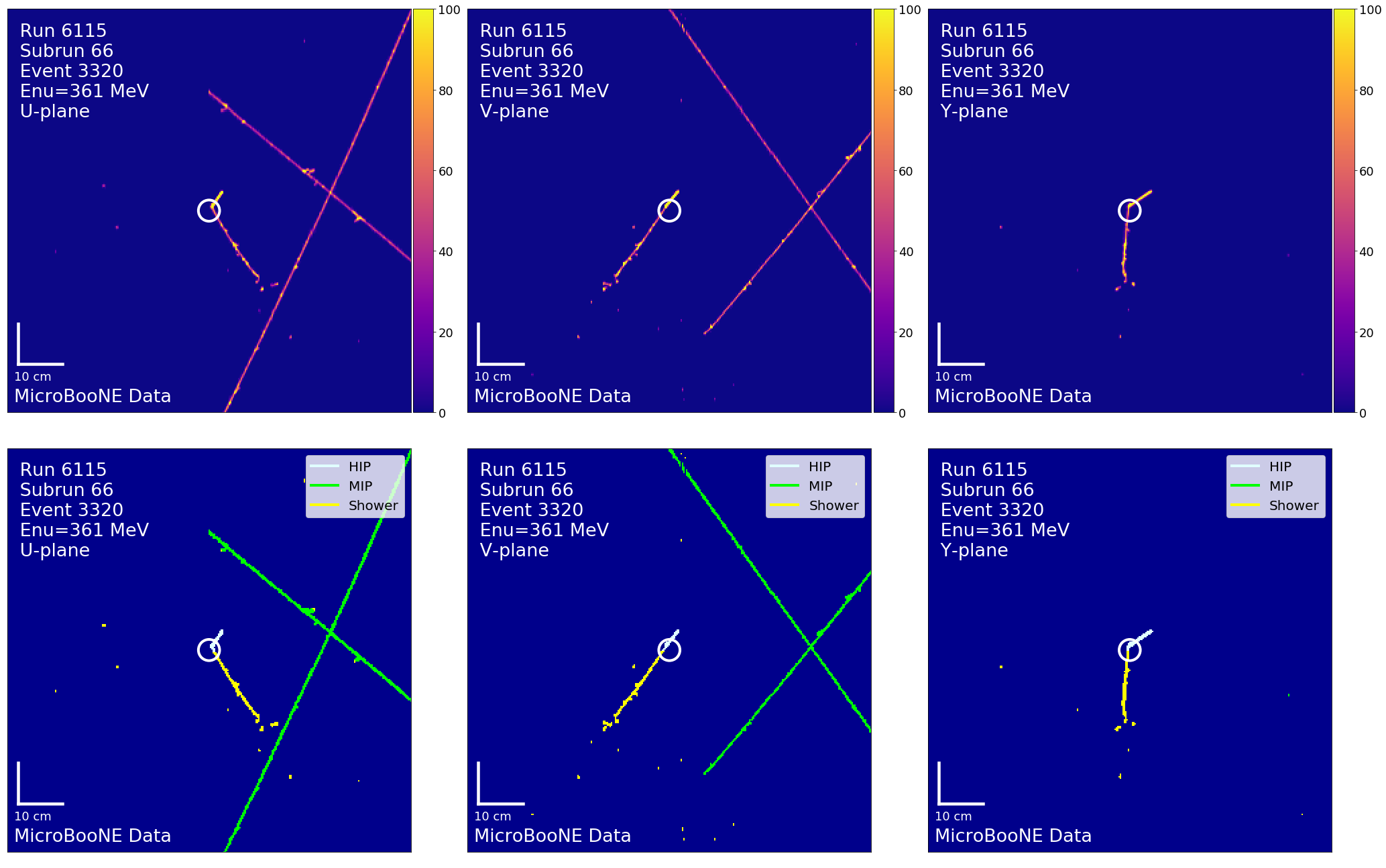}
    \caption{Top: pixel intensity; Bottom: \sssnet\ labels; Left to Right: U, V, Y, planes. The white circle indicates the reconstructed vertex.}
    \label{fig:evd1}
\end{figure*}

We provide the 2D event displays of all 25 events selected in the analysis range in Appendix~\ref{app:evtbyevt}. An example event is shown in Fig.~\ref{fig:evd1}.
In the event displays, each column shows the $U$, $V$ and $Y$ plane responses; on the top row are the pixel intensity images, as described in Sec.~\ref{sec:intro}, and on the bottom the \sssnet\ labeling.

\subsection{Tests of the LEE Model}
\label{subsec:lee_tests}

In this section, we provide the results of the other statistical tests described in Sec.~\ref{sec:stat_test}, which elaborate on the comparisons between the $1e1p$ data observation and our two benchmark hypotheses: $H_0$, the prediction which includes only the irreducible $\nu_e$ backgrounds and the misidentified $\nu_\mu$ backgrounds, and $H_1$, the prediction which additionally includes the LEE signal based on the model described in Sec.~\ref{sec:lee_model}.
In this way we test a purely phenomenological model in which the normalization can vary but the energy dependence is consistent with that obtained by unfolding the MiniBooNE result.

First, we consider the goodness-of-fit between the observation and the two hypotheses over over the full analysis range (200--1200\,MeV), as well as the LEE-enriched range (200--500\,MeV), which was motivated by the LEE signal model. These results are summarized in Table~\ref{tab:gof}.
After the $1\mu 1p$ constraint, the $\chi^2_\text{CNP}$ between the $1e1p$ data and the $H_1$ prediction is 36.35 and has a corresponding $p$-value of $5.0 \times 10^{-4}$.
This indicates that the observation over the full analysis range shows poor agreement with the LEE prediction and in particular significantly worse agreement than with the $H_0$ prediction. The result of the two-hypothesis test that more directly evaluates the comparison between the data and the two predictions is described below.

\begin{table}
    \centering
    \begin{tabular}{c|cc|cc}
    \hline \hline
        \multicolumn{5}{c}{Nominal Predictions} \\
    \hline
        \multirow{2}{*}{Range} & \multicolumn{2}{c|}{$H_0$} & \multicolumn{2}{c}{$H_1$} \\
         & $\chi^2_\text{CNP}/\text{dof}$ & $p$-value & $\chi^2_\text{CNP}/\text{dof}$ & $p$-value \\
    \hline
        200--500\,MeV & 6.06/3 & 0.138 & 8.30/3 & 0.053 \\
        200--1200\,MeV & 23.02/10 & 0.024 & 25.37/10 & 0.014 \\
    \hline \hline
        \multicolumn{5}{c}{Constrained Predictions} \\
    \hline
        \multirow{2}{*}{Range} & \multicolumn{2}{c|}{$H_0$} & \multicolumn{2}{c}{$H_1$} \\
         & $\chi^2_\text{CNP}/\text{dof}$ & $p$-value & $\chi^2_\text{CNP}/\text{dof}$ & $p$-value \\
    \hline
        200--500\,MeV & 7.91/3 & 0.075 & 17.3/3 & 0.002 \\
        200--1200\,MeV & 25.28/10 & 0.014 & 36.35/10 & $5.0 \times 10^{-4}$ \\
    \hline \hline
    \end{tabular}
    \caption{Summary of the results of the goodness-of-fit tests using the $\chi^2_\text{CNP}$ test statistic between the observed $1e1p$ data and the predictions for $H_0$ and $H_1$. The top portion of the table uses the nominal (unconstrained) predictions and uncertainties; the bottom portion uses the constrained predictions and uncertainties. All $p$-values are calculated using the frequentist method described in Sec.~\ref{sec:stat_test}.}
    \label{tab:gof}
\end{table}

\begin{figure}
    \centering
    \includegraphics[width=0.45\textwidth]{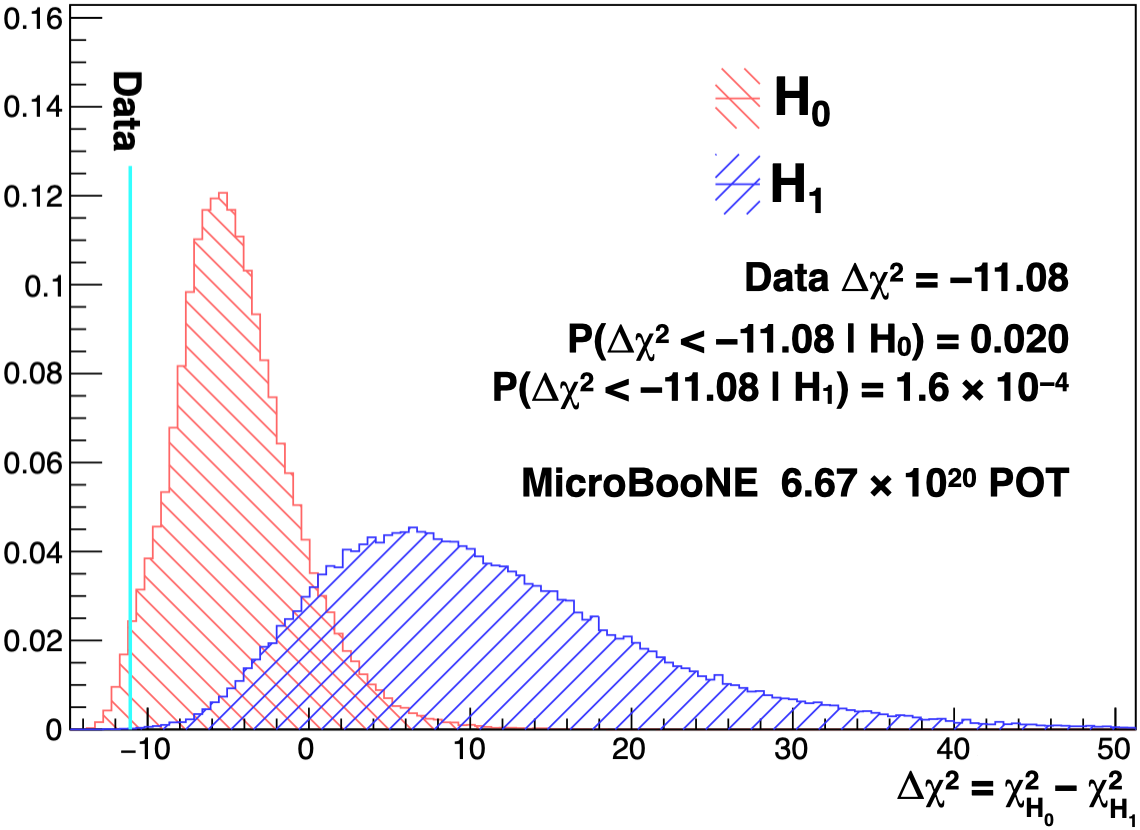}
    \caption{Results of the two-hypothesis test comparing the $1e1p$ observation to the constrained $H_0$ and $H_1$ predictions. The hatched red (blue) histogram shows the distribution of the $\Delta \chi^2$ test statistic under the $H_0$ ($H_1$) hypothesis. The cyan vertical line indicates the data value of $-11.08$.}
\label{fig:testB}
\end{figure}

Next, we consider the two-hypothesis test using the constrained predictions.
The $\Delta \chi^2$ for the data between the $H_0$ and $H_1$ predictions is $-11.08$. This value is shown relative to the distribution of the $\Delta \chi^2$ test statistic under each of the two hypotheses in Fig.~\ref{fig:testB}.
From this, we see that the observed $\Delta \chi^2$ is significantly below the median of both distributions. This is a result of the observation being generally lower than the $H_0$ prediction.
The probability that a $\Delta \chi^2$ value sampled from the $H_0$ distribution is smaller than the observed $\Delta \chi^2$ is 0.020. This indicates $2.1\sigma$ tension between the data and the $H_0$ hypothesis in the results of this statistical test.
The probability that a $\Delta \chi^2$ value sampled from the $H_1$ distribution is smaller than the observed value is $1.6 \times 10^{-4}$.
Using this method, we therefore reject the $H_1$ hypothesis with a significance of $3.6\sigma$.  %
Given the tension between the data and both hypotheses, we also apply the $\text{CL}_s$ method~\cite{Junk:1999,Read:2000}, which helps to mitigate issues related to observing fewer events than predicted by the background-only hypothesis.
The $\text{CL}_s$ statistic is calculated as follows.
\begin{equation}
    \text{CL}_s = \frac{p_{H_1}}{p_{H_0}} = \frac{1.6 \times 10^{-4}}{0.020} = 0.008
\end{equation}
This leads us to reject $H_1$ in favor of $H_0$ with a reduced significance of $2.4\sigma$.

\begin{figure}
    \centering
    \includegraphics[width=\columnwidth]{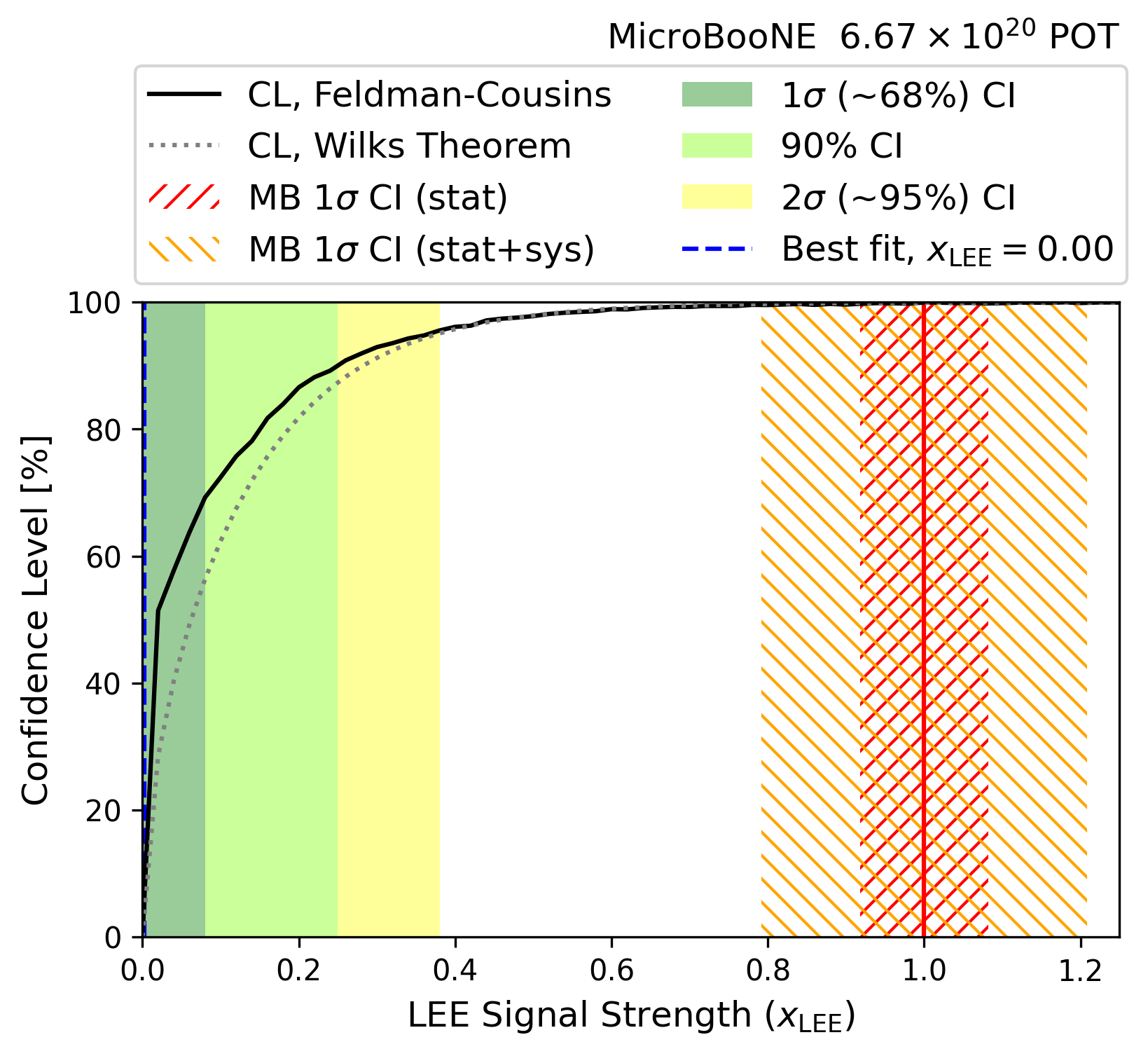}
    \caption{Confidence level at which we rule out values of $x_\text{LEE}$, based on the Feldman--Cousins procedure (solid black curve) and Wilks's theorem (dotted gray curve). The shaded dark green region is the $1\sigma$ confidence interval determined using the Feldman--Cousins procedure; the light green region is the 90\% interval; and the yellow region is the $2\sigma$ interval. The dashed blue vertical line indicates the best-fit value at $x_\text{LEE} = 0$. The solid red vertical line is the median MiniBooNE signal model at $x_\text{LEE} = 1$; the hatched red (orange) region is the estimated MiniBooNE $1\sigma$ confidence interval without (with) systematic uncertainties.}
    \label{fig:testC}
\end{figure}

Finally, we fit for the LEE signal strength, $x_\text{LEE}$.
The best-fit value is $x_\text{LEE} = 0$.
Using the Feldman--Cousins procedure, we also evaluate confidence intervals on the signal strength.
We find the 90\% confidence interval has an upper bound at $x_\text{LEE} \leq 0.25$, and the $2\sigma$ ($\sim$95\%) interval has an upper bound at 0.38.
Fig.~\ref{fig:testC} shows the confidence level at which we rule out various values of $x_\text{LEE}$.
Note that the upper limits on the LEE signal strength parameter obtained from the data are significantly lower than the expected limits based on the Asimov data set for $H_0$, which are 0.75 and 0.98 for the 90\% and $2\sigma$ confidence levels, respectively. This is another consequence of the observation showing a deficit compared to the $H_0$ prediction.

\section{Summary}
\label{sec:sum}
We have presented the results of the Deep-Learning-based analysis, one out of three MicroBooNE analyses targeting the electron-like LEE observed by the MiniBooNE experiment. 
This analysis treats the data as images, allowing the use of powerful computer vision techniques such as CNNs.  Also, unlike the other analyses, the analysis is an exclusive one, isolating events consistent with the two-body kinematics of CCQE scattering.   

The resulting analysis strongly rejects $\nu_\mu$ interaction misidentification,   
leaving a highly pure sample of CCQE $\nu_e$ events ($1e1p$).  These $\nu_e$ events are due to either irreducible intrinsic beam content, or, potentially, a signal that would explain the MiniBooNE electron-like excess.  The background rate (irreducible $\nu_e$ and misidentified $\nu_\mu$) is constrained by sideband measurements of $\nu_\mu$ interaction rates.

To describe the LEE signal, we make use of the MiniBooNE median LEE model, which is based on unfolding the MiniBooNE signal without consideration of uncertainties. This predicts 11.6 additional $\nu_e$ events on top of the background-only predictions. In the analysis range, 25 data events are observed, which is in excellent agreement with the final background-only prediction of $29.0 \pm 1.9_\text{(sys)} \pm 5.4_\text{(stat)}$ ($27.4 \pm 3.8_\text{(sys)} \pm 5.2_\text{(stat)}$) events for the analysis range with (without) the $1\mu 1p$ constraint.

We have reported three types of statistical tests based on this observation.
The first uses goodness-of-fit tests to compare the expected distribution to observed data across the analysis range. While there is excellent agreement between the observation and the prediction with no LEE signal in the total events, there is disagreement in the energy dependence of the observed events. A deficit is observed in the 300--500\,MeV range and an excess is observed in the 800--900\,MeV bin. Using the $\chi^2_\text{CNP}$ as a test statistic, this yields a $p$-value of 0.014 (0.0005) without (with) the contribution from the LEE model.
The second is a two-hypothesis $\Delta \chi^2_\text{CNP}$ test between the background-only and LEE hypotheses. We take into account the under-fluctuation of the data using $\text{CL}_s$, which results in a rejection of the LEE hypothesis with significance of $2.4\sigma$.
Lastly, we fit for an excess, retaining the energy dependence of the MiniBooNE median LEE model but allowing the normalization $x_\text{LEE}$ to vary. This allows us to reject values of $x_\text{LEE}$ greater than 0.25 (0.38) at the 90\% ($2\sigma$) confidence level, which can be compared to expected upper limit if the analysis had observed the background-only prediction of 0.75 (0.98).

In conclusion, the analysis reported in this paper is inconsistent with observation of an excess of $\nu_e$ events in the signal range.  Hence, these results disfavor explanations of the MiniBooNE low energy excess based purely on $\nu_e$ interactions. These results are in agreement with those of other $\nu_e$-based searches reported by the MicroBooNE experiment at this time, as reported in Ref.~\cite{jointMicroBooNEPRL}.

\section{Acknowledgments}
 This document was prepared by the MicroBooNE collaboration using the resources of the Fermi National Accelerator Laboratory (Fermilab), a U.S. Department of Energy, Office of Science, HEP User Facility. Fermilab is managed by Fermi Research Alliance, LLC (FRA), acting under Contract No. DE-AC02-07CH11359.  MicroBooNE is supported by the following: the U.S. Department of Energy, Office of Science, Offices of High Energy Physics and Nuclear Physics; the U.S. National Science Foundation; the Swiss National Science Foundation; the Science and Technology Facilities Council (STFC), part of the United Kingdom Research and Innovation; the Royal Society (United Kingdom); and The European Union’s Horizon 2020 Marie Sklodowska-Curie Actions. Additional support for the laser calibration system and cosmic ray tagger was provided by the Albert Einstein Center for Fundamental Physics, Bern, Switzerland. We also acknowledge the contributions of technical and scientific staff to the design, construction, and operation of the MicroBooNE detector as well as the contributions of past collaborators to the development of MicroBooNE analyses, without whom this work would not have been possible.

\FloatBarrier

\appendix

\section{Event-by-Event Information}
\label{app:evtbyevt}
\begin{figure*}
    \centering
    \includegraphics[width=0.92\linewidth]{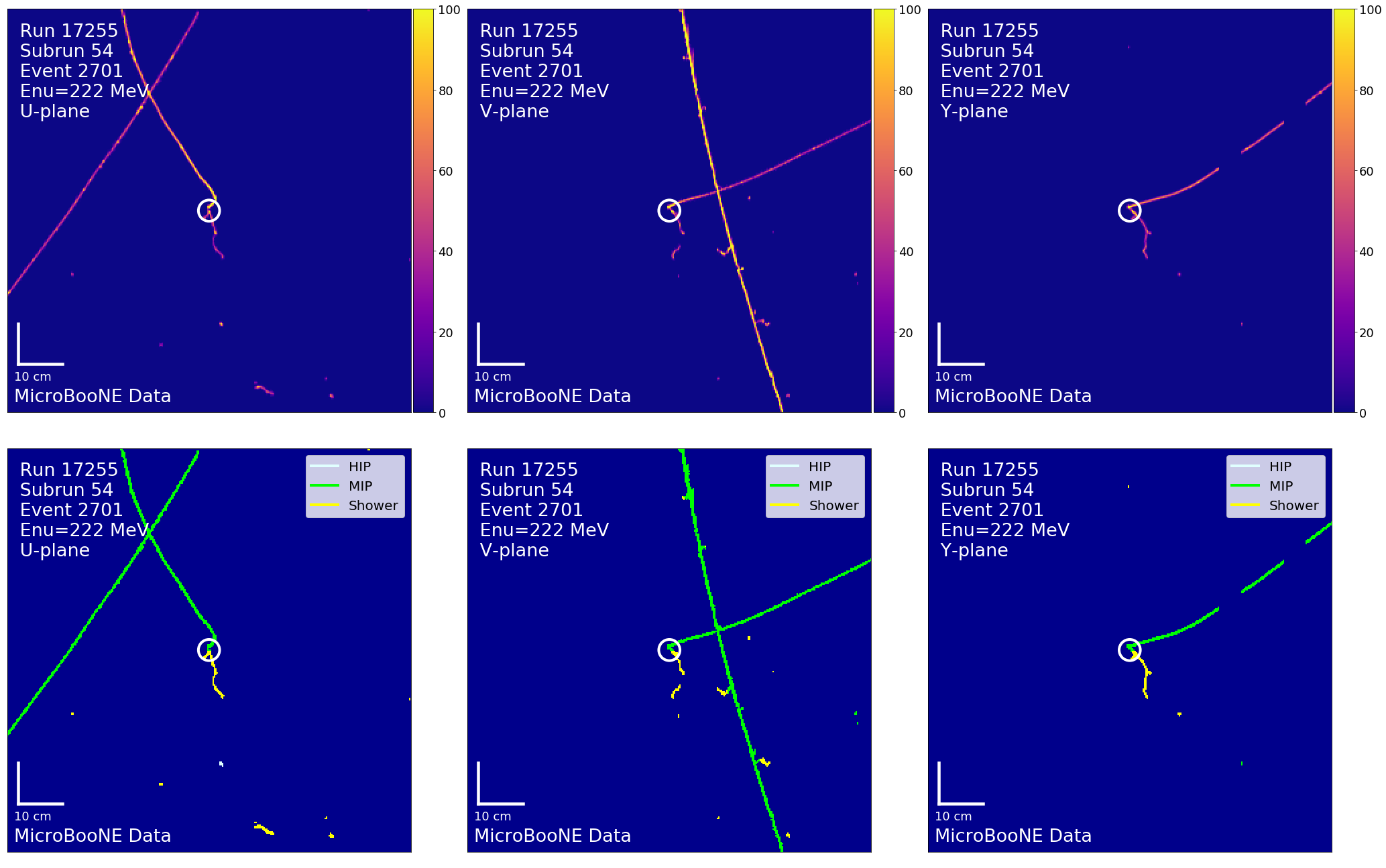}
    \caption{Top: pixel intensity; Bottom: \sssnet\ labels; Left to right: U, V, Y, planes. The white circle indicates the reconstructed vertex.}
    \label{fig:evd2}
\end{figure*}

\begin{figure*}
    \centering
    \includegraphics[width=0.92\linewidth]{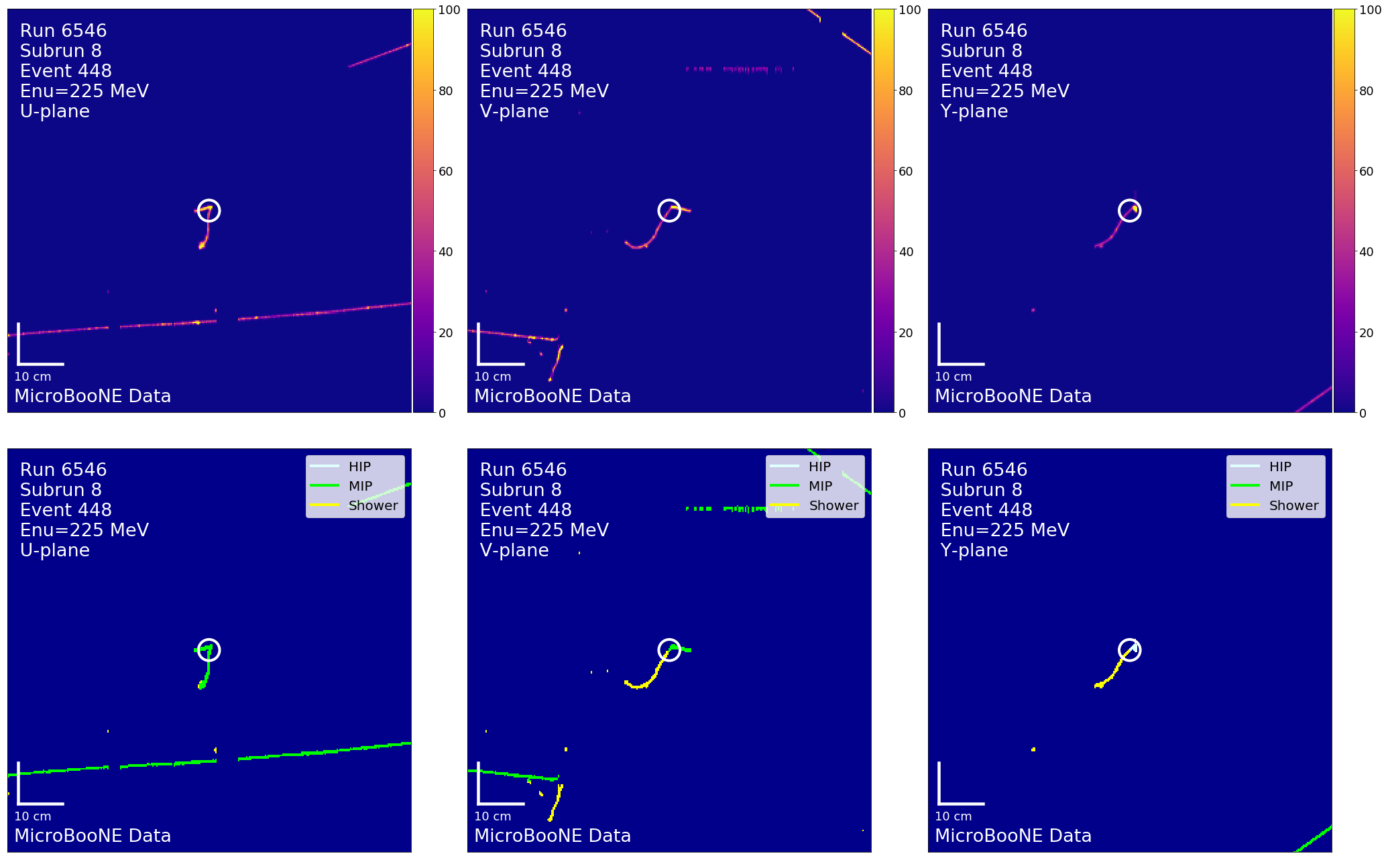}
    \caption{Top: pixel intensity; Bottom: \sssnet\ labels; Left to right: U, V, Y, planes. The white circle indicates the reconstructed vertex.}
    \label{fig:evd3}
\end{figure*}

\begin{figure*}
    \centering
    \includegraphics[width=0.92\linewidth]{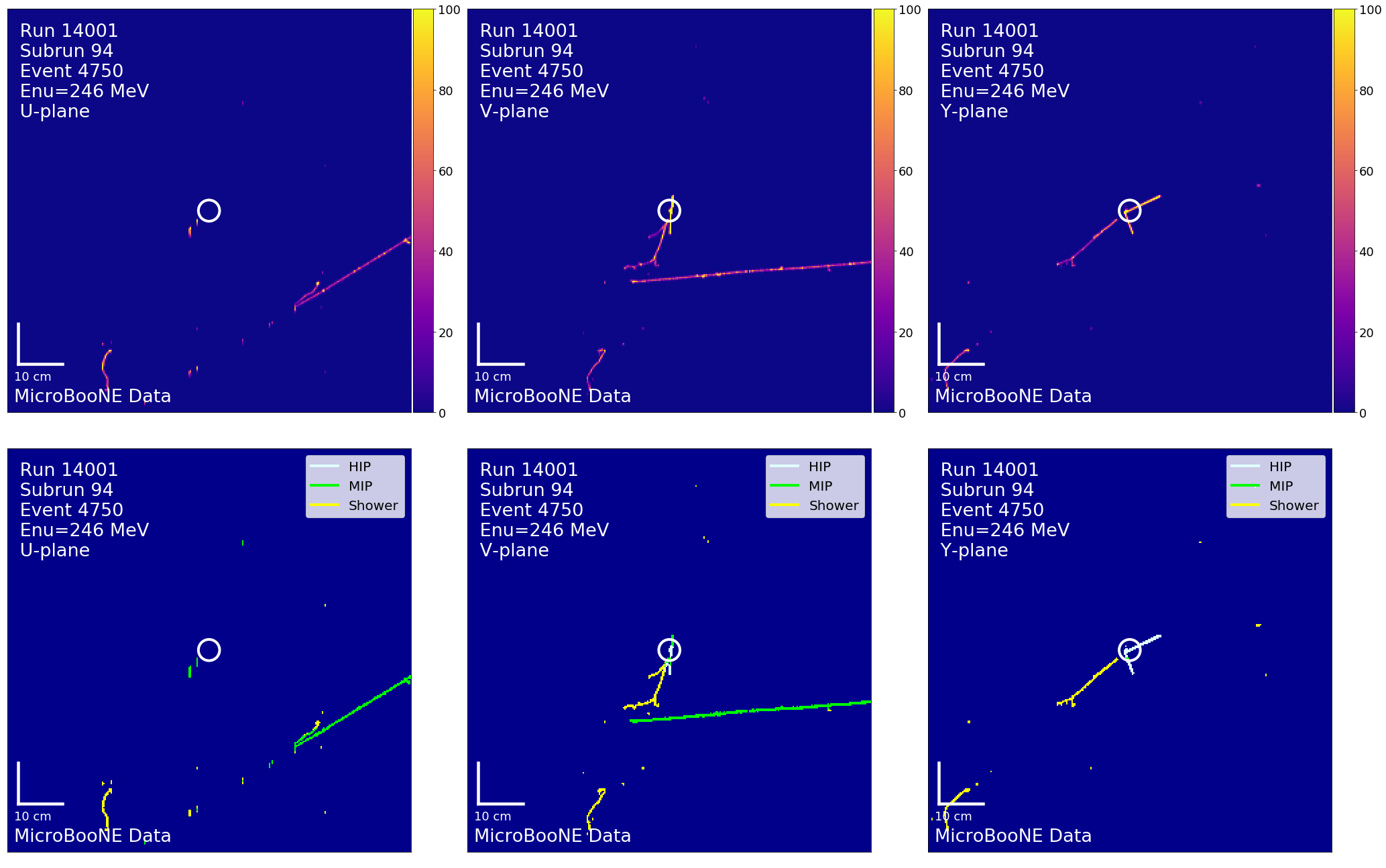}
    \caption{Top: pixel intensity; Bottom: \sssnet\ labels; Left to right: U, V, Y, planes. The white circle indicates the reconstructed vertex.}
    \label{fig:evd4}
\end{figure*}

\begin{figure*}
    \centering
    \includegraphics[width=0.92\linewidth]{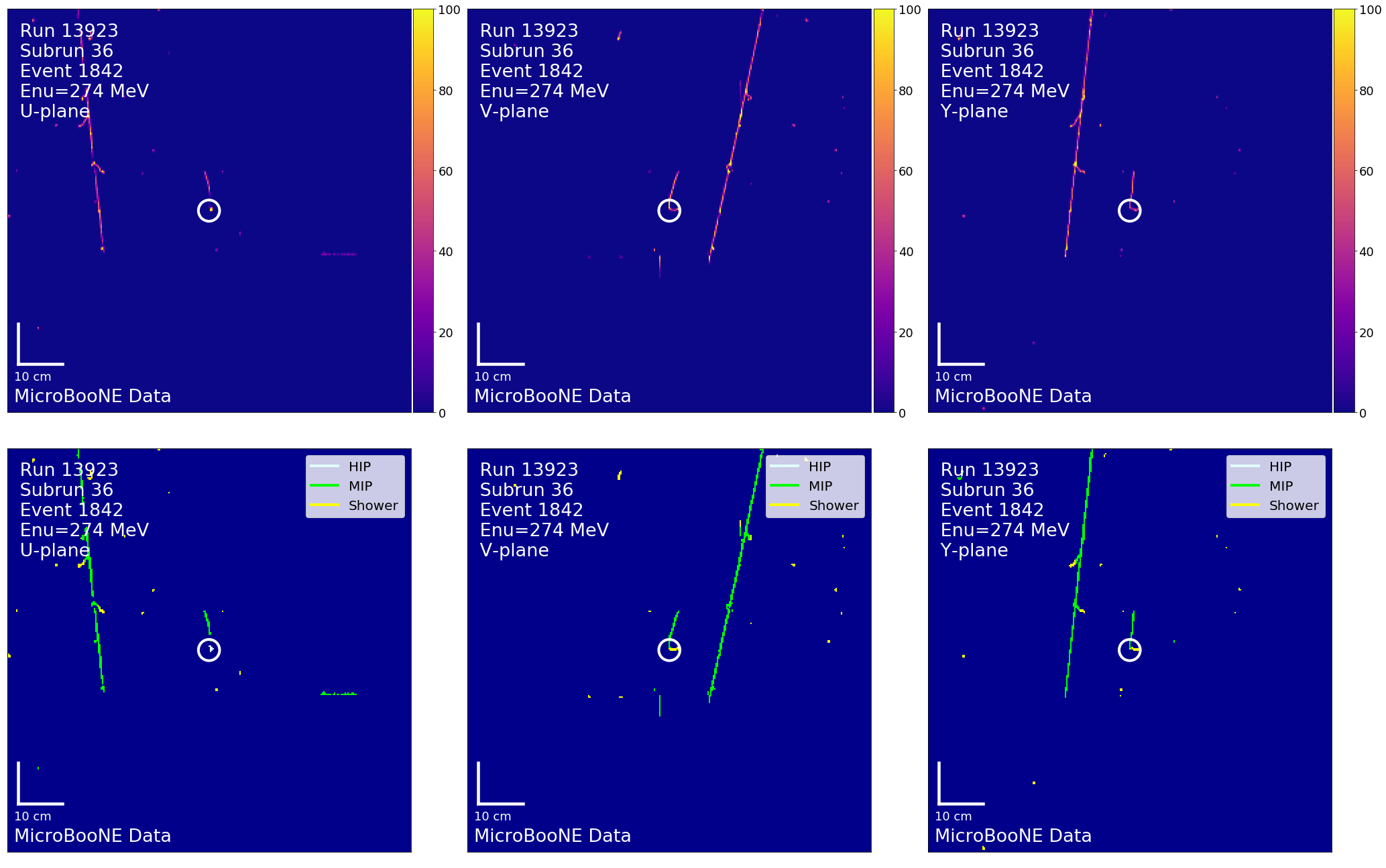}
    \caption{Top: pixel intensity; Bottom: \sssnet\ labels; Left to right: U, V, Y, planes. The white circle indicates the reconstructed vertex.}
    \label{fig:evd5}
\end{figure*}

\begin{figure*}
    \centering
    \includegraphics[width=0.92\linewidth]{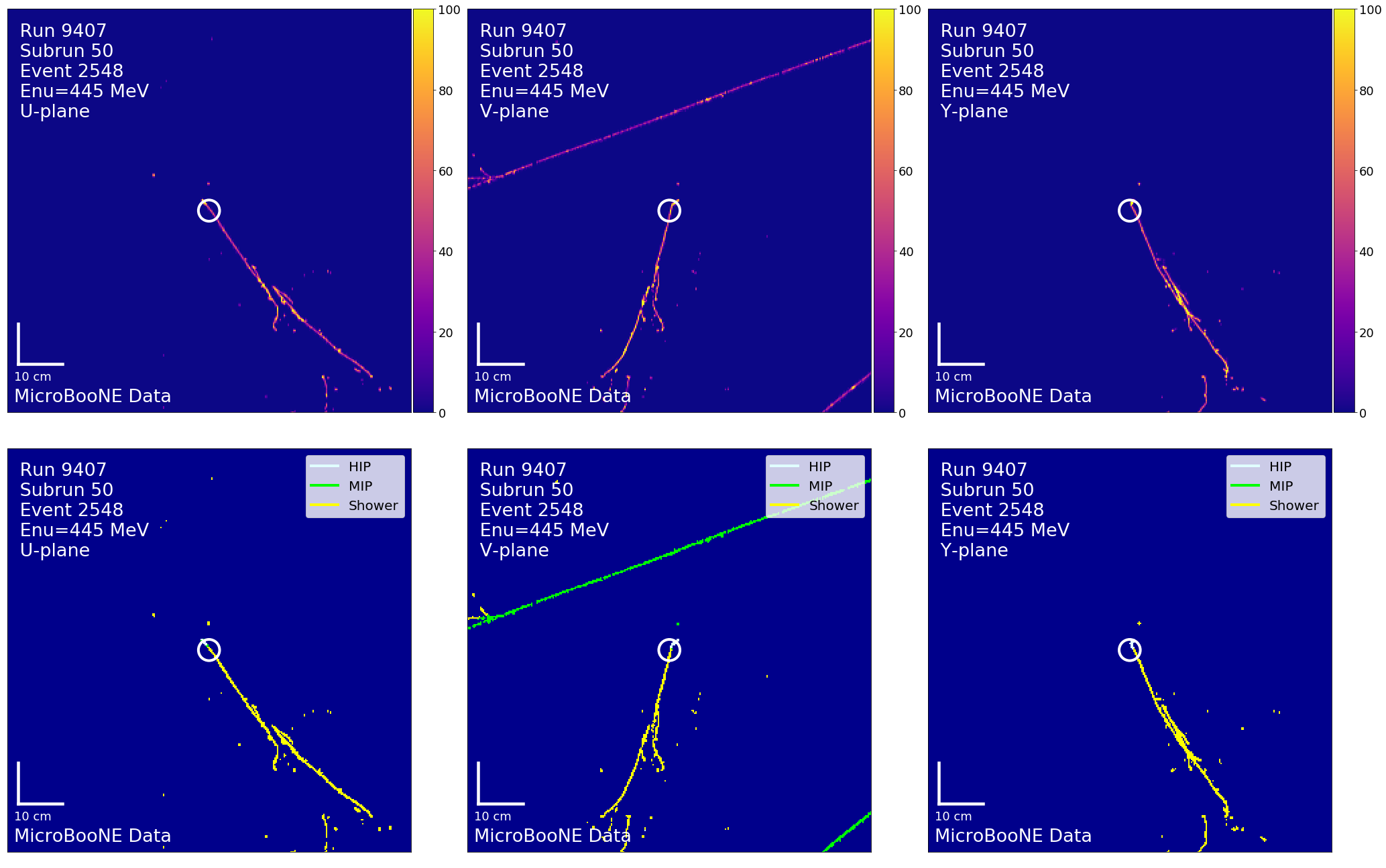}
    \caption{Top: pixel intensity; Bottom: \sssnet\ labels; Left to right: U, V, Y, planes. The white circle indicates the reconstructed vertex.}
    \label{fig:evd6}
\end{figure*}

\begin{figure*}
    \centering
    \includegraphics[width=0.92\linewidth]{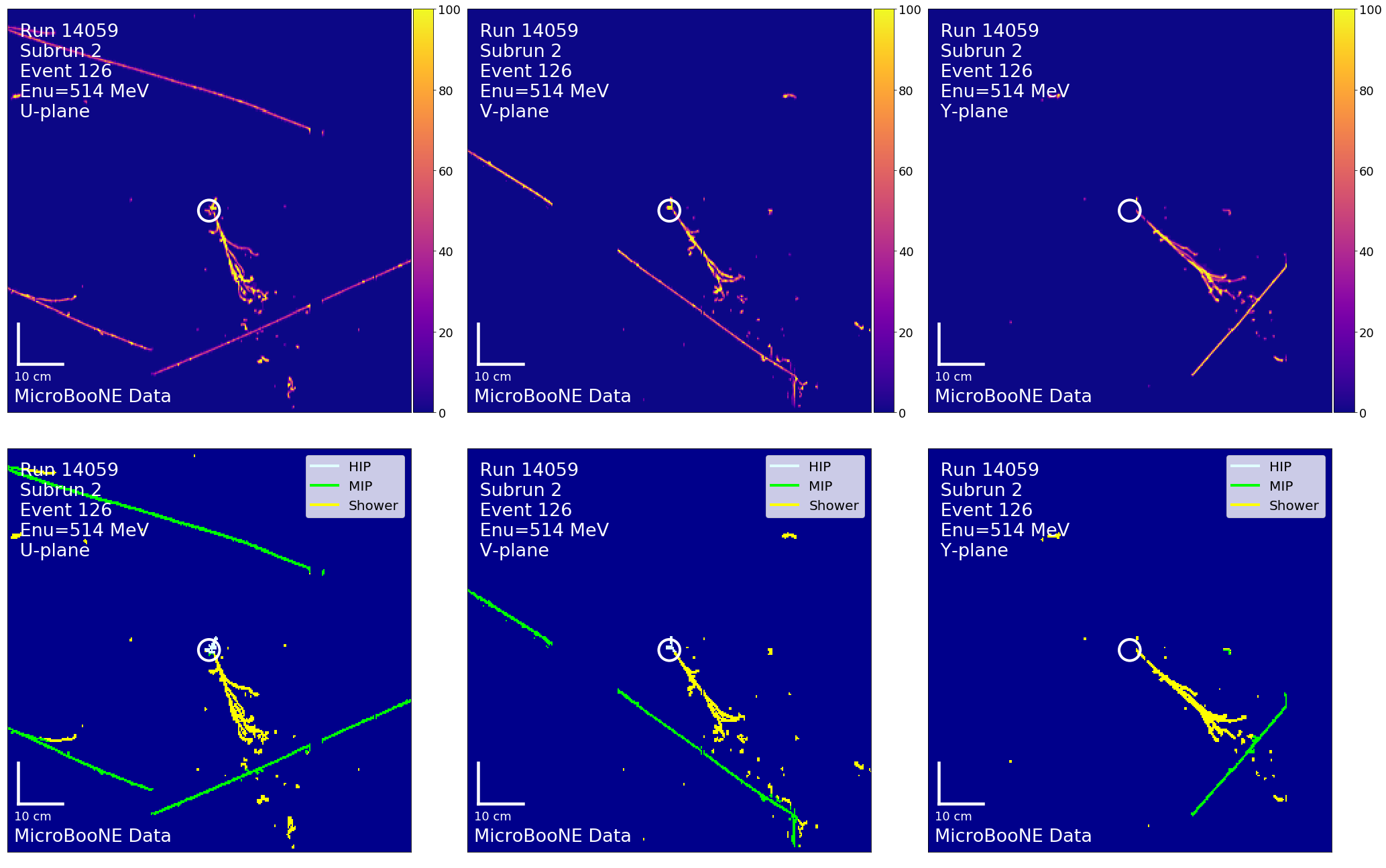}
    \caption{Top: pixel intensity; Bottom: \sssnet\ labels; Left to right: U, V, Y, planes. The white circle indicates the reconstructed vertex.}
    \label{fig:evd7}
\end{figure*}

\begin{figure*}
    \centering
    \includegraphics[width=0.92\linewidth]{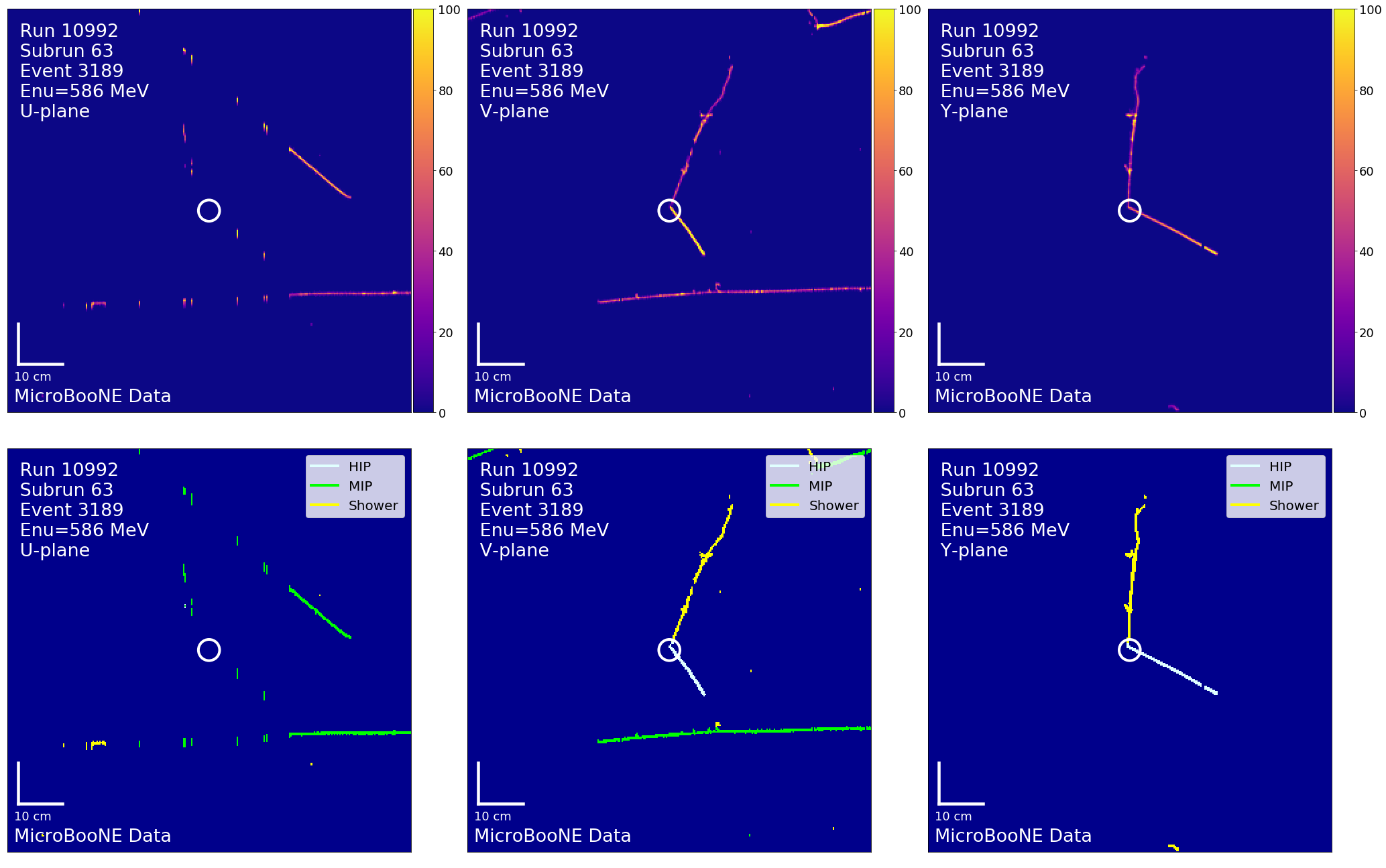}
    \caption{Top: pixel intensity; Bottom: \sssnet\ labels; Left to right: U, V, Y, planes. The white circle indicates the reconstructed vertex.}
    \label{fig:evd8}
\end{figure*}

\begin{figure*}
    \centering
    \includegraphics[width=0.92\linewidth]{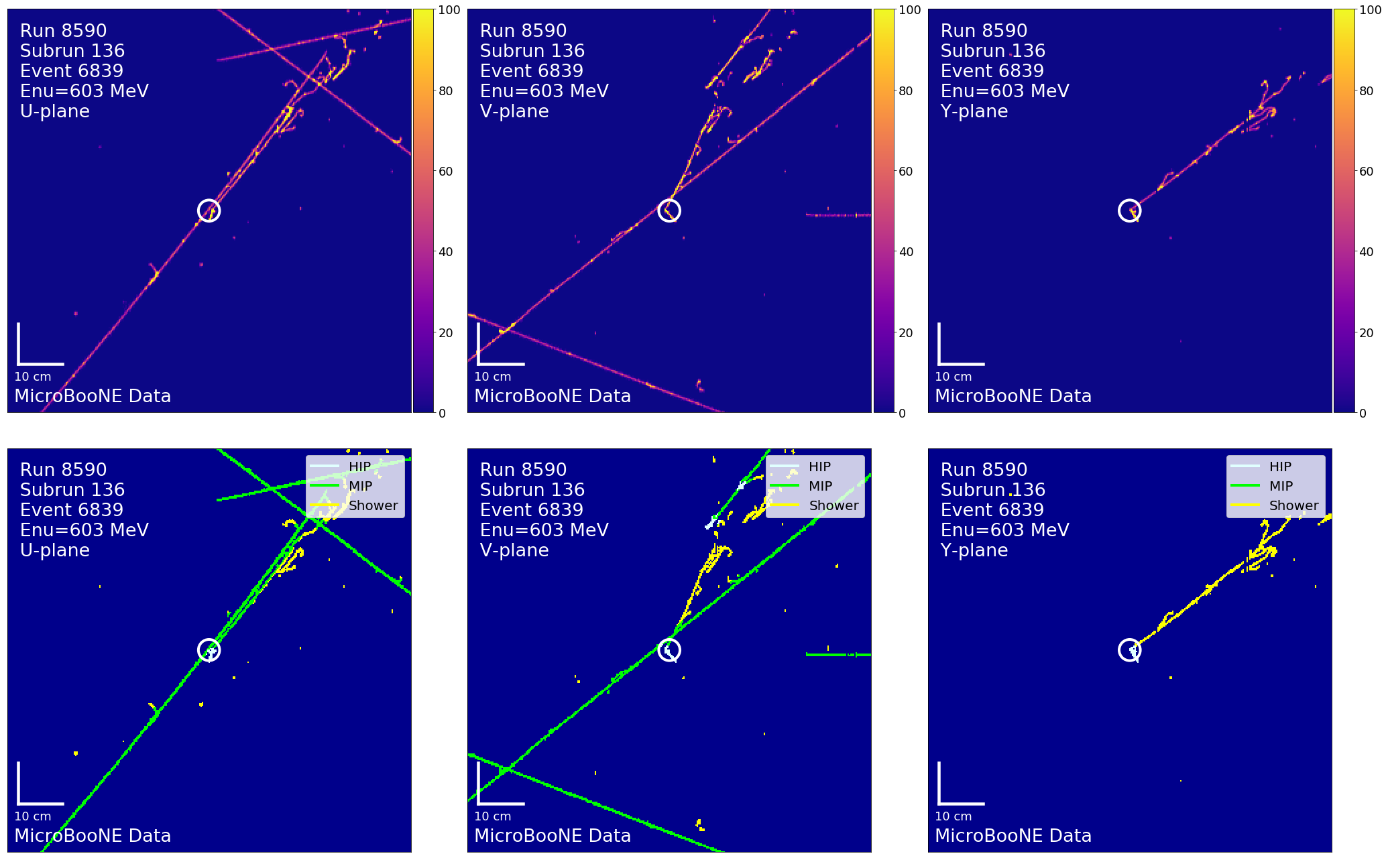}
    \caption{Top: pixel intensity; Bottom: \sssnet\ labels; Left to right: U, V, Y, planes. The white circle indicates the reconstructed vertex.}
    \label{fig:evd9}
\end{figure*}

\begin{figure*}
    \centering
    \includegraphics[width=0.92\linewidth]{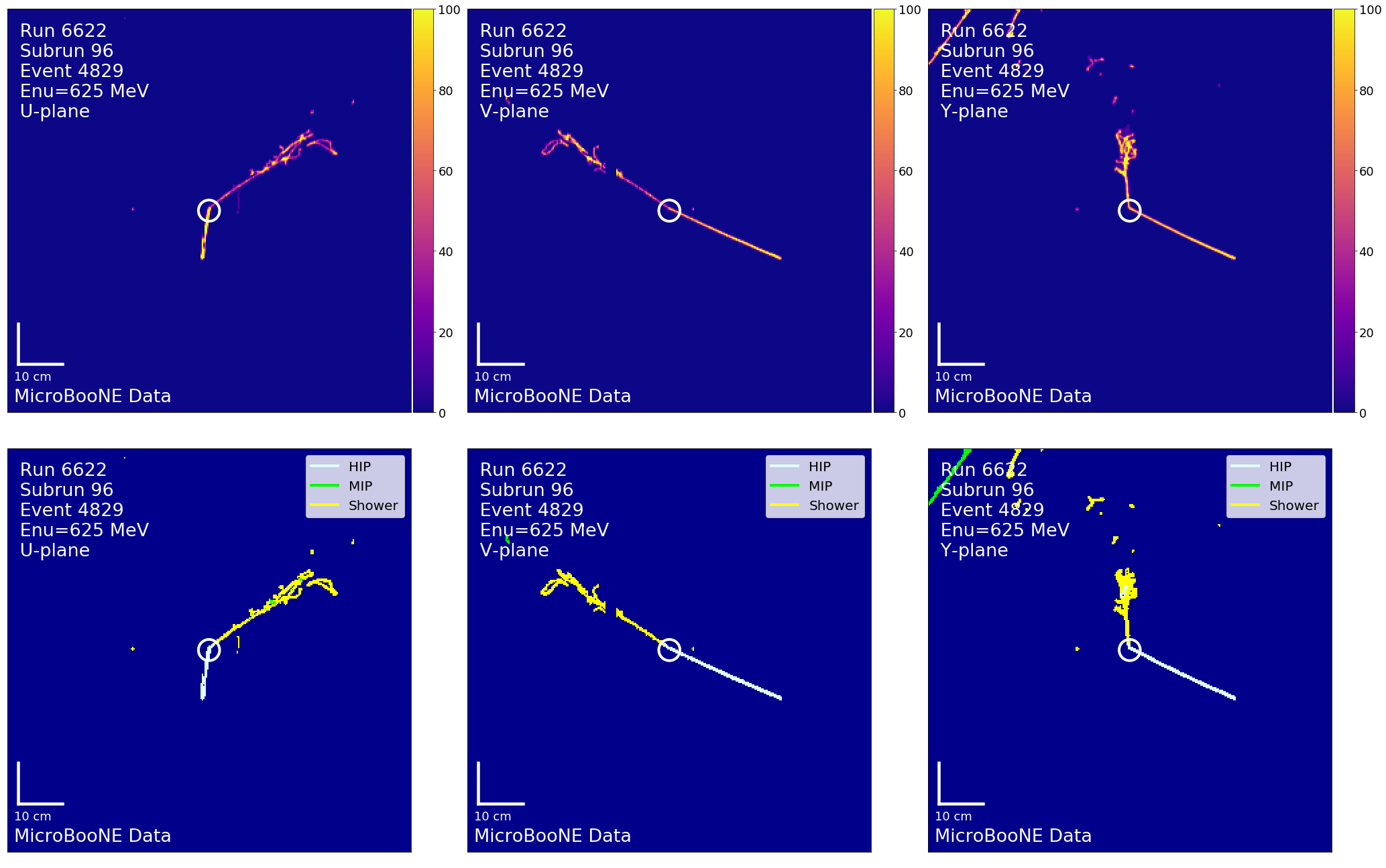}
    \caption{Top: pixel intensity; Bottom: \sssnet\ labels; Left to right: U, V, Y, planes. The white circle indicates the reconstructed vertex.}
    \label{fig:evd10}
\end{figure*}

\begin{figure*}
    \centering
    \includegraphics[width=0.92\linewidth]{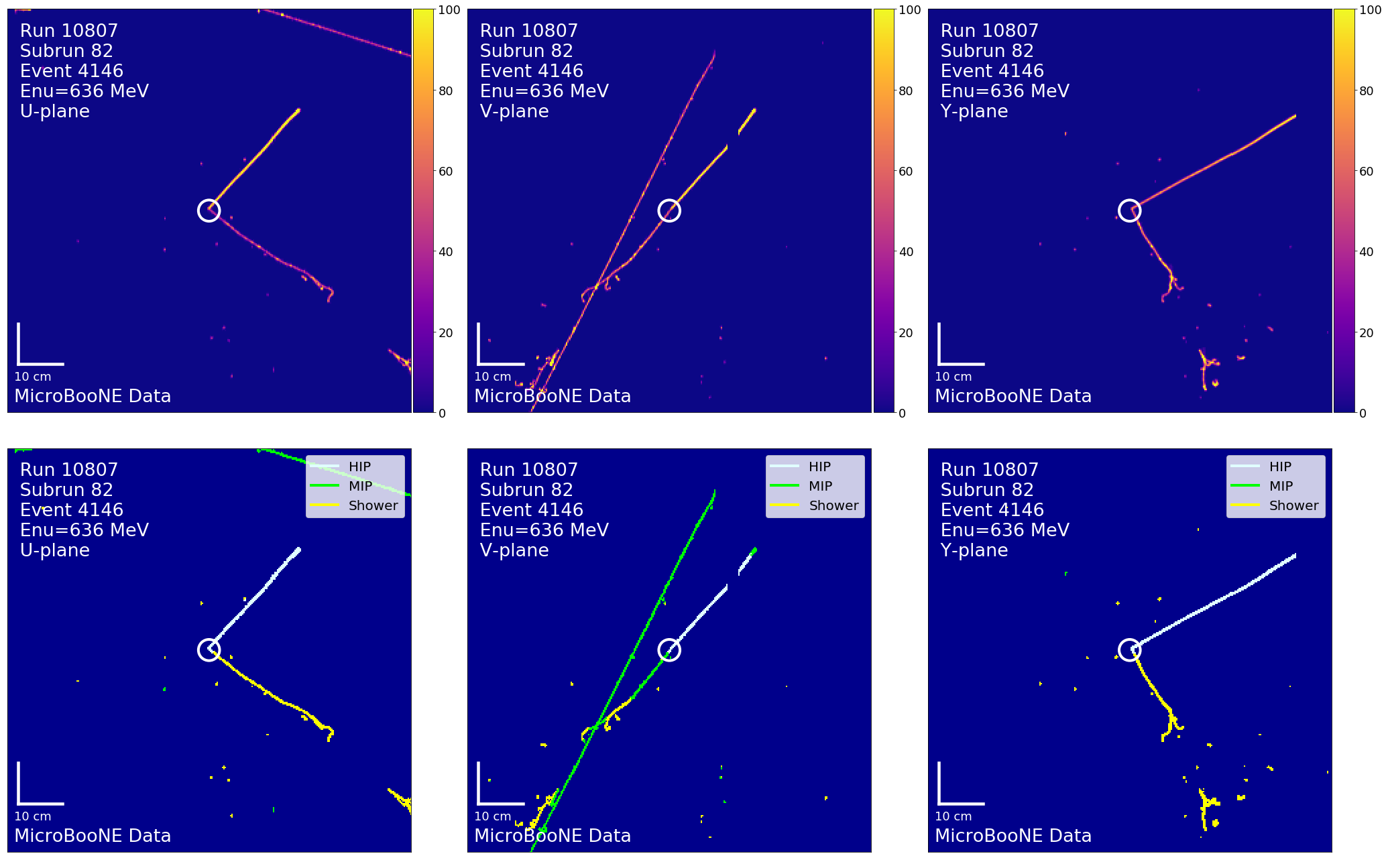}
    \caption{Top: pixel intensity; Bottom: \sssnet\ labels; Left to right: U, V, Y, planes. The white circle indicates the reconstructed vertex.}
    \label{fig:evd11}
\end{figure*}

\begin{figure*}
    \centering
    \includegraphics[width=0.92\linewidth]{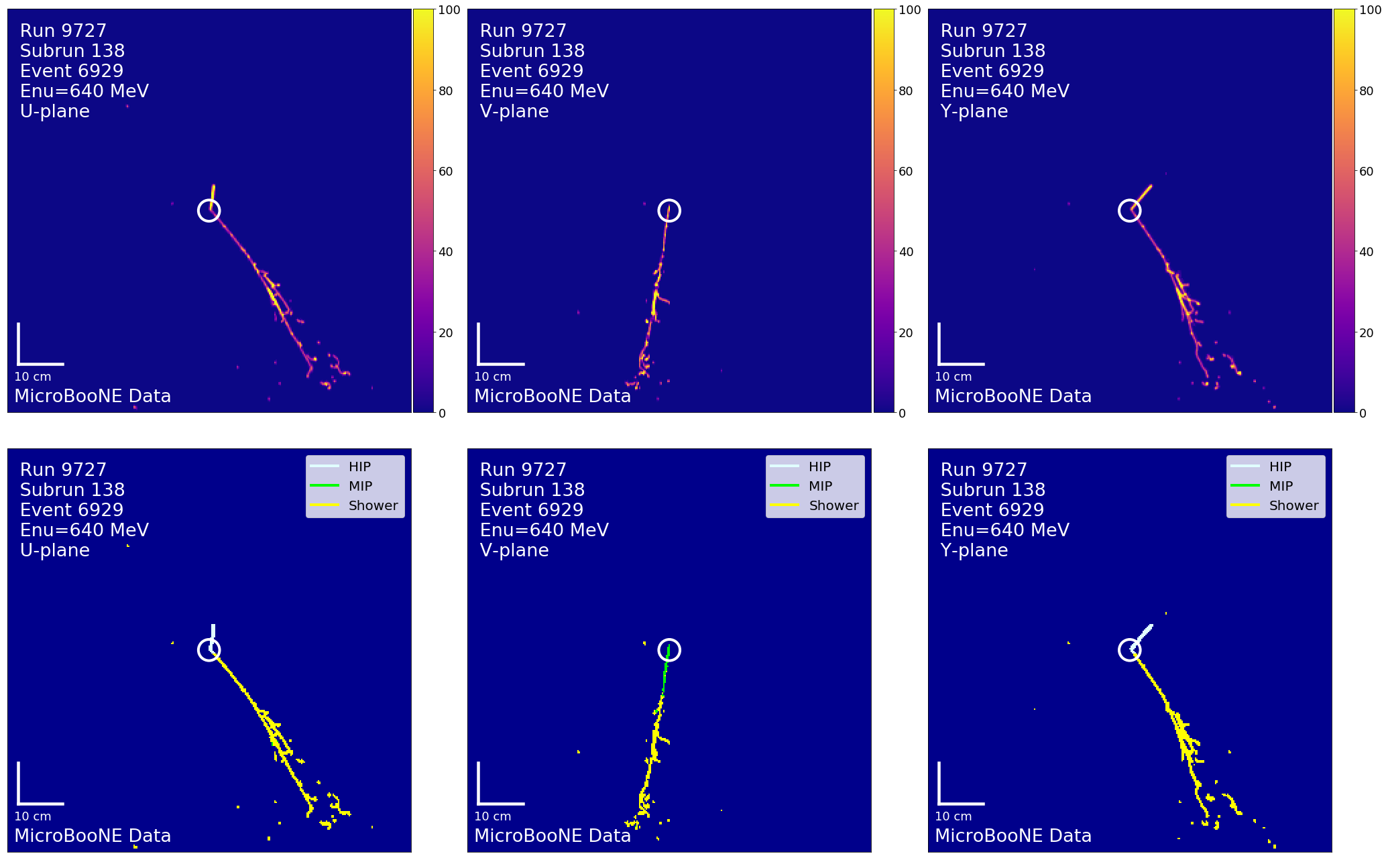}
    \caption{Top: pixel intensity; Bottom: \sssnet\ labels; Left to right: U, V, Y, planes. The white circle indicates the reconstructed vertex.}
    \label{fig:evd12}
\end{figure*}

\begin{figure*}
    \centering
    \includegraphics[width=0.92\linewidth]{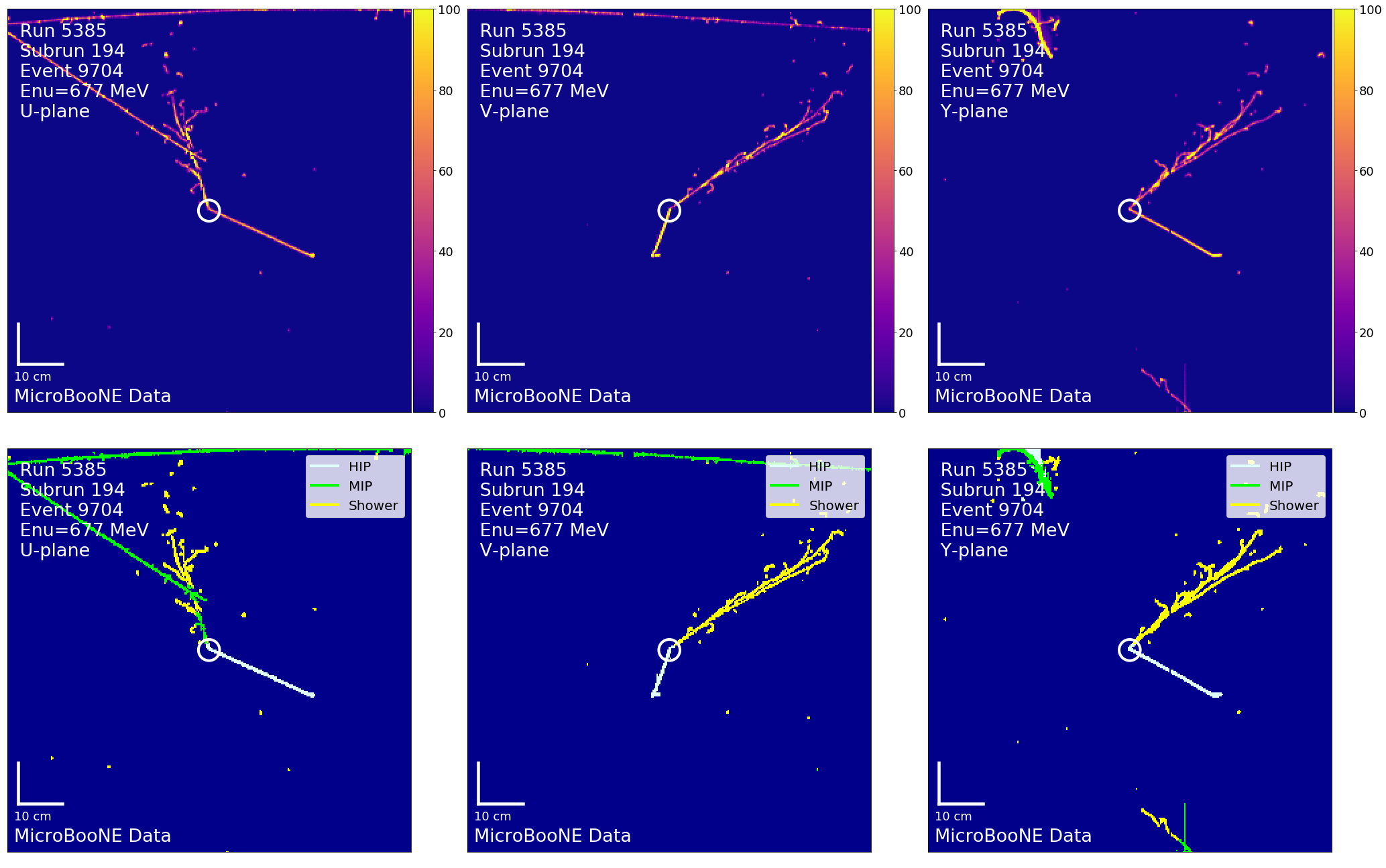}
    \caption{Top: pixel intensity; Bottom: \sssnet\ labels; Left to right: U, V, Y, planes. The white circle indicates the reconstructed vertex.}
    \label{fig:evd13}
\end{figure*}

\begin{figure*}
    \centering
    \includegraphics[width=0.92\linewidth]{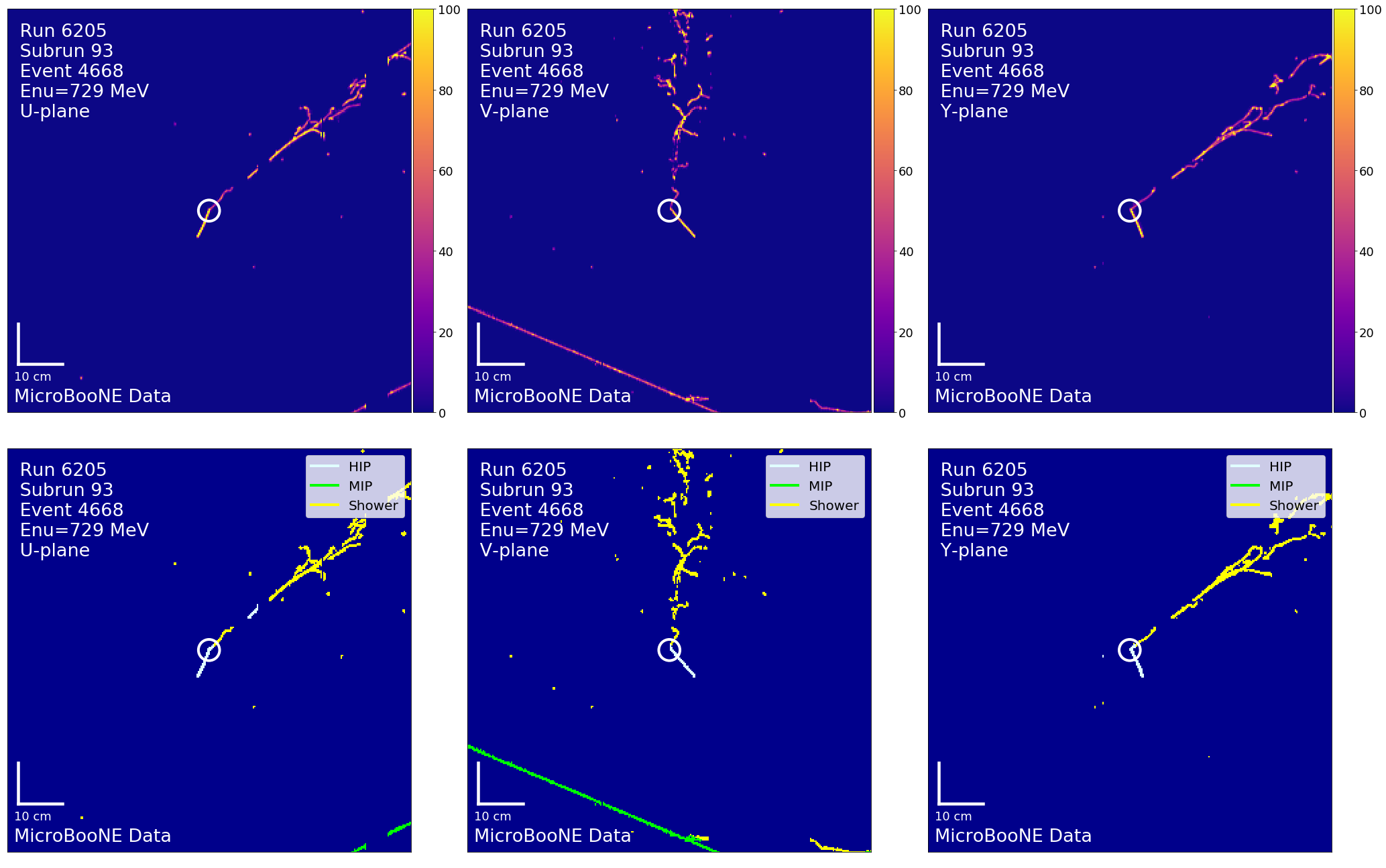}
    \caption{Top: pixel intensity; Bottom: \sssnet\ labels; Left to right: U, V, Y, planes. The white circle indicates the reconstructed vertex.}
    \label{fig:evd14}
\end{figure*}

\begin{figure*}
    \centering
    \includegraphics[width=0.92\linewidth]{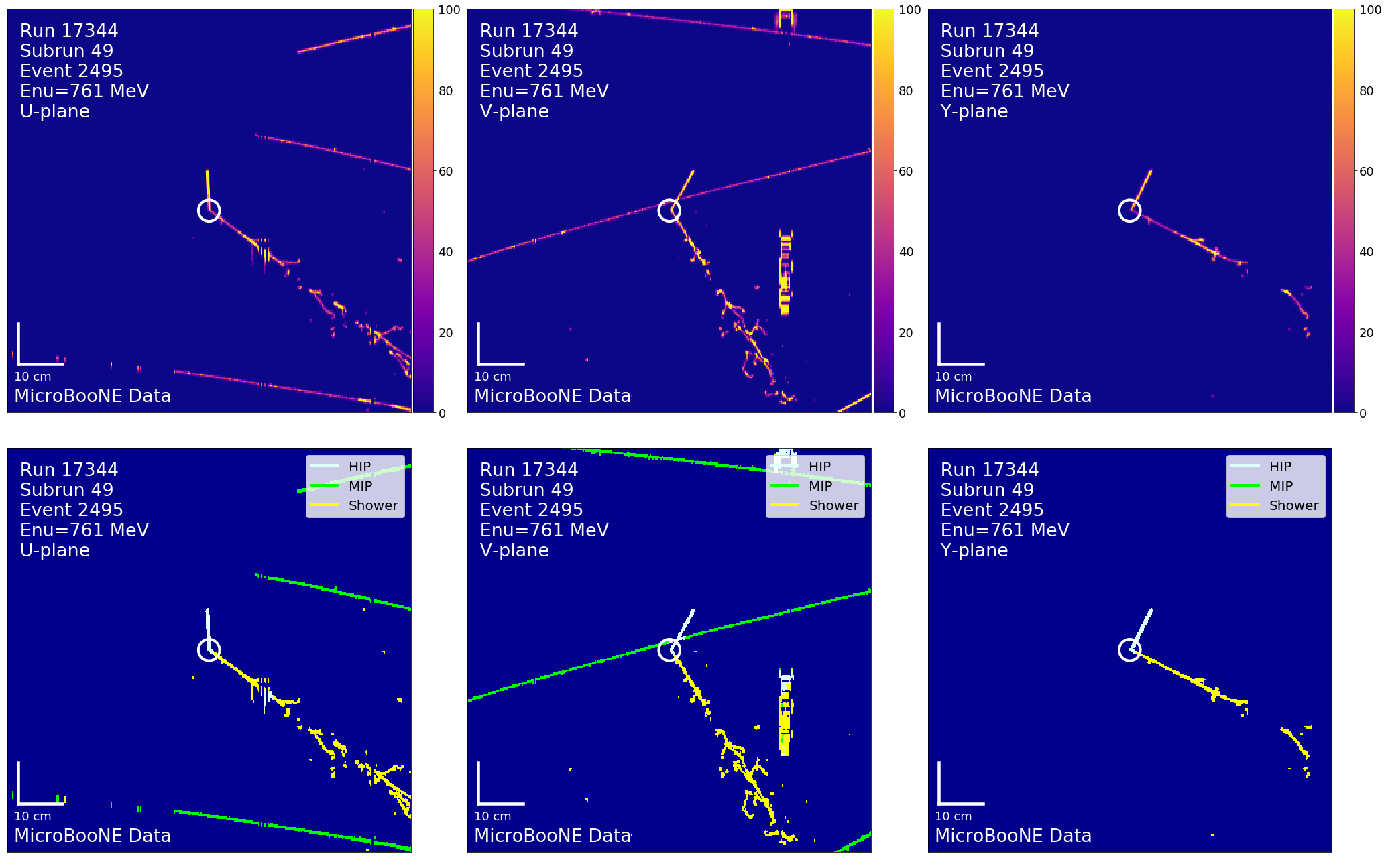}
    \caption{Top: pixel intensity; Bottom: \sssnet\ labels; Left to right: U, V, Y, planes. The white circle indicates the reconstructed vertex.}
    \label{fig:evd15}
\end{figure*}

\begin{figure*}
    \centering
    \includegraphics[width=0.92\linewidth]{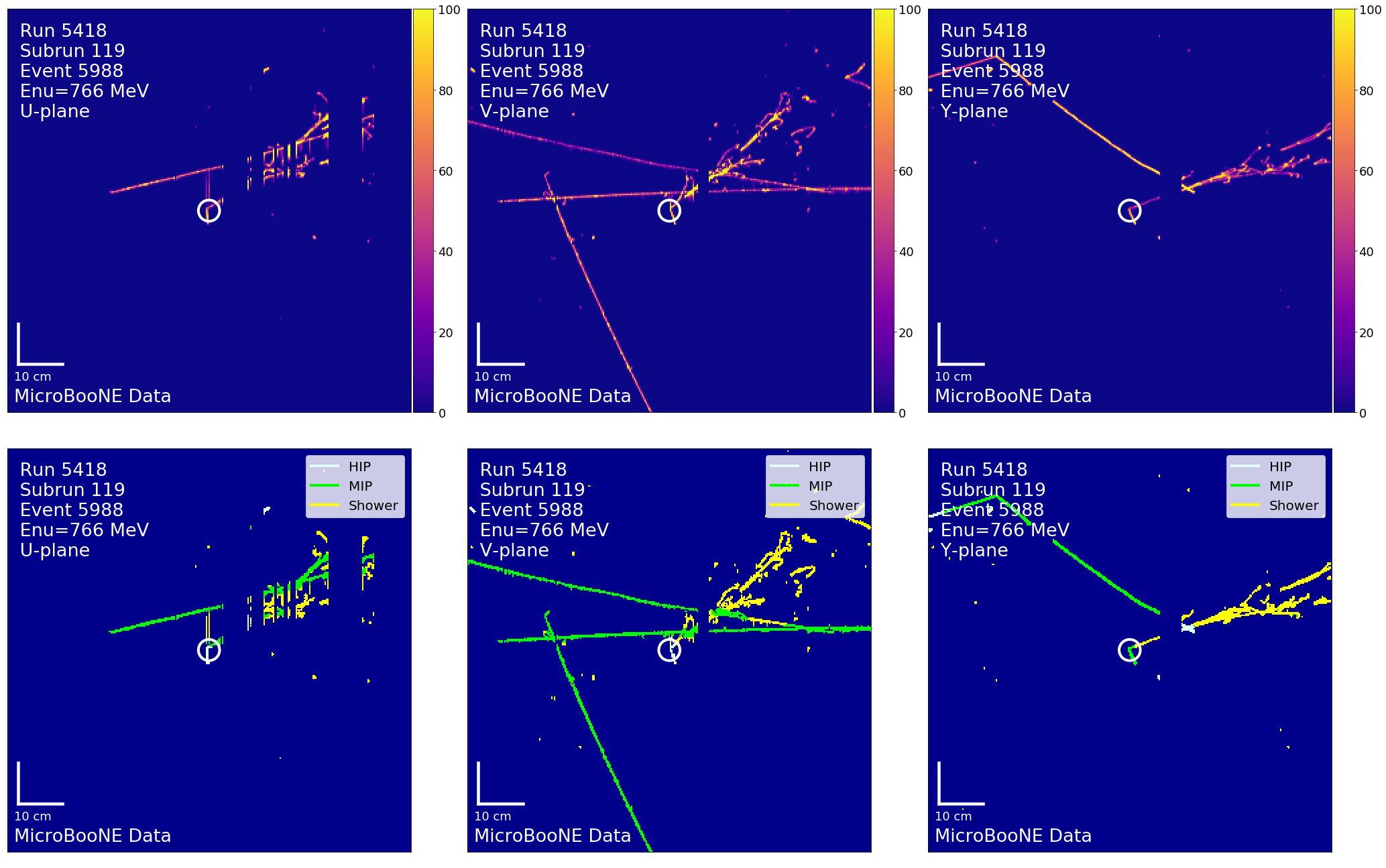}
    \caption{Top: pixel intensity; Bottom: \sssnet\ labels; Left to right: U, V, Y, planes. The white circle indicates the reconstructed vertex.}
    \label{fig:evd16}
\end{figure*}

\begin{figure*}
    \centering
    \includegraphics[width=0.92\linewidth]{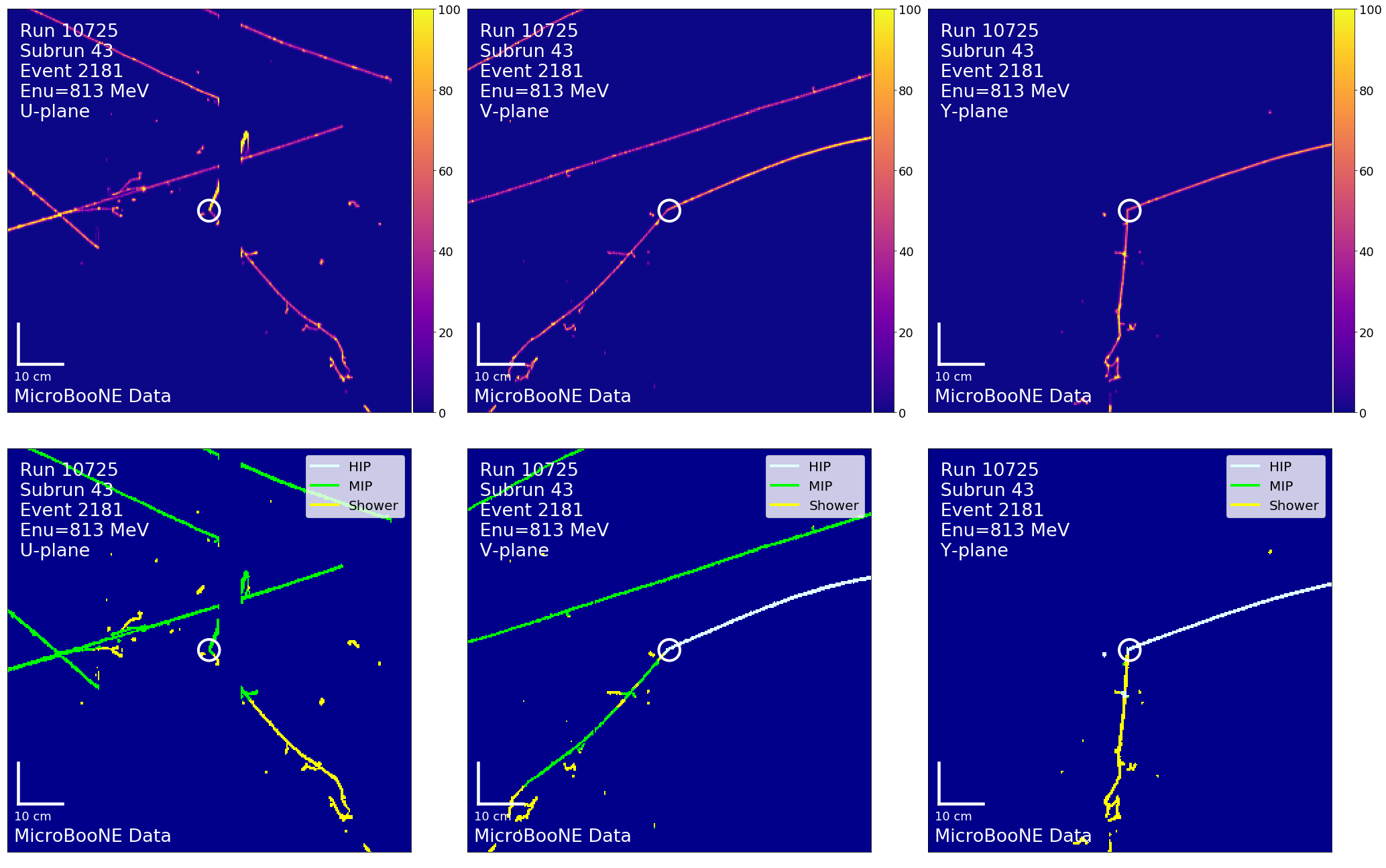}
    \caption{Top: pixel intensity; Bottom: \sssnet\ labels; Left to right: U, V, Y, planes. The white circle indicates the reconstructed vertex.}
    \label{fig:evd17}
\end{figure*}

\begin{figure*}
    \centering
    \includegraphics[width=0.92\linewidth]{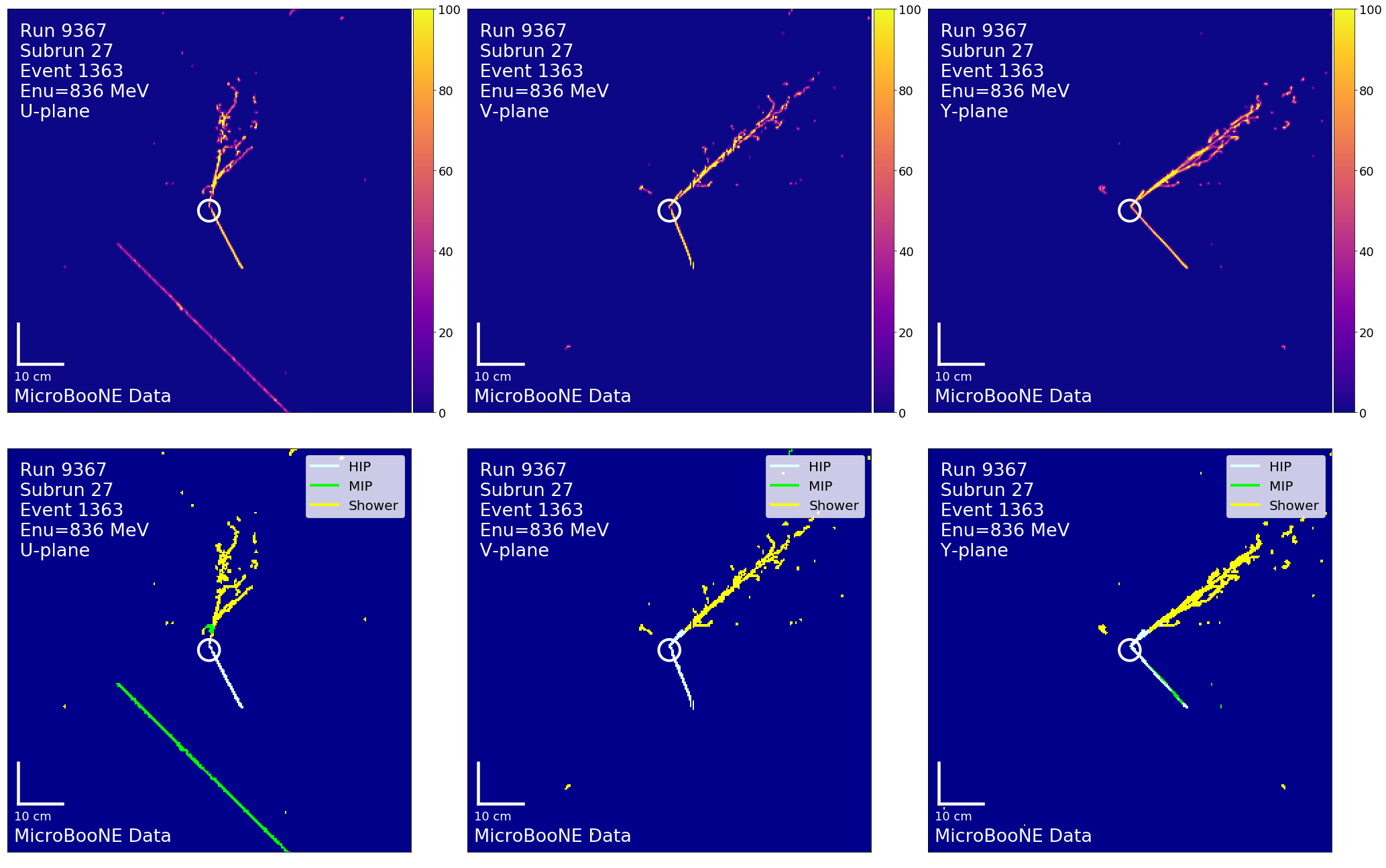}
    \caption{Top: pixel intensity; Bottom: \sssnet\ labels; Left to right: U, V, Y, planes. The white circle indicates the reconstructed vertex.}
    \label{fig:evd18}
\end{figure*}

\begin{figure*}
    \centering
    \includegraphics[width=0.92\linewidth]{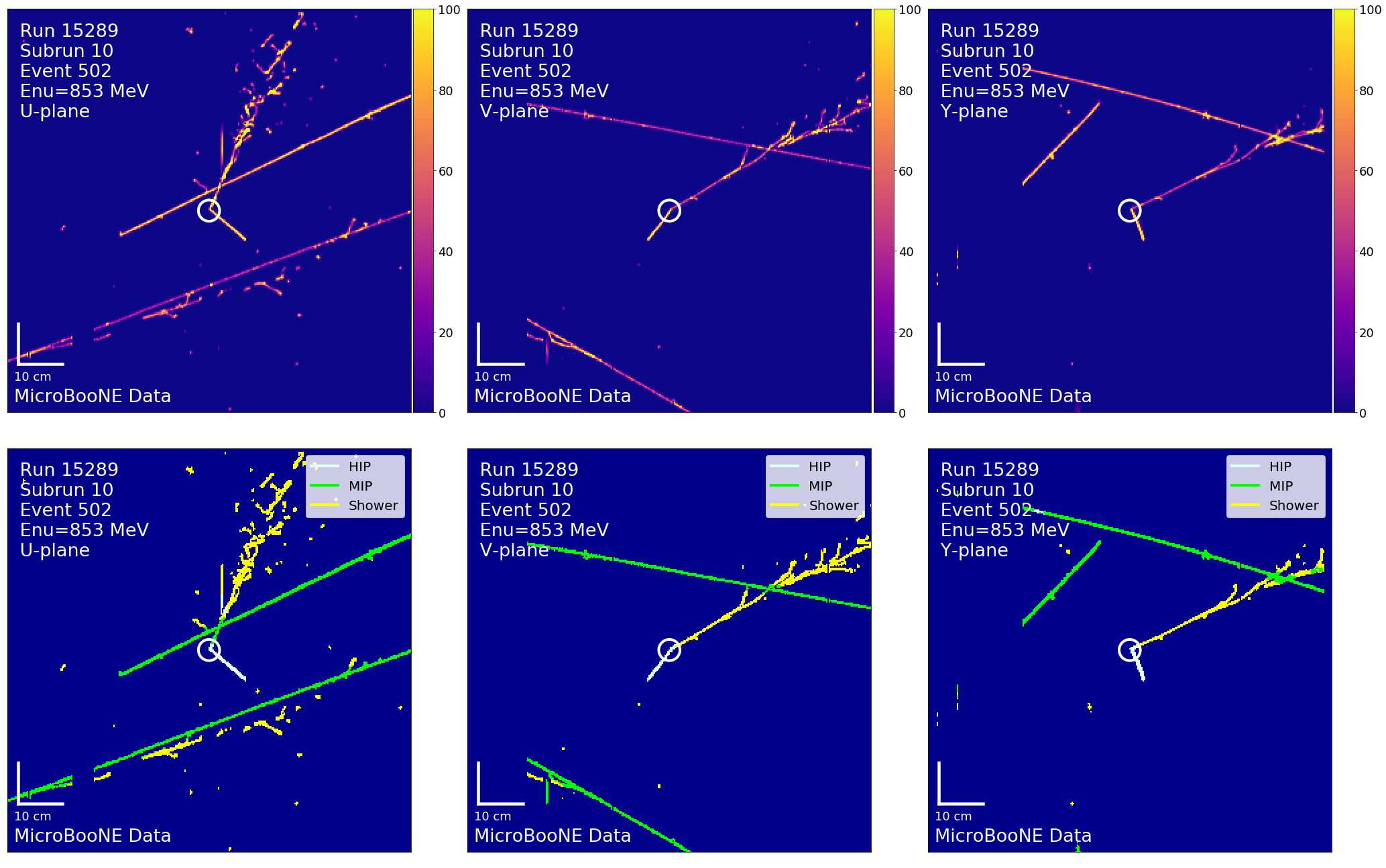}
    \caption{Top: pixel intensity; Bottom: \sssnet\ labels; Left to right: U, V, Y, planes. The white circle indicates the reconstructed vertex.}
    \label{fig:evd19}
\end{figure*}

\begin{figure*}
    \centering
    \includegraphics[width=0.92\linewidth]{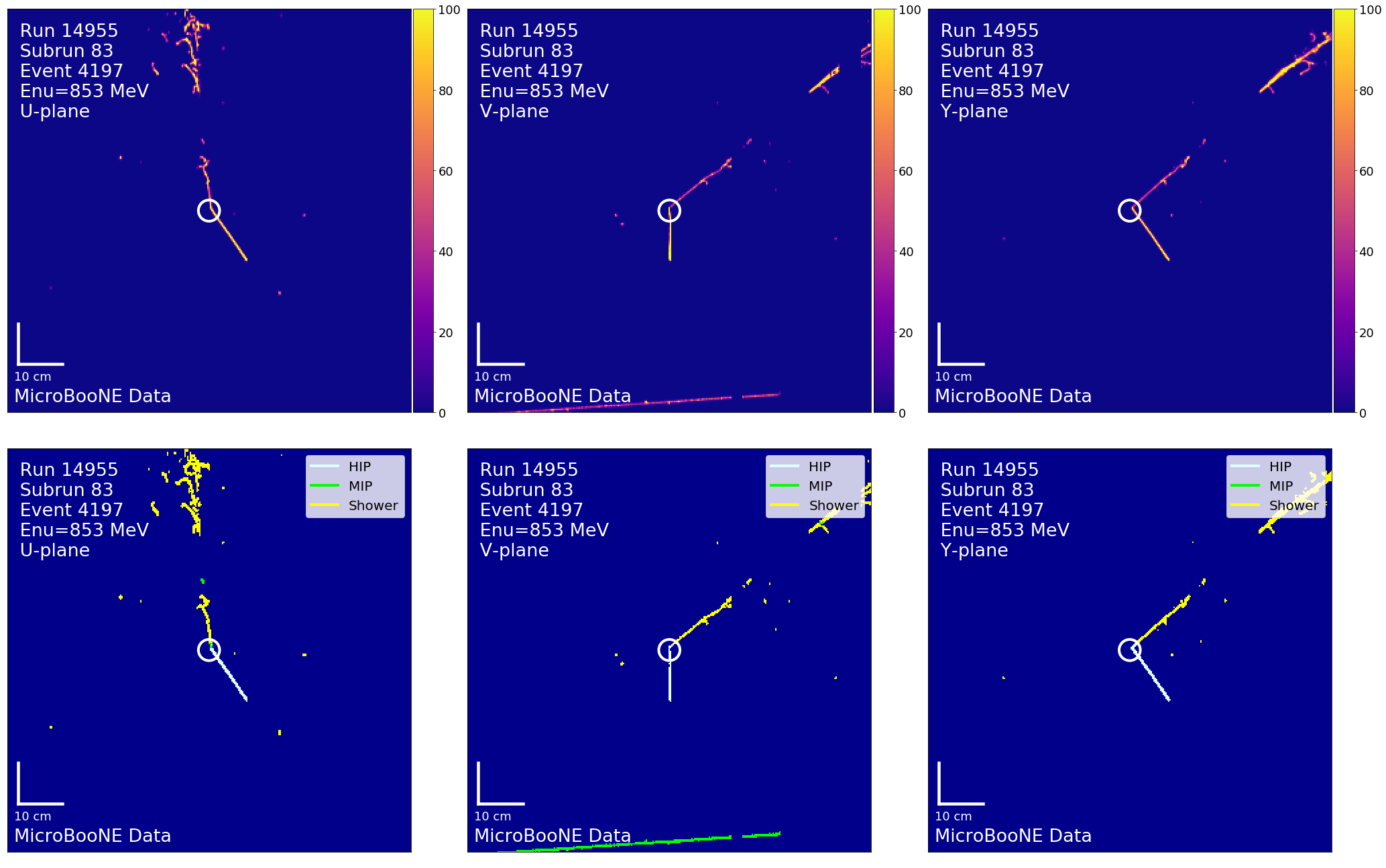}
    \caption{Top: pixel intensity; Bottom: \sssnet\ labels; Left to right: U, V, Y, planes. The white circle indicates the reconstructed vertex.}
    \label{fig:evd20}
\end{figure*}

\begin{figure*}
    \centering
    \includegraphics[width=0.92\linewidth]{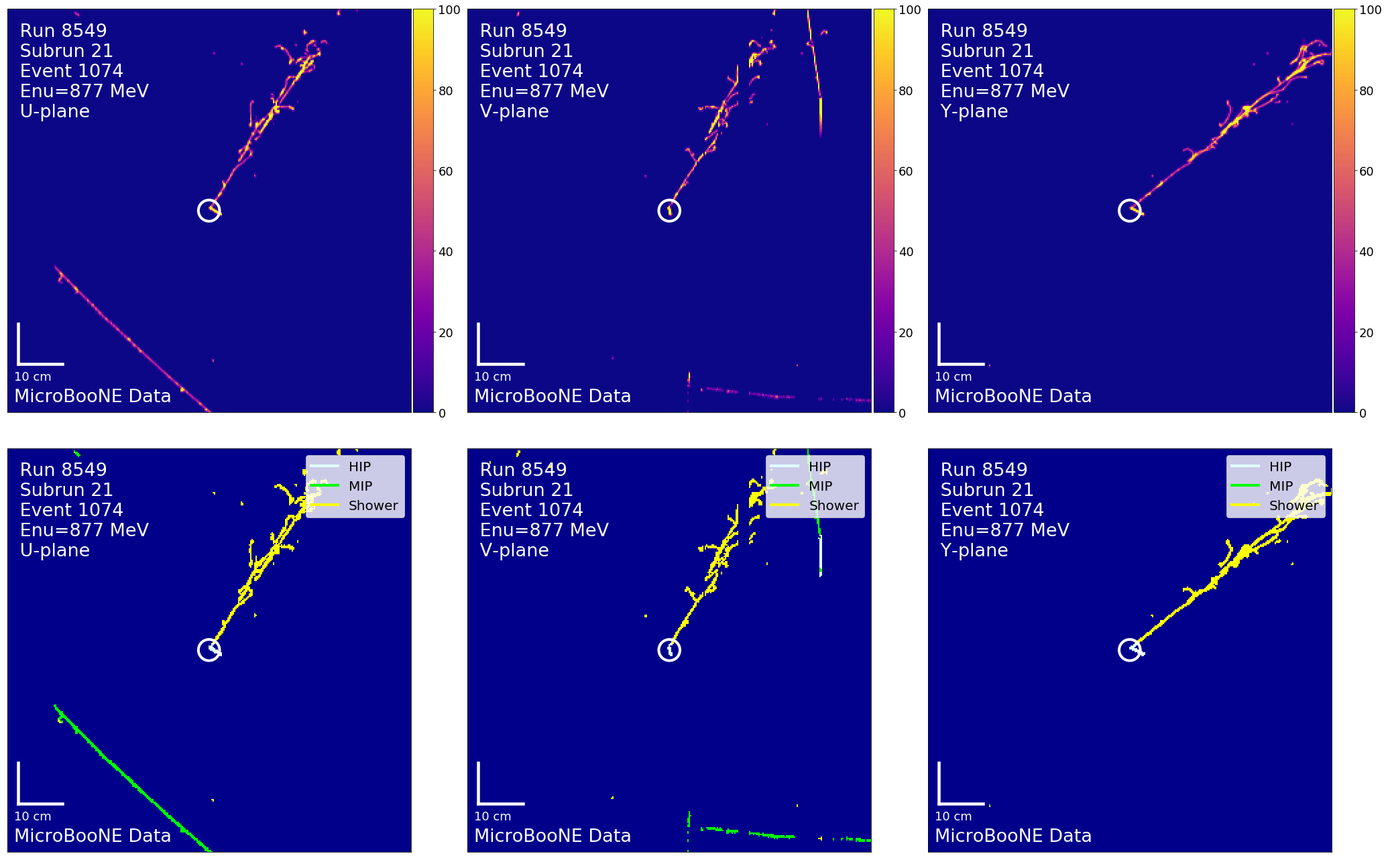}
    \caption{Top: pixel intensity; Bottom: \sssnet\ labels; Left to right: U, V, Y, planes. The white circle indicates the reconstructed vertex.}
    \label{fig:evd21}
\end{figure*}

\begin{figure*}
    \centering
    \includegraphics[width=0.92\linewidth]{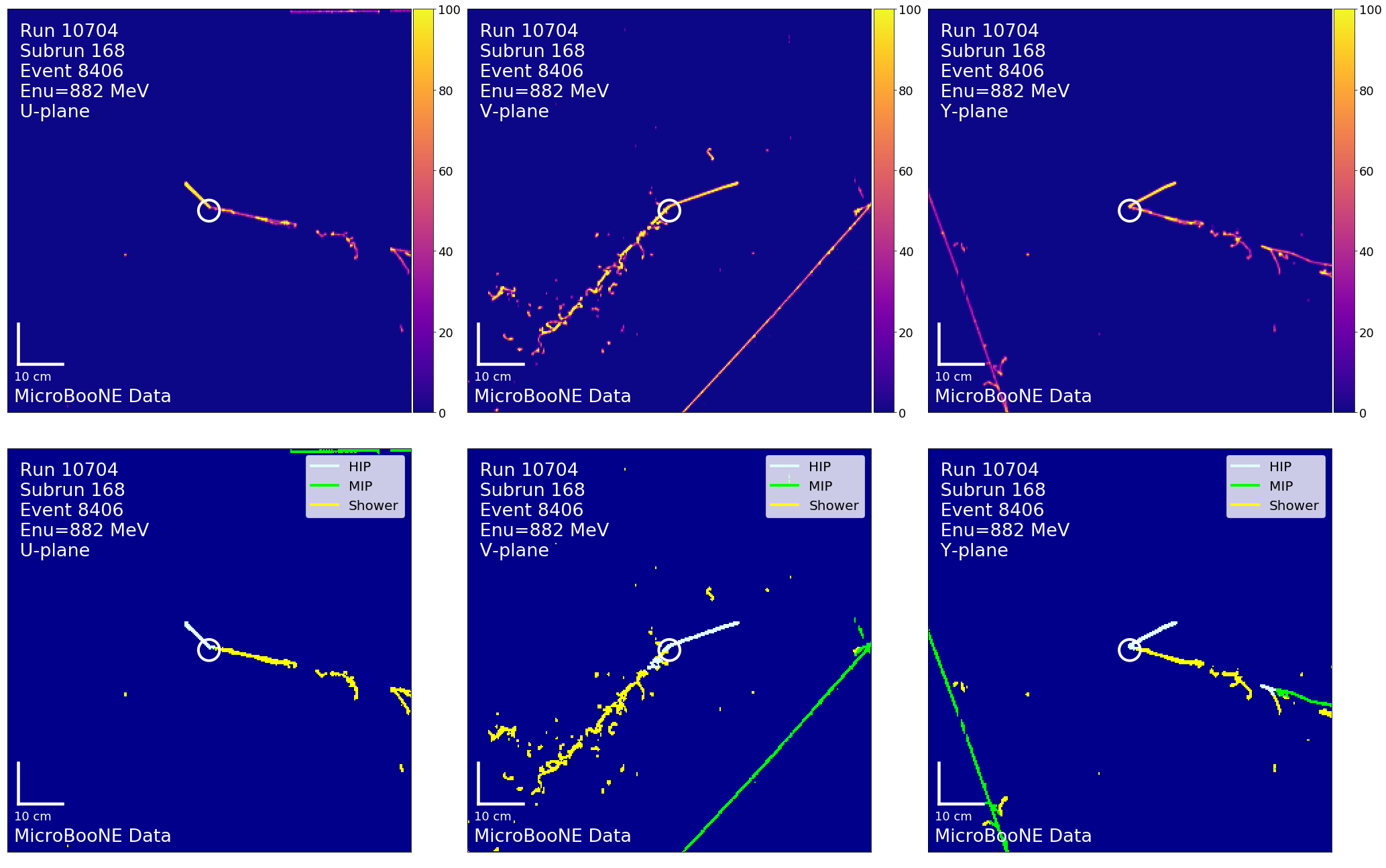}
    \caption{Top: pixel intensity; Bottom: \sssnet\ labels; Left to right: U, V, Y, planes. The white circle indicates the reconstructed vertex.}
    \label{fig:evd22}
\end{figure*}

\begin{figure*}
    \centering
    \includegraphics[width=0.92\linewidth]{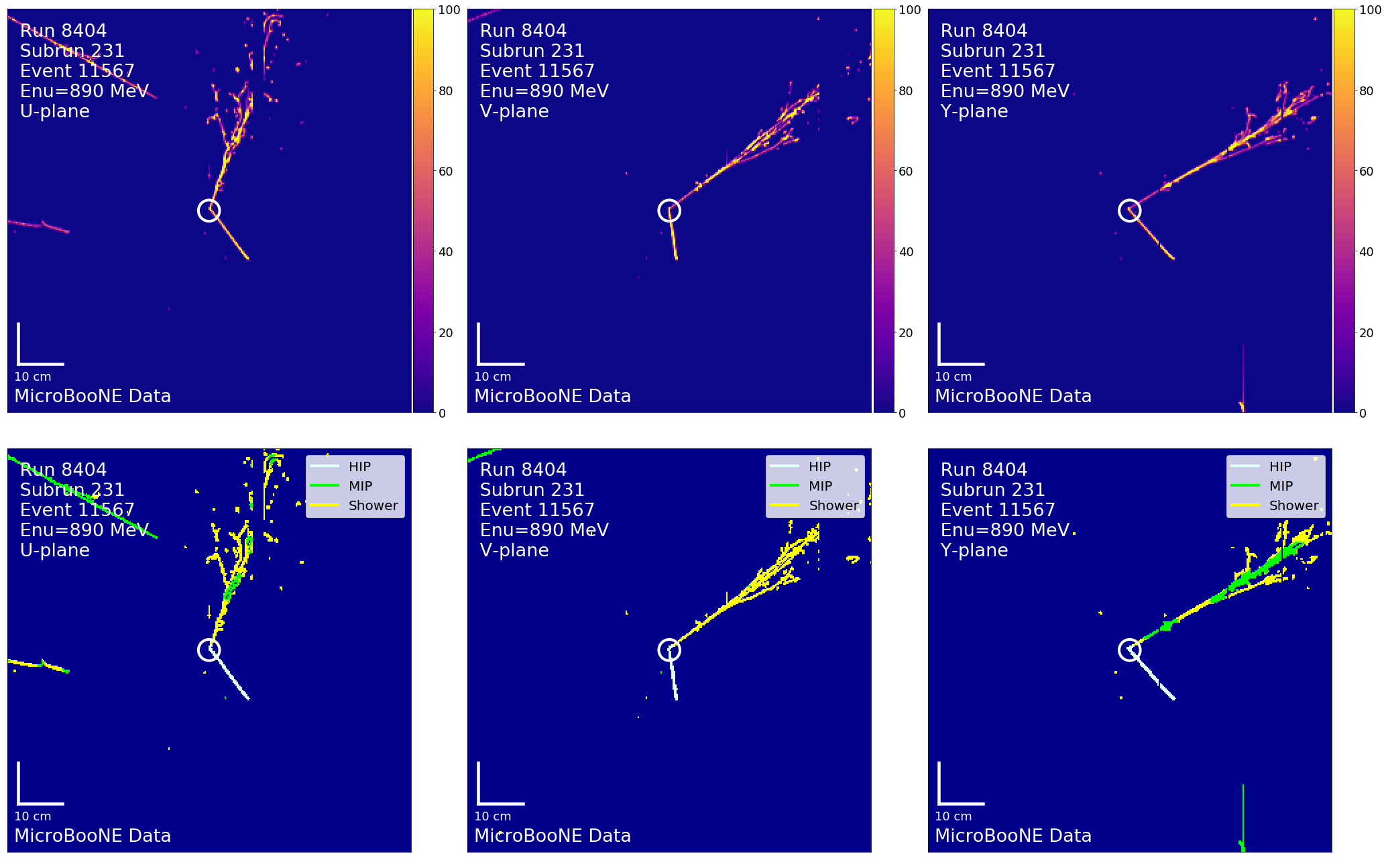}
    \caption{Top: pixel intensity; Bottom: \sssnet\ labels; Left to right: U, V, Y, planes. The white circle indicates the reconstructed vertex.}
    \label{fig:evd23}
\end{figure*}

\begin{figure*}
    \centering
    \includegraphics[width=0.92\linewidth]{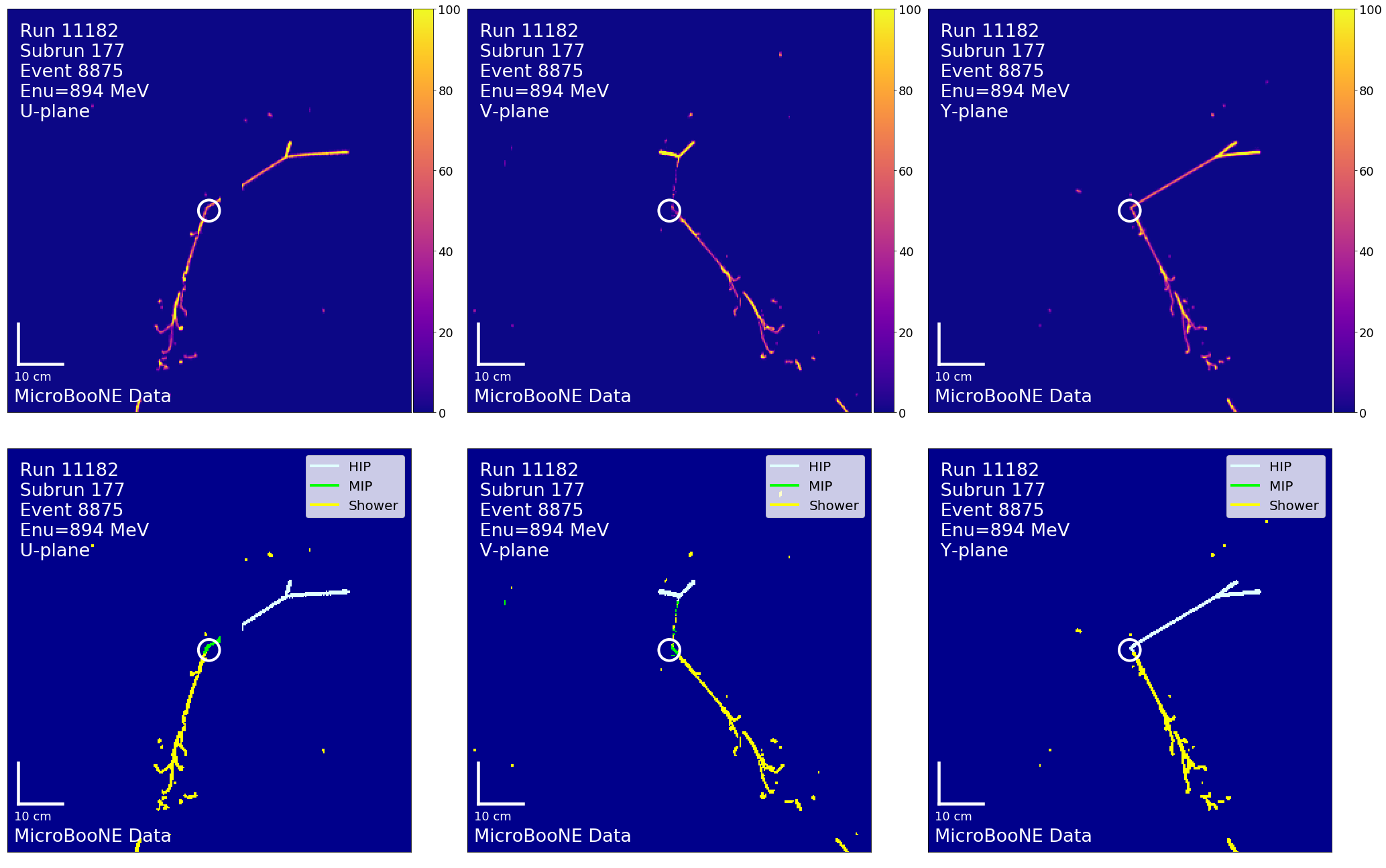}
    \caption{Top: pixel intensity; Bottom: \sssnet\ labels; Left to right: U, V, Y, planes. The white circle indicates the reconstructed vertex.}
    \label{fig:evd24}
\end{figure*}

\begin{figure*}
    \centering
    \includegraphics[width=0.92\linewidth]{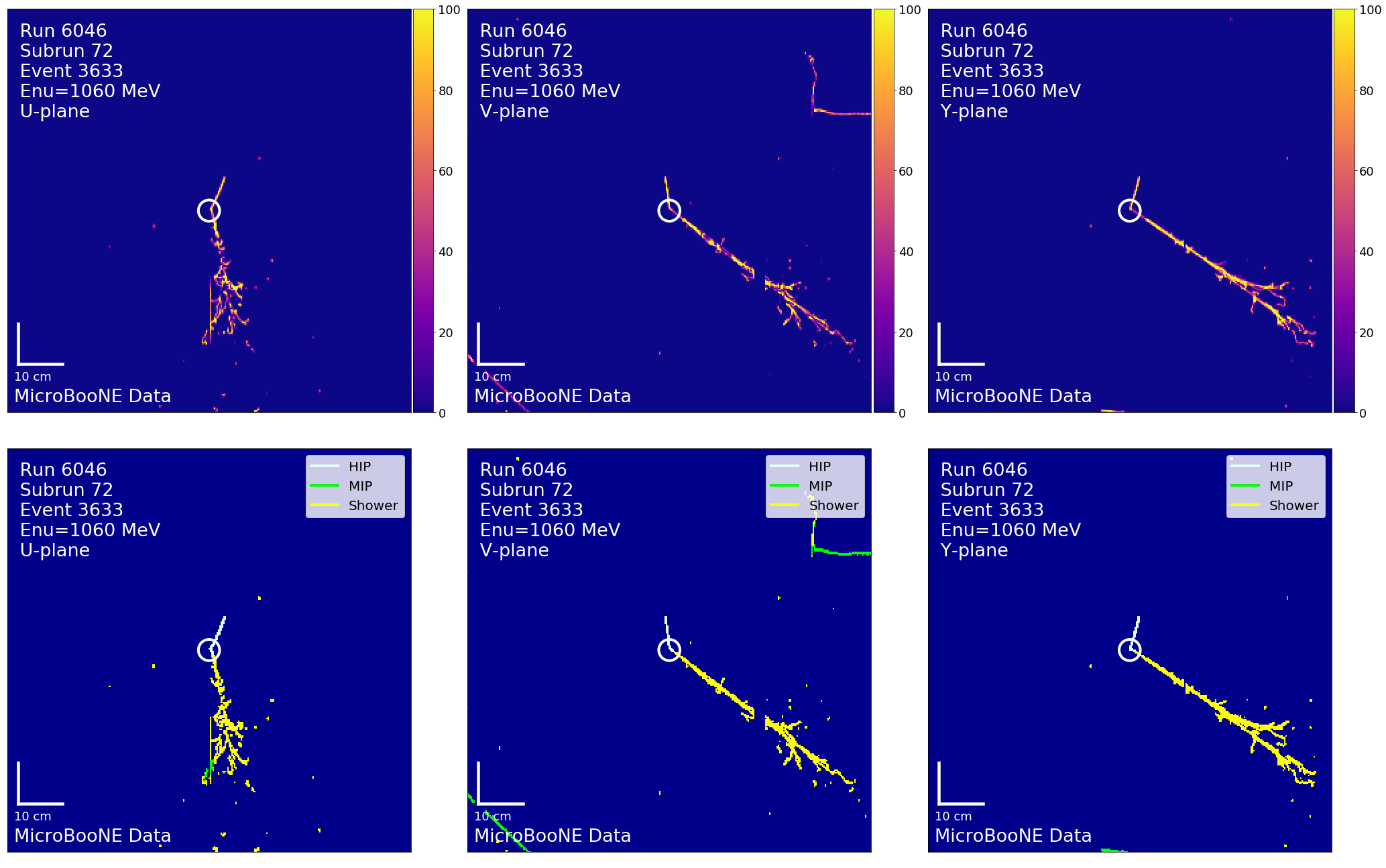}
    \caption{Top: pixel intensity; Bottom: \sssnet\ labels; Left to right: U, V, Y, planes. The white circle indicates the reconstructed vertex.}
    \label{fig:evd25}
\end{figure*}

We provide the event displays of the rest of the 24 $\nu_e$-candidate selected $\nu_e$ candidate events in Figs~\ref{fig:evd2}-\ref{fig:evd25} in ascending order in energy. For all displays, the top row is the pixel intensity image, while the bottom row is the \sssnet\ predictions. Each column is a different plane: U,V,Y from left to right.

\FloatBarrier

\FloatBarrier
\section{Selection Criteria and BDT Variables}

\bibliography{bib}

\begin{thebibliography}{67}%
\makeatletter
\providecommand \@ifxundefined [1]{%
 \@ifx{#1\undefined}
}%
\providecommand \@ifnum [1]{%
 \ifnum #1\expandafter \@firstoftwo
 \else \expandafter \@secondoftwo
 \fi
}%
\providecommand \@ifx [1]{%
 \ifx #1\expandafter \@firstoftwo
 \else \expandafter \@secondoftwo
 \fi
}%
\providecommand \natexlab [1]{#1}%
\providecommand \enquote  [1]{``#1''}%
\providecommand \bibnamefont  [1]{#1}%
\providecommand \bibfnamefont [1]{#1}%
\providecommand \citenamefont [1]{#1}%
\providecommand \href@noop [0]{\@secondoftwo}%
\providecommand \href [0]{\begingroup \@sanitize@url \@href}%
\providecommand \@href[1]{\@@startlink{#1}\@@href}%
\providecommand \@@href[1]{\endgroup#1\@@endlink}%
\providecommand \@sanitize@url [0]{\catcode `\\12\catcode `\$12\catcode
  `\&12\catcode `\#12\catcode `\^12\catcode `\_12\catcode `\%12\relax}%
\providecommand \@@startlink[1]{}%
\providecommand \@@endlink[0]{}%
\providecommand \url  [0]{\begingroup\@sanitize@url \@url }%
\providecommand \@url [1]{\endgroup\@href {#1}{\urlprefix }}%
\providecommand \urlprefix  [0]{URL }%
\providecommand \Eprint [0]{\href }%
\providecommand \doibase [0]{http://dx.doi.org/}%
\providecommand \selectlanguage [0]{\@gobble}%
\providecommand \bibinfo  [0]{\@secondoftwo}%
\providecommand \bibfield  [0]{\@secondoftwo}%
\providecommand \translation [1]{[#1]}%
\providecommand \BibitemOpen [0]{}%
\providecommand \bibitemStop [0]{}%
\providecommand \bibitemNoStop [0]{.\EOS\space}%
\providecommand \EOS [0]{\spacefactor3000\relax}%
\providecommand \BibitemShut  [1]{\csname bibitem#1\endcsname}%
\let\auto@bib@innerbib\@empty
\bibitem [{\citenamefont {Aguilar-Arevalo}\ \emph {et~al.}(2018)\citenamefont
  {Aguilar-Arevalo} \emph {et~al.}}]{MiniBooNE:2018esg}%
  \BibitemOpen
  \bibfield  {author} {\bibinfo {author} {\bibfnamefont {A.~A.}\ \bibnamefont
  {Aguilar-Arevalo}} \emph {et~al.} (\bibinfo {collaboration} {MiniBooNE}),\
  }\href {\doibase 10.1103/PhysRevLett.121.221801} {\bibfield  {journal}
  {\bibinfo  {journal} {Phys. Rev. Lett.}\ }\textbf {\bibinfo {volume} {121}},\
  \bibinfo {pages} {221801} (\bibinfo {year} {2018})},\ \Eprint
  {http://arxiv.org/abs/1805.12028} {arXiv:1805.12028 [hep-ex]} \BibitemShut
  {NoStop}%
\bibitem [{\citenamefont {Acciarri}\ \emph
  {et~al.}(2017{\natexlab{a}})\citenamefont {Acciarri} \emph
  {et~al.}}]{MicroBooNE:2016pwy}%
  \BibitemOpen
  \bibfield  {author} {\bibinfo {author} {\bibfnamefont {R.}~\bibnamefont
  {Acciarri}} \emph {et~al.} (\bibinfo {collaboration} {MicroBooNE}),\ }\href
  {\doibase 10.1088/1748-0221/12/02/P02017} {\bibfield  {journal} {\bibinfo
  {journal} {JINST}\ }\textbf {\bibinfo {volume} {12}},\ \bibinfo {pages}
  {P02017} (\bibinfo {year} {2017}{\natexlab{a}})},\ \Eprint
  {http://arxiv.org/abs/1612.05824} {arXiv:1612.05824 [physics.ins-det]}
  \BibitemShut {NoStop}%
\bibitem [{\citenamefont {Aguilar-Arevalo}\ \emph {et~al.}(2021)\citenamefont
  {Aguilar-Arevalo} \emph {et~al.}}]{MiniBooNE:2020pnu}%
  \BibitemOpen
  \bibfield  {author} {\bibinfo {author} {\bibfnamefont {A.~A.}\ \bibnamefont
  {Aguilar-Arevalo}} \emph {et~al.} (\bibinfo {collaboration} {MiniBooNE}),\
  }\href {\doibase 10.1103/PhysRevD.103.052002} {\bibfield  {journal} {\bibinfo
   {journal} {Phys. Rev. D}\ }\textbf {\bibinfo {volume} {103}},\ \bibinfo
  {pages} {052002} (\bibinfo {year} {2021})},\ \Eprint
  {http://arxiv.org/abs/2006.16883} {arXiv:2006.16883 [hep-ex]} \BibitemShut
  {NoStop}%
\bibitem [{\citenamefont {Diaz}\ \emph {et~al.}(2020)\citenamefont {Diaz},
  \citenamefont {Arg\"uelles}, \citenamefont {Collin}, \citenamefont {Conrad},\
  and\ \citenamefont {Shaevitz}}]{Diaz:2019fwt}%
  \BibitemOpen
  \bibfield  {author} {\bibinfo {author} {\bibfnamefont {A.}~\bibnamefont
  {Diaz}}, \bibinfo {author} {\bibfnamefont {C.~A.}\ \bibnamefont
  {Arg\"uelles}}, \bibinfo {author} {\bibfnamefont {G.~H.}\ \bibnamefont
  {Collin}}, \bibinfo {author} {\bibfnamefont {J.~M.}\ \bibnamefont {Conrad}},
  \ and\ \bibinfo {author} {\bibfnamefont {M.~H.}\ \bibnamefont {Shaevitz}},\
  }\href {\doibase 10.1016/j.physrep.2020.08.005} {\bibfield  {journal}
  {\bibinfo  {journal} {Phys. Rep.}\ }\textbf {\bibinfo {volume} {884}},\
  \bibinfo {pages} {1} (\bibinfo {year} {2020})},\ \Eprint
  {http://arxiv.org/abs/1906.00045} {arXiv:1906.00045 [hep-ex]} \BibitemShut
  {NoStop}%
\bibitem [{\citenamefont {Magill}\ \emph {et~al.}(2018)\citenamefont {Magill},
  \citenamefont {Plestid}, \citenamefont {Pospelov},\ and\ \citenamefont
  {Tsai}}]{Magill:2018jla}%
  \BibitemOpen
  \bibfield  {author} {\bibinfo {author} {\bibfnamefont {G.}~\bibnamefont
  {Magill}}, \bibinfo {author} {\bibfnamefont {R.}~\bibnamefont {Plestid}},
  \bibinfo {author} {\bibfnamefont {M.}~\bibnamefont {Pospelov}}, \ and\
  \bibinfo {author} {\bibfnamefont {Y.-D.}\ \bibnamefont {Tsai}},\ }\href
  {\doibase 10.1103/PhysRevD.98.115015} {\bibfield  {journal} {\bibinfo
  {journal} {Phys. Rev. D}\ }\textbf {\bibinfo {volume} {98}},\ \bibinfo
  {pages} {115015} (\bibinfo {year} {2018})},\ \Eprint
  {http://arxiv.org/abs/1803.03262} {arXiv:1803.03262 [hep-ph]} \BibitemShut
  {NoStop}%
\bibitem [{\citenamefont {Fujikawa}\ and\ \citenamefont
  {Shrock}(1980)}]{Fujikawa:1980yx}%
  \BibitemOpen
  \bibfield  {author} {\bibinfo {author} {\bibfnamefont {K.}~\bibnamefont
  {Fujikawa}}\ and\ \bibinfo {author} {\bibfnamefont {R.}~\bibnamefont
  {Shrock}},\ }\href {\doibase 10.1103/PhysRevLett.45.963} {\bibfield
  {journal} {\bibinfo  {journal} {Phys. Rev. Lett.}\ }\textbf {\bibinfo
  {volume} {45}},\ \bibinfo {pages} {963} (\bibinfo {year} {1980})}\BibitemShut
  {NoStop}%
\bibitem [{\citenamefont {Palazzo}\ \emph {et~al.}(2007)\citenamefont
  {Palazzo}, \citenamefont {Cumberbatch}, \citenamefont {Slosar},\ and\
  \citenamefont {Silk}}]{Palazzo:2007gz}%
  \BibitemOpen
  \bibfield  {author} {\bibinfo {author} {\bibfnamefont {A.}~\bibnamefont
  {Palazzo}}, \bibinfo {author} {\bibfnamefont {D.}~\bibnamefont
  {Cumberbatch}}, \bibinfo {author} {\bibfnamefont {A.}~\bibnamefont {Slosar}},
  \ and\ \bibinfo {author} {\bibfnamefont {J.}~\bibnamefont {Silk}},\ }\href
  {\doibase 10.1103/PhysRevD.76.103511} {\bibfield  {journal} {\bibinfo
  {journal} {Phys. Rev. D}\ }\textbf {\bibinfo {volume} {76}},\ \bibinfo
  {pages} {103511} (\bibinfo {year} {2007})},\ \Eprint
  {http://arxiv.org/abs/0707.1495} {arXiv:0707.1495 [astro-ph]} \BibitemShut
  {NoStop}%
\bibitem [{\citenamefont {Shrock}(1982)}]{Shrock:1982sc}%
  \BibitemOpen
  \bibfield  {author} {\bibinfo {author} {\bibfnamefont {R.~E.}\ \bibnamefont
  {Shrock}},\ }\href {\doibase 10.1016/0550-3213(82)90273-5} {\bibfield
  {journal} {\bibinfo  {journal} {Nucl. Phys. B}\ }\textbf {\bibinfo {volume}
  {206}},\ \bibinfo {pages} {359} (\bibinfo {year} {1982})}\BibitemShut
  {NoStop}%
\bibitem [{\citenamefont {Dvornikov}\ and\ \citenamefont
  {Studenikin}(2004)}]{Dvornikov:2003js}%
  \BibitemOpen
  \bibfield  {author} {\bibinfo {author} {\bibfnamefont {M.}~\bibnamefont
  {Dvornikov}}\ and\ \bibinfo {author} {\bibfnamefont {A.}~\bibnamefont
  {Studenikin}},\ }\href {\doibase 10.1103/PhysRevD.69.073001} {\bibfield
  {journal} {\bibinfo  {journal} {Phys. Rev. D}\ }\textbf {\bibinfo {volume}
  {69}},\ \bibinfo {pages} {073001} (\bibinfo {year} {2004})},\ \Eprint
  {http://arxiv.org/abs/hep-ph/0305206} {arXiv:hep-ph/0305206} \BibitemShut
  {NoStop}%
\bibitem [{\citenamefont {Giunti}\ and\ \citenamefont
  {Studenikin}(2015)}]{Giunti:2014ixa}%
  \BibitemOpen
  \bibfield  {author} {\bibinfo {author} {\bibfnamefont {C.}~\bibnamefont
  {Giunti}}\ and\ \bibinfo {author} {\bibfnamefont {A.}~\bibnamefont
  {Studenikin}},\ }\href {\doibase 10.1103/RevModPhys.87.531} {\bibfield
  {journal} {\bibinfo  {journal} {Rev. Mod. Phys.}\ }\textbf {\bibinfo {volume}
  {87}},\ \bibinfo {pages} {531} (\bibinfo {year} {2015})},\ \Eprint
  {http://arxiv.org/abs/1403.6344} {arXiv:1403.6344 [hep-ph]} \BibitemShut
  {NoStop}%
\bibitem [{\citenamefont {Brdar}\ \emph {et~al.}(2021)\citenamefont {Brdar},
  \citenamefont {Greljo}, \citenamefont {Kopp},\ and\ \citenamefont
  {Opferkuch}}]{Brdar:2020quo}%
  \BibitemOpen
  \bibfield  {author} {\bibinfo {author} {\bibfnamefont {V.}~\bibnamefont
  {Brdar}}, \bibinfo {author} {\bibfnamefont {A.}~\bibnamefont {Greljo}},
  \bibinfo {author} {\bibfnamefont {J.}~\bibnamefont {Kopp}}, \ and\ \bibinfo
  {author} {\bibfnamefont {T.}~\bibnamefont {Opferkuch}},\ }\href {\doibase
  10.1088/1475-7516/2021/01/039} {\bibfield  {journal} {\bibinfo  {journal}
  {JCAP}\ }\textbf {\bibinfo {volume} {01}},\ \bibinfo {pages} {039} (\bibinfo
  {year} {2021})},\ \Eprint {http://arxiv.org/abs/2007.15563} {arXiv:2007.15563
  [hep-ph]} \BibitemShut {NoStop}%
\bibitem [{\citenamefont {Vergani}\ \emph {et~al.}(2021)\citenamefont
  {Vergani}, \citenamefont {Kamp}, \citenamefont {Diaz}, \citenamefont
  {Arg\"uelles}, \citenamefont {Conrad}, \citenamefont {Shaevitz},\ and\
  \citenamefont {Uchida}}]{Vergani:2021tgc}%
  \BibitemOpen
  \bibfield  {author} {\bibinfo {author} {\bibfnamefont {S.}~\bibnamefont
  {Vergani}}, \bibinfo {author} {\bibfnamefont {N.~W.}\ \bibnamefont {Kamp}},
  \bibinfo {author} {\bibfnamefont {A.}~\bibnamefont {Diaz}}, \bibinfo {author}
  {\bibfnamefont {C.~A.}\ \bibnamefont {Arg\"uelles}}, \bibinfo {author}
  {\bibfnamefont {J.~M.}\ \bibnamefont {Conrad}}, \bibinfo {author}
  {\bibfnamefont {M.~H.}\ \bibnamefont {Shaevitz}}, \ and\ \bibinfo {author}
  {\bibfnamefont {M.~A.}\ \bibnamefont {Uchida}},\ }\href@noop {} {\  (\bibinfo
  {year} {2021})},\ \Eprint {http://arxiv.org/abs/2105.06470} {arXiv:2105.06470
  [hep-ph]} \BibitemShut {NoStop}%
\bibitem [{\citenamefont {Ericson}\ \emph {et~al.}(2016)\citenamefont
  {Ericson}, \citenamefont {Garzelli}, \citenamefont {Giunti},\ and\
  \citenamefont {Martini}}]{Ericson:2016yjn}%
  \BibitemOpen
  \bibfield  {author} {\bibinfo {author} {\bibfnamefont {M.}~\bibnamefont
  {Ericson}}, \bibinfo {author} {\bibfnamefont {M.~V.}\ \bibnamefont
  {Garzelli}}, \bibinfo {author} {\bibfnamefont {C.}~\bibnamefont {Giunti}}, \
  and\ \bibinfo {author} {\bibfnamefont {M.}~\bibnamefont {Martini}},\ }\href
  {\doibase 10.1103/PhysRevD.93.073008} {\bibfield  {journal} {\bibinfo
  {journal} {Phys. Rev. D}\ }\textbf {\bibinfo {volume} {93}},\ \bibinfo
  {pages} {073008} (\bibinfo {year} {2016})},\ \Eprint
  {http://arxiv.org/abs/1602.01390} {arXiv:1602.01390 [hep-ph]} \BibitemShut
  {NoStop}%
\bibitem [{\citenamefont {Alvarez-Ruso}\ \emph {et~al.}(2015)\citenamefont
  {Alvarez-Ruso}, \citenamefont {Nieves},\ and\ \citenamefont
  {Wang}}]{Alvarez-Ruso:2015kua}%
  \BibitemOpen
  \bibfield  {author} {\bibinfo {author} {\bibfnamefont {L.}~\bibnamefont
  {Alvarez-Ruso}}, \bibinfo {author} {\bibfnamefont {J.}~\bibnamefont
  {Nieves}}, \ and\ \bibinfo {author} {\bibfnamefont {E.}~\bibnamefont
  {Wang}},\ }\href {\doibase 10.1063/1.4931861} {\bibfield  {journal} {\bibinfo
   {journal} {AIP Conf. Proc.}\ }\textbf {\bibinfo {volume} {1680}},\ \bibinfo
  {pages} {020002} (\bibinfo {year} {2015})},\ \Eprint
  {http://arxiv.org/abs/1501.05995} {arXiv:1501.05995 [nucl-th]} \BibitemShut
  {NoStop}%
\bibitem [{\citenamefont {Abratenko}\ \emph
  {et~al.}({\natexlab{a}})\citenamefont {Abratenko} \emph {et~al.}}]{PeLEE}%
  \BibitemOpen
  \bibfield  {author} {\bibinfo {author} {\bibfnamefont {P.}~\bibnamefont
  {Abratenko}} \emph {et~al.} (\bibinfo {collaboration} {MicroBooNE}),\
  }\Eprint {http://arxiv.org/abs/2110.14065} {arXiv:2110.14065 [hep-ex]}
  \BibitemShut {NoStop}%
\bibitem [{\citenamefont {Abratenko}\ \emph
  {et~al.}({\natexlab{b}})\citenamefont {Abratenko} \emph {et~al.}}]{WCLEE}%
  \BibitemOpen
  \bibfield  {author} {\bibinfo {author} {\bibfnamefont {P.}~\bibnamefont
  {Abratenko}} \emph {et~al.} (\bibinfo {collaboration} {MicroBooNE}),\
  }\Eprint {http://arxiv.org/abs/2110.13978} {arXiv:2110.13978 [hep-ex]}
  \BibitemShut {NoStop}%
\bibitem [{\citenamefont {Abratenko}\ \emph
  {et~al.}(2021{\natexlab{a}})\citenamefont {Abratenko} \emph
  {et~al.}}]{MicroBooNE:2021zai}%
  \BibitemOpen
  \bibfield  {author} {\bibinfo {author} {\bibfnamefont {P.}~\bibnamefont
  {Abratenko}} \emph {et~al.} (\bibinfo {collaboration} {MicroBooNE}),\
  }\href@noop {} {\  (\bibinfo {year} {2021}{\natexlab{a}})},\ \Eprint
  {http://arxiv.org/abs/2110.00409} {arXiv:2110.00409 [hep-ex]} \BibitemShut
  {NoStop}%
\bibitem [{\citenamefont {Abratenko}\ \emph
  {et~al.}({\natexlab{c}})\citenamefont {Abratenko} \emph
  {et~al.}}]{jointMicroBooNEPRL}%
  \BibitemOpen
  \bibfield  {author} {\bibinfo {author} {\bibfnamefont {P.}~\bibnamefont
  {Abratenko}} \emph {et~al.} (\bibinfo {collaboration} {MicroBooNE}),\
  }\Eprint {http://arxiv.org/abs/2110.14054} {arXiv:2110.14054 [hep-ex]}
  \BibitemShut {NoStop}%
\bibitem [{\citenamefont {Acciarri}\ \emph
  {et~al.}(2017{\natexlab{b}})\citenamefont {Acciarri} \emph
  {et~al.}}]{MicroBooNE:2017qiu}%
  \BibitemOpen
  \bibfield  {author} {\bibinfo {author} {\bibfnamefont {R.}~\bibnamefont
  {Acciarri}} \emph {et~al.} (\bibinfo {collaboration} {MicroBooNE}),\ }\href
  {\doibase 10.1088/1748-0221/12/08/P08003} {\bibfield  {journal} {\bibinfo
  {journal} {JINST}\ }\textbf {\bibinfo {volume} {12}},\ \bibinfo {pages}
  {P08003} (\bibinfo {year} {2017}{\natexlab{b}})},\ \Eprint
  {http://arxiv.org/abs/1705.07341} {arXiv:1705.07341 [physics.ins-det]}
  \BibitemShut {NoStop}%
\bibitem [{\citenamefont {Adams}\ \emph
  {et~al.}(2018{\natexlab{a}})\citenamefont {Adams} \emph
  {et~al.}}]{MicroBooNE:2018swd}%
  \BibitemOpen
  \bibfield  {author} {\bibinfo {author} {\bibfnamefont {C.}~\bibnamefont
  {Adams}} \emph {et~al.} (\bibinfo {collaboration} {MicroBooNE}),\ }\href
  {\doibase 10.1088/1748-0221/13/07/P07006} {\bibfield  {journal} {\bibinfo
  {journal} {JINST}\ }\textbf {\bibinfo {volume} {13}},\ \bibinfo {pages}
  {P07006} (\bibinfo {year} {2018}{\natexlab{a}})},\ \Eprint
  {http://arxiv.org/abs/1802.08709} {arXiv:1802.08709 [physics.ins-det]}
  \BibitemShut {NoStop}%
\bibitem [{\citenamefont {Adams}\ \emph
  {et~al.}(2018{\natexlab{b}})\citenamefont {Adams} \emph
  {et~al.}}]{MicroBooNE:2018vro}%
  \BibitemOpen
  \bibfield  {author} {\bibinfo {author} {\bibfnamefont {C.}~\bibnamefont
  {Adams}} \emph {et~al.} (\bibinfo {collaboration} {MicroBooNE}),\ }\href
  {\doibase 10.1088/1748-0221/13/07/P07007} {\bibfield  {journal} {\bibinfo
  {journal} {JINST}\ }\textbf {\bibinfo {volume} {13}},\ \bibinfo {pages}
  {P07007} (\bibinfo {year} {2018}{\natexlab{b}})},\ \Eprint
  {http://arxiv.org/abs/1804.02583} {arXiv:1804.02583 [physics.ins-det]}
  \BibitemShut {NoStop}%
\bibitem [{\citenamefont {Adams}\ \emph {et~al.}(2019)\citenamefont {Adams}
  \emph {et~al.}}]{MicroBooNE:2018kka}%
  \BibitemOpen
  \bibfield  {author} {\bibinfo {author} {\bibfnamefont {C.}~\bibnamefont
  {Adams}} \emph {et~al.} (\bibinfo {collaboration} {MicroBooNE}),\ }\href
  {\doibase 10.1103/PhysRevD.99.092001} {\bibfield  {journal} {\bibinfo
  {journal} {Phys. Rev. D}\ }\textbf {\bibinfo {volume} {99}},\ \bibinfo
  {pages} {092001} (\bibinfo {year} {2019})},\ \Eprint
  {http://arxiv.org/abs/1808.07269} {arXiv:1808.07269 [hep-ex]} \BibitemShut
  {NoStop}%
\bibitem [{\citenamefont {Abratenko}\ \emph
  {et~al.}(2021{\natexlab{b}})\citenamefont {Abratenko} \emph
  {et~al.}}]{MicroBooNE:2020yze}%
  \BibitemOpen
  \bibfield  {author} {\bibinfo {author} {\bibfnamefont {P.}~\bibnamefont
  {Abratenko}} \emph {et~al.} (\bibinfo {collaboration} {MicroBooNE}),\ }\href
  {\doibase 10.1103/PhysRevD.103.052012} {\bibfield  {journal} {\bibinfo
  {journal} {Phys. Rev. D}\ }\textbf {\bibinfo {volume} {103}},\ \bibinfo
  {pages} {052012} (\bibinfo {year} {2021}{\natexlab{b}})},\ \Eprint
  {http://arxiv.org/abs/2012.08513} {arXiv:2012.08513 [physics.ins-det]}
  \BibitemShut {NoStop}%
\bibitem [{\citenamefont {Acciarri}\ \emph
  {et~al.}(2017{\natexlab{c}})\citenamefont {Acciarri} \emph
  {et~al.}}]{MicroBooNE:2016dpb}%
  \BibitemOpen
  \bibfield  {author} {\bibinfo {author} {\bibfnamefont {R.}~\bibnamefont
  {Acciarri}} \emph {et~al.} (\bibinfo {collaboration} {MicroBooNE}),\ }\href
  {\doibase 10.1088/1748-0221/12/03/P03011} {\bibfield  {journal} {\bibinfo
  {journal} {JINST}\ }\textbf {\bibinfo {volume} {12}},\ \bibinfo {pages}
  {P03011} (\bibinfo {year} {2017}{\natexlab{c}})},\ \Eprint
  {http://arxiv.org/abs/1611.05531} {arXiv:1611.05531 [physics.ins-det]}
  \BibitemShut {NoStop}%
\bibitem [{\citenamefont {Abratenko}\ \emph
  {et~al.}(2021{\natexlab{c}})\citenamefont {Abratenko} \emph
  {et~al.}}]{MicroBooNE:2020sar}%
  \BibitemOpen
  \bibfield  {author} {\bibinfo {author} {\bibfnamefont {P.}~\bibnamefont
  {Abratenko}} \emph {et~al.} (\bibinfo {collaboration} {MicroBooNE}),\ }\href
  {\doibase 10.1088/1748-0221/16/02/P02017} {\bibfield  {journal} {\bibinfo
  {journal} {JINST}\ }\textbf {\bibinfo {volume} {16}},\ \bibinfo {pages}
  {P02017} (\bibinfo {year} {2021}{\natexlab{c}})},\ \Eprint
  {http://arxiv.org/abs/2002.09375} {arXiv:2002.09375 [physics.ins-det]}
  \BibitemShut {NoStop}%
\bibitem [{\citenamefont {Abratenko}\ \emph
  {et~al.}(2021{\natexlab{d}})\citenamefont {Abratenko} \emph
  {et~al.}}]{MicroBooNE:2021nss}%
  \BibitemOpen
  \bibfield  {author} {\bibinfo {author} {\bibfnamefont {P.}~\bibnamefont
  {Abratenko}} \emph {et~al.} (\bibinfo {collaboration} {MicroBooNE}),\
  }\href@noop {} {\  (\bibinfo {year} {2021}{\natexlab{d}})},\ \Eprint
  {http://arxiv.org/abs/2110.11874} {arXiv:2110.11874 [hep-ex]} \BibitemShut
  {NoStop}%
\bibitem [{\citenamefont {Abratenko}\ \emph
  {et~al.}(2021{\natexlab{e}})\citenamefont {Abratenko} \emph
  {et~al.}}]{MicroBooNE:2020hho}%
  \BibitemOpen
  \bibfield  {author} {\bibinfo {author} {\bibfnamefont {P.}~\bibnamefont
  {Abratenko}} \emph {et~al.} (\bibinfo {collaboration} {MicroBooNE}),\ }\href
  {\doibase 10.1103/PhysRevD.103.092003} {\bibfield  {journal} {\bibinfo
  {journal} {Phys. Rev. D}\ }\textbf {\bibinfo {volume} {103}},\ \bibinfo
  {pages} {092003} (\bibinfo {year} {2021}{\natexlab{e}})},\ \Eprint
  {http://arxiv.org/abs/2010.08653} {arXiv:2010.08653 [hep-ex]} \BibitemShut
  {NoStop}%
\bibitem [{\citenamefont {Chen}\ and\ \citenamefont
  {Guestrin}(2016)}]{Chen_2016}%
  \BibitemOpen
  \bibfield  {author} {\bibinfo {author} {\bibfnamefont {T.}~\bibnamefont
  {Chen}}\ and\ \bibinfo {author} {\bibfnamefont {C.}~\bibnamefont
  {Guestrin}},\ }in\ \href {\doibase 10.1145/2939672.2939785} {\emph {\bibinfo
  {booktitle} {Proceedings of the 22nd ACM SIGKDD International Conference on
  Knowledge Discovery and Data Mining}}},\ \bibinfo {series and number} {KDD
  '16}\ (\bibinfo  {publisher} {Association for Computing Machinery},\ \bibinfo
  {address} {New York, NY, USA},\ \bibinfo {year} {2016})\ p.\ \bibinfo {pages}
  {785–794}\BibitemShut {NoStop}%
\bibitem [{\citenamefont {Yuan}\ \emph {et~al.}(2019)\citenamefont {Yuan},
  \citenamefont {Wang}, \citenamefont {Han}, \citenamefont {Liu}, \citenamefont
  {Chen}, \citenamefont {Zhang},\ and\ \citenamefont {Ye}}]{8855897}%
  \BibitemOpen
  \bibfield  {author} {\bibinfo {author} {\bibfnamefont {X.}~\bibnamefont
  {Yuan}}, \bibinfo {author} {\bibfnamefont {X.}~\bibnamefont {Wang}}, \bibinfo
  {author} {\bibfnamefont {J.}~\bibnamefont {Han}}, \bibinfo {author}
  {\bibfnamefont {J.}~\bibnamefont {Liu}}, \bibinfo {author} {\bibfnamefont
  {H.}~\bibnamefont {Chen}}, \bibinfo {author} {\bibfnamefont {K.}~\bibnamefont
  {Zhang}}, \ and\ \bibinfo {author} {\bibfnamefont {Q.}~\bibnamefont {Ye}},\
  }in\ \href {\doibase 10.1109/ICCChina.2019.8855897} {\emph {\bibinfo
  {booktitle} {2019 IEEE/CIC International Conference on Communications in
  China (ICCC)}}}\ (\bibinfo {year} {2019})\ pp.\ \bibinfo {pages}
  {467--471}\BibitemShut {NoStop}%
\bibitem [{\citenamefont {Liu}\ \emph {et~al.}(2019)\citenamefont {Liu},
  \citenamefont {Tan},\ and\ \citenamefont {Tang}}]{Liu_2019}%
  \BibitemOpen
  \bibfield  {author} {\bibinfo {author} {\bibfnamefont {X.}~\bibnamefont
  {Liu}}, \bibinfo {author} {\bibfnamefont {W.}~\bibnamefont {Tan}}, \ and\
  \bibinfo {author} {\bibfnamefont {S.}~\bibnamefont {Tang}},\ }\href {\doibase
  10.1088/1755-1315/237/2/022027} {\bibfield  {journal} {\bibinfo  {journal}
  {{IOP} Conference Series: Earth and Environmental Science}\ }\textbf
  {\bibinfo {volume} {237}},\ \bibinfo {pages} {022027} (\bibinfo {year}
  {2019})}\BibitemShut {NoStop}%
\bibitem [{\citenamefont {Alvarez-Ruso}\ \emph {et~al.}(2018)\citenamefont
  {Alvarez-Ruso} \emph {et~al.}}]{NuSTEC:2017hzk}%
  \BibitemOpen
  \bibfield  {author} {\bibinfo {author} {\bibfnamefont {L.}~\bibnamefont
  {Alvarez-Ruso}} \emph {et~al.} (\bibinfo {collaboration} {NuSTEC}),\ }\href
  {\doibase 10.1016/j.ppnp.2018.01.006} {\bibfield  {journal} {\bibinfo
  {journal} {Prog. Part. Nucl. Phys.}\ }\textbf {\bibinfo {volume} {100}},\
  \bibinfo {pages} {1} (\bibinfo {year} {2018})},\ \Eprint
  {http://arxiv.org/abs/1706.03621} {arXiv:1706.03621 [hep-ph]} \BibitemShut
  {NoStop}%
\bibitem [{\citenamefont {Abratenko}\ \emph
  {et~al.}(2021{\natexlab{f}})\citenamefont {Abratenko}, \citenamefont
  {Alrashed}, \citenamefont {An}, \citenamefont {Anthony}, \citenamefont
  {Asaadi}, \citenamefont {Ashkenazi}, \citenamefont {Balasubramanian},
  \citenamefont {Baller}, \citenamefont {Barnes}, \citenamefont {Barr},\ and\
  \citenamefont {et~al.}}]{MicroBooNE:2021ldh}%
  \BibitemOpen
  \bibfield  {author} {\bibinfo {author} {\bibfnamefont {P.}~\bibnamefont
  {Abratenko}}, \bibinfo {author} {\bibfnamefont {M.}~\bibnamefont {Alrashed}},
  \bibinfo {author} {\bibfnamefont {R.}~\bibnamefont {An}}, \bibinfo {author}
  {\bibfnamefont {J.}~\bibnamefont {Anthony}}, \bibinfo {author} {\bibfnamefont
  {J.}~\bibnamefont {Asaadi}}, \bibinfo {author} {\bibfnamefont
  {A.}~\bibnamefont {Ashkenazi}}, \bibinfo {author} {\bibfnamefont
  {S.}~\bibnamefont {Balasubramanian}}, \bibinfo {author} {\bibfnamefont
  {B.}~\bibnamefont {Baller}}, \bibinfo {author} {\bibfnamefont
  {C.}~\bibnamefont {Barnes}}, \bibinfo {author} {\bibfnamefont
  {G.}~\bibnamefont {Barr}}, \ and\ \bibinfo {author} {\bibnamefont {et~al.}}
  (\bibinfo {collaboration} {MicroBooNE}),\ }\href {\doibase
  10.1103/physrevd.104.052002} {\bibfield  {journal} {\bibinfo  {journal}
  {Physical Review D}\ }\textbf {\bibinfo {volume} {104}} (\bibinfo {year}
  {2021}{\natexlab{f}}),\ 10.1103/physrevd.104.052002}\BibitemShut {NoStop}%
\bibitem [{\citenamefont {Abratenko}\ \emph
  {et~al.}({\natexlab{d}})\citenamefont {Abratenko} \emph
  {et~al.}}]{MicroBooNE:2021genie_tune}%
  \BibitemOpen
  \bibfield  {author} {\bibinfo {author} {\bibfnamefont {P.}~\bibnamefont
  {Abratenko}} \emph {et~al.} (\bibinfo {collaboration} {MicroBooNE}),\
  }\Eprint {http://arxiv.org/abs/2110.14028} {arXiv:2110.14028 [hep-ex]}
  \BibitemShut {NoStop}%
\bibitem [{\citenamefont {Aguilar-Arevalo}\ \emph {et~al.}(2009)\citenamefont
  {Aguilar-Arevalo} \emph {et~al.}}]{MiniBooNE:2008hfu}%
  \BibitemOpen
  \bibfield  {author} {\bibinfo {author} {\bibfnamefont {A.~A.}\ \bibnamefont
  {Aguilar-Arevalo}} \emph {et~al.} (\bibinfo {collaboration} {MiniBooNE}),\
  }\href {\doibase 10.1103/PhysRevD.79.072002} {\bibfield  {journal} {\bibinfo
  {journal} {Phys. Rev. D}\ }\textbf {\bibinfo {volume} {79}},\ \bibinfo
  {pages} {072002} (\bibinfo {year} {2009})},\ \Eprint
  {http://arxiv.org/abs/0806.1449} {arXiv:0806.1449 [hep-ex]} \BibitemShut
  {NoStop}%
\bibitem [{\citenamefont {Abratenko}\ \emph {et~al.}(2019)\citenamefont
  {Abratenko} \emph {et~al.}}]{MicroBooNE:2019nio}%
  \BibitemOpen
  \bibfield  {author} {\bibinfo {author} {\bibfnamefont {P.}~\bibnamefont
  {Abratenko}} \emph {et~al.} (\bibinfo {collaboration} {MicroBooNE}),\ }\href
  {\doibase 10.1103/PhysRevLett.123.131801} {\bibfield  {journal} {\bibinfo
  {journal} {Phys. Rev. Lett.}\ }\textbf {\bibinfo {volume} {123}},\ \bibinfo
  {pages} {131801} (\bibinfo {year} {2019})},\ \Eprint
  {http://arxiv.org/abs/1905.09694} {arXiv:1905.09694 [hep-ex]} \BibitemShut
  {NoStop}%
\bibitem [{\citenamefont {Aguilar-Arevalo}\ \emph {et~al.}(2007)\citenamefont
  {Aguilar-Arevalo} \emph {et~al.}}]{MiniBooNE:2007uho}%
  \BibitemOpen
  \bibfield  {author} {\bibinfo {author} {\bibfnamefont {A.~A.}\ \bibnamefont
  {Aguilar-Arevalo}} \emph {et~al.} (\bibinfo {collaboration} {MiniBooNE}),\
  }\href {\doibase 10.1103/PhysRevLett.98.231801} {\bibfield  {journal}
  {\bibinfo  {journal} {Phys. Rev. Lett.}\ }\textbf {\bibinfo {volume} {98}},\
  \bibinfo {pages} {231801} (\bibinfo {year} {2007})},\ \Eprint
  {http://arxiv.org/abs/0704.1500} {arXiv:0704.1500 [hep-ex]} \BibitemShut
  {NoStop}%
\bibitem [{\citenamefont {Andreopoulos}\ \emph {et~al.}(2015)\citenamefont
  {Andreopoulos}, \citenamefont {Barry}, \citenamefont {Dytman}, \citenamefont
  {Gallagher}, \citenamefont {Golan}, \citenamefont {Hatcher}, \citenamefont
  {Perdue},\ and\ \citenamefont {Yarba}}]{Andreopoulos:2015wxa}%
  \BibitemOpen
  \bibfield  {author} {\bibinfo {author} {\bibfnamefont {C.}~\bibnamefont
  {Andreopoulos}}, \bibinfo {author} {\bibfnamefont {C.}~\bibnamefont {Barry}},
  \bibinfo {author} {\bibfnamefont {S.}~\bibnamefont {Dytman}}, \bibinfo
  {author} {\bibfnamefont {H.}~\bibnamefont {Gallagher}}, \bibinfo {author}
  {\bibfnamefont {T.}~\bibnamefont {Golan}}, \bibinfo {author} {\bibfnamefont
  {R.}~\bibnamefont {Hatcher}}, \bibinfo {author} {\bibfnamefont
  {G.}~\bibnamefont {Perdue}}, \ and\ \bibinfo {author} {\bibfnamefont
  {J.}~\bibnamefont {Yarba}},\ }\href@noop {} {\  (\bibinfo {year} {2015})},\
  \Eprint {http://arxiv.org/abs/1510.05494} {arXiv:1510.05494 [hep-ph]}
  \BibitemShut {NoStop}%
\bibitem [{\citenamefont {Abratenko}\ \emph
  {et~al.}({\natexlab{e}})\citenamefont {Abratenko} \emph
  {et~al.}}]{MicroBooNE:2021wiremod}%
  \BibitemOpen
  \bibfield  {author} {\bibinfo {author} {\bibfnamefont {P.}~\bibnamefont
  {Abratenko}} \emph {et~al.} (\bibinfo {collaboration} {MicroBooNE}),\
  }\bibinfo {note}
  {\url{https://microboone.fnal.gov/wp-content/uploads/MICROBOONE-NOTE-1075-PUB.pdf}}\BibitemShut
  {NoStop}%
\bibitem [{\citenamefont {Scott}(2018)}]{KDEcite}%
  \BibitemOpen
  \bibfield  {author} {\bibinfo {author} {\bibfnamefont {D.~W.}\ \bibnamefont
  {Scott}},\ }\enquote {\bibinfo {title} {Kernel density estimation},}\ in\
  \href {\doibase https://doi.org/10.1002/9781118445112.stat07186.pub2} {\emph
  {\bibinfo {booktitle} {Wiley StatsRef: Statistics Reference Online}}}\
  (\bibinfo  {publisher} {American Cancer Society},\ \bibinfo {year} {2018})\
  pp.\ \bibinfo {pages} {1--7}\BibitemShut {NoStop}%
\bibitem [{\citenamefont {Ji}\ \emph {et~al.}(2020)\citenamefont {Ji},
  \citenamefont {Gu}, \citenamefont {Qian}, \citenamefont {Wei},\ and\
  \citenamefont {Zhang}}]{Ji:2019yca}%
  \BibitemOpen
  \bibfield  {author} {\bibinfo {author} {\bibfnamefont {X.}~\bibnamefont
  {Ji}}, \bibinfo {author} {\bibfnamefont {W.}~\bibnamefont {Gu}}, \bibinfo
  {author} {\bibfnamefont {X.}~\bibnamefont {Qian}}, \bibinfo {author}
  {\bibfnamefont {H.}~\bibnamefont {Wei}}, \ and\ \bibinfo {author}
  {\bibfnamefont {C.}~\bibnamefont {Zhang}},\ }\href {\doibase
  10.1016/j.nima.2020.163677} {\bibfield  {journal} {\bibinfo  {journal} {Nucl.
  Instrum. Meth. A}\ }\textbf {\bibinfo {volume} {961}},\ \bibinfo {pages}
  {163677} (\bibinfo {year} {2020})},\ \Eprint
  {http://arxiv.org/abs/1903.07185} {arXiv:1903.07185 [physics.data-an]}
  \BibitemShut {NoStop}%
\bibitem [{\citenamefont {Adams}\ \emph {et~al.}(2020)\citenamefont {Adams}
  \emph {et~al.}}]{MicroBooNE:2019rgx}%
  \BibitemOpen
  \bibfield  {author} {\bibinfo {author} {\bibfnamefont {C.}~\bibnamefont
  {Adams}} \emph {et~al.} (\bibinfo {collaboration} {MicroBooNE}),\ }\href
  {\doibase 10.1088/1748-0221/15/02/P02007} {\bibfield  {journal} {\bibinfo
  {journal} {JINST}\ }\textbf {\bibinfo {volume} {15}},\ \bibinfo {pages}
  {P02007} (\bibinfo {year} {2020})},\ \Eprint
  {http://arxiv.org/abs/1910.02166} {arXiv:1910.02166 [hep-ex]} \BibitemShut
  {NoStop}%
\bibitem [{\citenamefont {Andreopoulos}\ \emph {et~al.}(2010)\citenamefont
  {Andreopoulos} \emph {et~al.}}]{Andreopoulos:2009rq}%
  \BibitemOpen
  \bibfield  {author} {\bibinfo {author} {\bibfnamefont {C.}~\bibnamefont
  {Andreopoulos}} \emph {et~al.},\ }\href {\doibase 10.1016/j.nima.2009.12.009}
  {\bibfield  {journal} {\bibinfo  {journal} {Nucl. Instrum. Meth. A}\ }\textbf
  {\bibinfo {volume} {614}},\ \bibinfo {pages} {87} (\bibinfo {year} {2010})},\
  \Eprint {http://arxiv.org/abs/0905.2517} {arXiv:0905.2517 [hep-ph]}
  \BibitemShut {NoStop}%
\bibitem [{\citenamefont {Alvarez-Ruso}\ \emph {et~al.}(2021)\citenamefont
  {Alvarez-Ruso} \emph {et~al.}}]{GENIE:2021npt}%
  \BibitemOpen
  \bibfield  {author} {\bibinfo {author} {\bibfnamefont {L.}~\bibnamefont
  {Alvarez-Ruso}} \emph {et~al.} (\bibinfo {collaboration} {GENIE}),\
  }\href@noop {} {\  (\bibinfo {year} {2021})},\ \Eprint
  {http://arxiv.org/abs/2106.09381} {arXiv:2106.09381 [hep-ph]} \BibitemShut
  {NoStop}%
\bibitem [{\citenamefont {Tena-Vidal}\ \emph {et~al.}(2021)\citenamefont
  {Tena-Vidal} \emph {et~al.}}]{Tena-Vidal:2021rpu}%
  \BibitemOpen
  \bibfield  {author} {\bibinfo {author} {\bibfnamefont {J.}~\bibnamefont
  {Tena-Vidal}} \emph {et~al.} (\bibinfo {collaboration} {GENIE}),\ }\href@noop
  {} {\  (\bibinfo {year} {2021})},\ \Eprint {http://arxiv.org/abs/2104.09179}
  {arXiv:2104.09179 [hep-ph]} \BibitemShut {NoStop}%
\bibitem [{\citenamefont {Nieves}\ \emph {et~al.}(2011)\citenamefont {Nieves},
  \citenamefont {Simo},\ and\ \citenamefont {Vacas}}]{ValenciaModel1}%
  \BibitemOpen
  \bibfield  {author} {\bibinfo {author} {\bibfnamefont {J.}~\bibnamefont
  {Nieves}}, \bibinfo {author} {\bibfnamefont {I.~R.}\ \bibnamefont {Simo}}, \
  and\ \bibinfo {author} {\bibfnamefont {M.~J.~V.}\ \bibnamefont {Vacas}},\
  }\href {\doibase 10.1103/PhysRevC.83.045501} {\bibfield  {journal} {\bibinfo
  {journal} {Phys. Rev. C}\ }\textbf {\bibinfo {volume} {83}},\ \bibinfo
  {pages} {045501} (\bibinfo {year} {2011})},\ \Eprint
  {http://arxiv.org/abs/1102.2777} {1102.2777} \BibitemShut {NoStop}%
\bibitem [{\citenamefont {Aguilar-Arevalo}\ \emph {et~al.}(2010)\citenamefont
  {Aguilar-Arevalo} \emph {et~al.}}]{miniboone-ccqe}%
  \BibitemOpen
  \bibfield  {author} {\bibinfo {author} {\bibfnamefont {A.}~\bibnamefont
  {Aguilar-Arevalo}} \emph {et~al.} (\bibinfo {collaboration} {MiniBooNE}),\
  }\href {\doibase 10.1103/PhysRevD.81.092005} {\bibfield  {journal} {\bibinfo
  {journal} {Phys. Rev. D}\ }\textbf {\bibinfo {volume} {81}},\ \bibinfo
  {pages} {092005} (\bibinfo {year} {2010})},\ \Eprint
  {http://arxiv.org/abs/1002.2680} {arXiv:1002.2680 [hep-ex]} \BibitemShut
  {NoStop}%
\bibitem [{\citenamefont {Abe}\ \emph {et~al.}(2016)\citenamefont {Abe},
  \citenamefont {Andreopoulos}, \citenamefont {Antonova}, \citenamefont {Aoki},
  \citenamefont {Ariga}, \citenamefont {Assylbekov}, \citenamefont {Autiero},
  \citenamefont {Barbi}, \citenamefont {Barker}, \citenamefont {Barr} \emph
  {et~al.}}]{t2k2016}%
  \BibitemOpen
  \bibfield  {author} {\bibinfo {author} {\bibfnamefont {K.}~\bibnamefont
  {Abe}}, \bibinfo {author} {\bibfnamefont {C.}~\bibnamefont {Andreopoulos}},
  \bibinfo {author} {\bibfnamefont {M.}~\bibnamefont {Antonova}}, \bibinfo
  {author} {\bibfnamefont {S.}~\bibnamefont {Aoki}}, \bibinfo {author}
  {\bibfnamefont {A.}~\bibnamefont {Ariga}}, \bibinfo {author} {\bibfnamefont
  {S.}~\bibnamefont {Assylbekov}}, \bibinfo {author} {\bibfnamefont
  {D.}~\bibnamefont {Autiero}}, \bibinfo {author} {\bibfnamefont
  {M.}~\bibnamefont {Barbi}}, \bibinfo {author} {\bibfnamefont {G.~J.}\
  \bibnamefont {Barker}}, \bibinfo {author} {\bibfnamefont {G.}~\bibnamefont
  {Barr}},  \emph {et~al.} (\bibinfo {collaboration} {{T2K}}),\ }\href
  {\doibase 10.1103/PhysRevD.93.112012} {\bibfield  {journal} {\bibinfo
  {journal} {Phys. Rev. D}\ }\textbf {\bibinfo {volume} {93}},\ \bibinfo
  {pages} {112012} (\bibinfo {year} {2016})}\BibitemShut {NoStop}%
\bibitem [{\citenamefont {Agostinelli}\ \emph {et~al.}(2003)\citenamefont
  {Agostinelli} \emph {et~al.}}]{Agostinelli:2002hh}%
  \BibitemOpen
  \bibfield  {author} {\bibinfo {author} {\bibfnamefont {S.}~\bibnamefont
  {Agostinelli}} \emph {et~al.} (\bibinfo {collaboration} {GEANT4}),\ }\href
  {\doibase 10.1016/S0168-9002(03)01368-8} {\bibfield  {journal} {\bibinfo
  {journal} {Nucl. Instrum. Meth. A}\ }\textbf {\bibinfo {volume} {506}},\
  \bibinfo {pages} {250} (\bibinfo {year} {2003})}\BibitemShut {NoStop}%
\bibitem [{\citenamefont {Ankowski}\ and\ \citenamefont
  {Sobczyk}(2006)}]{Ankowski:2005wi}%
  \BibitemOpen
  \bibfield  {author} {\bibinfo {author} {\bibfnamefont {A.~M.}\ \bibnamefont
  {Ankowski}}\ and\ \bibinfo {author} {\bibfnamefont {J.~T.}\ \bibnamefont
  {Sobczyk}},\ }\href {\doibase 10.1103/PhysRevC.74.054316} {\bibfield
  {journal} {\bibinfo  {journal} {Phys. Rev. C}\ }\textbf {\bibinfo {volume}
  {74}},\ \bibinfo {pages} {054316} (\bibinfo {year} {2006})},\ \Eprint
  {http://arxiv.org/abs/nucl-th/0512004} {arXiv:nucl-th/0512004} \BibitemShut
  {NoStop}%
\bibitem [{\citenamefont {Abratenko}\ \emph
  {et~al.}(2021{\natexlab{g}})\citenamefont {Abratenko} \emph
  {et~al.}}]{WCJINST}%
  \BibitemOpen
  \bibfield  {author} {\bibinfo {author} {\bibfnamefont {P.}~\bibnamefont
  {Abratenko}} \emph {et~al.},\ }\href {\doibase
  10.1088/1748-0221/16/06/p06043} {\bibfield  {journal} {\bibinfo  {journal}
  {Journal of Instrumentation}\ }\textbf {\bibinfo {volume} {16}},\ \bibinfo
  {pages} {P06043} (\bibinfo {year} {2021}{\natexlab{g}})}\BibitemShut
  {NoStop}%
\bibitem [{\citenamefont {Abratenko}\ \emph
  {et~al.}(2021{\natexlab{h}})\citenamefont {Abratenko} \emph
  {et~al.}}]{MicroBooNE:2021zul}%
  \BibitemOpen
  \bibfield  {author} {\bibinfo {author} {\bibfnamefont {P.}~\bibnamefont
  {Abratenko}} \emph {et~al.} (\bibinfo {collaboration} {MicroBooNE}),\ }\href
  {\doibase 10.1103/PhysRevApplied.15.064071} {\bibfield  {journal} {\bibinfo
  {journal} {Phys. Rev. Applied}\ }\textbf {\bibinfo {volume} {15}},\ \bibinfo
  {pages} {064071} (\bibinfo {year} {2021}{\natexlab{h}})},\ \Eprint
  {http://arxiv.org/abs/2101.05076} {arXiv:2101.05076 [physics.ins-det]}
  \BibitemShut {NoStop}%
\bibitem [{\citenamefont {Genty}(2019)}]{rVicThesis}%
  \BibitemOpen
  \bibfield  {author} {\bibinfo {author} {\bibfnamefont {V.}~\bibnamefont
  {Genty}},\ }\emph {\bibinfo {title} {The MicroBooNE Search For Anomalous
  Electron Neutrino Appearance Using Image Based Data Reconstruction}},\
  \href@noop {} {Ph.D. thesis},\ \bibinfo  {school} {Columbia University}
  (\bibinfo {year} {2019})\BibitemShut {NoStop}%
\bibitem [{pdg()}]{pdgMuonAr}%
  \BibitemOpen
  \href@noop {} {}\bibinfo {howpublished}
  {\url{https://pdg.lbl.gov/2017/AtomicNuclearProperties/HTML/liquid_argon.html}}\BibitemShut
  {NoStop}%
\bibitem [{Nis()}]{NistAr}%
  \BibitemOpen
  \href@noop {} {}\bibinfo {howpublished}
  {\url{https://physics.nist.gov/PhysRefData/Star/Text/PSTAR.html}}\BibitemShut
  {NoStop}%
\bibitem [{Sup()}]{Supp}%
  \BibitemOpen
  \href@noop {} {\enquote {\bibinfo {title} {Supplemental materials},}\
  }\bibinfo {howpublished} {for more validation and detailed studies regarding
  this analysis see
  \url{https://journals.aps.org/prd/abstract/10.1103/PhysRevD.105.112003##supplemental}}\BibitemShut
  {NoStop}%
\bibitem [{\citenamefont {Ganjisaffar}\ \emph {et~al.}(2011)\citenamefont
  {Ganjisaffar}, \citenamefont {Caruana},\ and\ \citenamefont
  {Lopes}}]{ganjisaffar2011bagging}%
  \BibitemOpen
  \bibfield  {author} {\bibinfo {author} {\bibfnamefont {Y.}~\bibnamefont
  {Ganjisaffar}}, \bibinfo {author} {\bibfnamefont {R.}~\bibnamefont
  {Caruana}}, \ and\ \bibinfo {author} {\bibfnamefont {C.~V.}\ \bibnamefont
  {Lopes}},\ }in\ \href@noop {} {\emph {\bibinfo {booktitle} {Proceedings of
  the 34th international ACM SIGIR conference on Research and development in
  Information Retrieval}}}\ (\bibinfo {year} {2011})\ pp.\ \bibinfo {pages}
  {85--94}\BibitemShut {NoStop}%
\bibitem [{\citenamefont {Moyal}(1949)}]{moyalapprox}%
  \BibitemOpen
  \bibfield  {author} {\bibinfo {author} {\bibfnamefont {J.~E.}\ \bibnamefont
  {Moyal}},\ }\href {\doibase
  https://doi.org/10.1111/j.2517-6161.1949.tb00030.x} {\bibfield  {journal}
  {\bibinfo  {journal} {Journal of the Royal Statistical Society: Series B
  (Methodological)}\ }\textbf {\bibinfo {volume} {11}},\ \bibinfo {pages} {150}
  (\bibinfo {year} {1949})},\ \Eprint
  {http://arxiv.org/abs/https://rss.onlinelibrary.wiley.com/doi/pdf/10.1111/j.2517-6161.1949.tb00030.x}
  {https://rss.onlinelibrary.wiley.com/doi/pdf/10.1111/j.2517-6161.1949.tb00030.x}
  \BibitemShut {NoStop}%
\bibitem [{\citenamefont {Higham}(2009)}]{cholesky}%
  \BibitemOpen
  \bibfield  {author} {\bibinfo {author} {\bibfnamefont {N.}~\bibnamefont
  {Higham}},\ }\href {\doibase 10.1002/wics.18} {\bibfield  {journal} {\bibinfo
   {journal} {Wiley Interdisciplinary Reviews: Computational Statistics}\
  }\textbf {\bibinfo {volume} {1}},\ \bibinfo {pages} {251 } (\bibinfo {year}
  {2009})}\BibitemShut {NoStop}%
\bibitem [{\citenamefont {Nguyen}(2008)}]{ref:MinibooneReweight}%
  \BibitemOpen
  \bibfield  {author} {\bibinfo {author} {\bibfnamefont {V.}~\bibnamefont
  {Nguyen}} (\bibinfo {collaboration} {MiniBooNE}),\ }\href {\doibase
  10.1063/1.2898950} {\bibfield  {journal} {\bibinfo  {journal} {AIP
  Conf.Proc}\ }\textbf {\bibinfo {volume} {981}},\ \bibinfo {pages} {250}
  (\bibinfo {year} {2008})},\ \Eprint {http://arxiv.org/abs/0801.0628}
  {arXiv:0801.0628 [hep-ex]} \BibitemShut {NoStop}%
\bibitem [{\citenamefont {Day}\ and\ \citenamefont
  {McFarland}(2012)}]{Day:2012}%
  \BibitemOpen
  \bibfield  {author} {\bibinfo {author} {\bibfnamefont {M.}~\bibnamefont
  {Day}}\ and\ \bibinfo {author} {\bibfnamefont {K.~S.}\ \bibnamefont
  {McFarland}},\ }\href {\doibase 10.1103/PhysRevD.86.053003} {\bibfield
  {journal} {\bibinfo  {journal} {Phys. Rev. D}\ }\textbf {\bibinfo {volume}
  {86}},\ \bibinfo {pages} {053003} (\bibinfo {year} {2012})},\ \Eprint
  {http://arxiv.org/abs/1206.6745} {arXiv:1206.6745 [hep-ph]} \BibitemShut
  {NoStop}%
\bibitem [{\citenamefont {Calcutt}\ \emph {et~al.}(2021)\citenamefont
  {Calcutt}, \citenamefont {Thorpe}, \citenamefont {Mahn},\ and\ \citenamefont
  {Fields}}]{geant4reweight:2021}%
  \BibitemOpen
  \bibfield  {author} {\bibinfo {author} {\bibfnamefont {J.}~\bibnamefont
  {Calcutt}}, \bibinfo {author} {\bibfnamefont {C.}~\bibnamefont {Thorpe}},
  \bibinfo {author} {\bibfnamefont {K.}~\bibnamefont {Mahn}}, \ and\ \bibinfo
  {author} {\bibfnamefont {L.}~\bibnamefont {Fields}},\ }\href {\doibase
  10.1088/1748-0221/16/08/p08042} {\bibfield  {journal} {\bibinfo  {journal}
  {JINST}\ }\textbf {\bibinfo {volume} {16}},\ \bibinfo {pages} {P08042}
  (\bibinfo {year} {2021})},\ \Eprint {http://arxiv.org/abs/2105.01744}
  {arXiv:2105.01744 [physics.data-an]} \BibitemShut {NoStop}%
\bibitem [{\citenamefont {Epanechnikov}(1969)}]{epach}%
  \BibitemOpen
  \bibfield  {author} {\bibinfo {author} {\bibfnamefont {V.~A.}\ \bibnamefont
  {Epanechnikov}},\ }\href {\doibase 10.1137/1114019} {\bibfield  {journal}
  {\bibinfo  {journal} {Theory of Probability \& Its Applications}\ }\textbf
  {\bibinfo {volume} {14}},\ \bibinfo {pages} {153} (\bibinfo {year} {1969})},\
  \Eprint {http://arxiv.org/abs/https://doi.org/10.1137/1114019}
  {https://doi.org/10.1137/1114019} \BibitemShut {NoStop}%
\bibitem [{\citenamefont {Sheather}\ and\ \citenamefont {Jones}(1991)}]{SJ}%
  \BibitemOpen
  \bibfield  {author} {\bibinfo {author} {\bibfnamefont {S.~J.}\ \bibnamefont
  {Sheather}}\ and\ \bibinfo {author} {\bibfnamefont {M.~C.}\ \bibnamefont
  {Jones}},\ }\href {\doibase
  https://doi.org/10.1111/j.2517-6161.1991.tb01857.x} {\bibfield  {journal}
  {\bibinfo  {journal} {Journal of the Royal Statistical Society: Series B
  (Methodological)}\ }\textbf {\bibinfo {volume} {53}},\ \bibinfo {pages} {683}
  (\bibinfo {year} {1991})},\ \Eprint
  {http://arxiv.org/abs/https://rss.onlinelibrary.wiley.com/doi/pdf/10.1111/j.2517-6161.1991.tb01857.x}
  {https://rss.onlinelibrary.wiley.com/doi/pdf/10.1111/j.2517-6161.1991.tb01857.x}
  \BibitemShut {NoStop}%
\bibitem [{\citenamefont {Cianci}\ and\ \citenamefont
  {Ross-Lonergan}()}]{sbnfit}%
  \BibitemOpen
  \bibfield  {author} {\bibinfo {author} {\bibfnamefont {D.}~\bibnamefont
  {Cianci}}\ and\ \bibinfo {author} {\bibfnamefont {M.}~\bibnamefont
  {Ross-Lonergan}},\ }\href@noop {} {\enquote {\bibinfo {title}
  {\uppercase{SBN}fit},}\ }\bibinfo {howpublished} {Available at
  \url{https://github.com/NevisUB/whipping_star}(2021)}\BibitemShut {NoStop}%
\bibitem [{\citenamefont {Feldman}\ and\ \citenamefont
  {Cousins}(1998)}]{Feldman:1998}%
  \BibitemOpen
  \bibfield  {author} {\bibinfo {author} {\bibfnamefont {G.}~\bibnamefont
  {Feldman}}\ and\ \bibinfo {author} {\bibfnamefont {R.}~\bibnamefont
  {Cousins}},\ }\href@noop {} {\bibfield  {journal} {\bibinfo  {journal} {Phys.
  Rev. D}\ }\textbf {\bibinfo {volume} {57}} (\bibinfo {year} {1998})},\
  \Eprint {http://arxiv.org/abs/physics/9711021} {arXiv:physics/9711021
  [physics.data-an]} \BibitemShut {NoStop}%
\bibitem [{\citenamefont {Junk}(1999)}]{Junk:1999}%
  \BibitemOpen
  \bibfield  {author} {\bibinfo {author} {\bibfnamefont {T.}~\bibnamefont
  {Junk}},\ }\href {\doibase https://doi.org/10.1016/S0168-9002(99)00498-2}
  {\bibfield  {journal} {\bibinfo  {journal} {Nucl.\ Instrum.\ Meth.\ A}\
  }\textbf {\bibinfo {volume} {434}},\ \bibinfo {pages} {435} (\bibinfo {year}
  {1999})},\ \Eprint {http://arxiv.org/abs/hep-ex/9902006}
  {arXiv:hep-ex/9902006 [hep-ex]} \BibitemShut {NoStop}%
\bibitem [{\citenamefont {Read}(2000)}]{Read:2000}%
  \BibitemOpen
  \bibfield  {author} {\bibinfo {author} {\bibfnamefont {A.~L.}\ \bibnamefont
  {Read}},\ }in\ \href {\doibase https://doi.org/10.5170/CERN-2000-005.81}
  {\emph {\bibinfo {booktitle} {Workshop on Confidence Limits}}}\ (\bibinfo
  {year} {2000})\ pp.\ \bibinfo {pages} {81--101}\BibitemShut {NoStop}%
\end{thebibliography}%

\end{document}